\documentclass[11pt, a4paper, oneside]{Thesis} % Paper size, default font size and one-sided paper

\usepackage{wrapfig}
\usepackage{lscape}
\usepackage{rotating}
\usepackage{graphicx}
\usepackage{caption}
\usepackage{subfigure}
\usepackage{amsmath}
\DeclareMathOperator*{\argmax}{arg\,max}
\DeclareMathOperator*{\argmin}{arg\,min}
\usepackage{gensymb}
\usepackage[section]{placeins}
\usepackage{relsize}
\usepackage{mathtools}
\DeclarePairedDelimiterX{\norm}[1]{\lVert}{\rVert}{#1}
\DeclarePairedDelimiter{\ceil}{\lceil}{\rceil}
\usepackage{mydefs}
\usepackage{hyperref}
\usepackage[table,xcdraw]{xcolor}
\usepackage{tabularx}
\usepackage{listings}
\usepackage{color}
\usepackage{float}
\usepackage{csquotes}
\usepackage{tikz}
\usepackage{rotating}
\usepackage{algorithmicx}
\usepackage{algorithm}
\usepackage{pdfpages}
\usepackage{multirow}
\usepackage[noend]{algpseudocode}
\definecolor{codegreen}{rgb}{0,0.6,0}
\definecolor{codegray}{rgb}{0.5,0.5,0.5}
\definecolor{codepurple}{rgb}{0.58,0,0.82}
\definecolor{backcolour}{rgb}{0.99,0.99,0.99}

\lstdefinestyle{mystyle}{
    backgroundcolor=\color{backcolour},   
    commentstyle=\color{codegreen},
    keywordstyle=\color{magenta},
    numberstyle=\tiny\color{codegray},
    stringstyle=\color{codepurple},
    basicstyle=\footnotesize,
    breakatwhitespace=false,         
    breaklines=true,                 
    captionpos=b,                    
    keepspaces=true,                 
    numbers=left,                    
    numbersep=5pt,                  
    showspaces=false,                
    showstringspaces=false,
    showtabs=false,                  
    tabsize=2
}
 
\graphicspath{{Pictures/}} % Specifies the directory where pictures are stored

\hypersetup{urlcolor=black, colorlinks=true} % Colors hyperlinks in blue - change to black if annoyingv`	
\title{\ttitle} % Defines the thesis title - don't touch this

\begin{document}

% \makeatletter
% \renewcommand*{\NAT@nmfmt}[1]{\textsc{#1}}
% \makeatother

% prints author names as small caps

\frontmatter % Use roman page numbering style (i, ii, iii, iv...) for the pre-content pages

\setstretch{1.6} % Line spacing of 1.6 (double line spacing)

% Define the page headers using the FancyHdr package and set up for one-sided printing
\fancyhead{} % Clears all page headers and footers
\rhead{\thepage} % Sets the right side header to show the page number
\lhead{} % Clears the left side page header

\pagestyle{fancy} % Finally, use the "fancy" page style to implement the FancyHdr headers

\newcommand{\HRule}{\rule{\linewidth}{0.5mm}} % New command to make the lines in the title page

% PDF meta-data
\hypersetup{pdftitle={\ttitle}}
\hypersetup{pdfsubject=\subjectname}
\hypersetup{pdfauthor=\authornames}
\hypersetup{pdfkeywords=\keywordnames}

%----------------------------------------------------------------------------------------
%	TITLE PAGE
%----------------------------------------------------------------------------------------

\begin{titlepage}
\begin{center}

\HRule \\[0.4cm] % Horizontal line
{\huge \bfseries \ttitle}\\[0.4cm] % Thesis title
\HRule \\[1.5cm] % Horizontal line
 
\large \textit{A thesis submitted in fulfillment of the requirements\\ for the degree of \degreename}\\[0.3cm] % University requirement text
\textit{by}\\[0.4cm]

\href{https://ee.iitk.ac.in/users/mandal/}{\textbf{\authornames}}\\

\vfill
\graphicspath{{./Figures/}}
\begin{figure}[hb]
  \centering
  \includegraphics[width=0.4\linewidth]{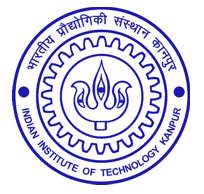}
\end{figure}

\DEPTNAME\\ % Research group name and department name
\textsc{\UNIVNAME}\\[1.5cm] % University name
\large {August 2019}\\[2cm] % Date

\end{center}

\end{titlepage}

%----------------------------------------------------------------------------------------
%	DECLARATION PAGE
%	Your institution may give you a different text to place here
%----------------------------------------------------------------------------------------
% \newpage
% \thispagestyle{empty}
% \mbox{}
% \includepdf{certificate.pdf}
% \includepdf{declaration.pdf}
% \newpage
% \thispagestyle{empty}
% \mbox{}

% \newpage
% \thispagestyle{empty}
% \includepdf[pages=-,offset=75 -75]{certificate_seal.pdf}

% \begin{center}
% {\bf CERTIFICATE}
% \end{center}
% %\vspace{2.5cm}
% \par         It is  certified  that the work contained in the thesis entitled
%     ``{\em Optimum Methods for Quasi-Orthographic
% Surface Imaging }''
%      by {\em Maniratnam Mandal } has been carried out under
%     my supervision and that this work has not been submitted elsewhere for a degree.
% \vspace*{2.5cm}\\
% August 2019
% \hspace*{5.8cm}Dr. K. S. Venkatesh\\
% \hspace*{8cm}Professor,\\
% \hspace*{8cm}Department of Electrical Engineering,\\
% \hspace*{8cm}Indian Institute of Technology,\\
% \hspace*{8cm}Kanpur-208016. \\
\clearpage % Start a new page

%----------------------------------------------------------------------------------------
%	ABSTRACT PAGE
%----------------------------------------------------------------------------------------

\addtotoc{Abstract} % Add the "Abstract" page entry to the Contents

\abstract{\addtocontents{toc}{\vspace{1em}} % Add a gap in the Contents, for aesthetics
 
}

\clearpage % Start a new page

%----------------------------------------------------------------------------------------
%	ACKNOWLEDGEMENTS
%----------------------------------------------------------------------------------------

\setstretch{1.3} % Reset the line-spacing to 1.3 for body text (if it has changed)

\acknowledgements{\addtocontents{toc}{\vspace{1em}} % Add a gap in the Contents, for aesthetics
I would like to express my sincerest gratitude to Prof K S Venkatesh of the Department of Electrical Engineering, who mentored me throughout my Masters and without whom, this thesis would not have been possible. I am extremely grateful for his supervision and enthusiastic guidance in both academic and non-academic aspects of my study. I tried to imbibe every ounce of knowledge that he imparted on my academic journey here at IIT Kanpur.
\\\\
I would like to thank all the professors here in Electrical Engineering Department who introduced us to the field and provided the technical knowledge to navigate it, which will remain an asset throughout our lifetime. I specially thank Prof Ketan Rajawat for his invaluable input in my thesis. 
\\\\
I would like to express my gratitude to my family who always had faith in me and supported  me in each of my endeavours throughout my life, and it is because of them I have the privilege of standing here today. I thank all of my CV Lab colleagues for creating a jovial and productive atmosphere for work. I would also like to thank my classmates for making my stay at IIT Kanpur productive and memorable. 
\begin{flushright}
Maniratnam Mandal
\end{flushright}

}

\clearpage % Start a new page

%----------------------------------------------------------------------------------------
%	LIST OF CONTENTS/FIGURES/TABLES PAGES
%----------------------------------------------------------------------------------------

\pagestyle{fancy} % The page style headers have been "empty" all this time, now use the "fancy" headers as defined before to bring them back

\lhead{\emph{Contents}} % Set the left side page header to "Contents"
\tableofcontents % Write out the Table of Contents

\lhead{\emph{List of Figures}} % Set the left side page header to "List of Figures"
\listoffigures % Write out the List of Figures

\lhead{\emph{List of Tables}} % Set the left side page header to "List of Tables"
\listoftables % Write out the List of Tables

\chapter*{List of Abbreviations}\label{RefSymbols}
  \begin{longtable}{lp{0.75\textwidth}}
    DEM & Digital Elevation Model \\
    GPS & Global Positioning System \\
    FOV & Field Of View \\
    WD  & Working Dsitance \\
    DOF  & Depth Of Field \\
    PMAG  & Primary MAGnification \\
	COP  & Centre Of Projection\\
    DTEM  & Digital Terrain Elevation Map \\
    BF  & Batch Filling \\
    SF & Sequential Filling \\
    IH  & Imaging Height \\
    MS  & Multi Start \\
    GS  & Global Search \\
\end{longtable}
\addcontentsline{toc}{chapter}{List of Abbreviations}
\clearpage
%------------------------------------------------------------------------
%	PHYSICAL CONSTANTS/OTHER DEFINITIONS
%----------------------------------------------------------------------------------------
%
%\clearpage % Start a new page
%
%\lhead{\emph{Physical Constants}} % Set the left side page header to "Physical Constants"
%
%\listofconstants{lrcl} % Include a list of Physical Constants (a four column table)
%{
%Speed of Light & $c$ & $=$ & $2.997\ 924\ 58\times10^{8}\ \mbox{ms}^{-\mbox{s}}$ (exact)\\
%% Constant Name & Symbol & = & Constant Value (with units) \\
%}

%----------------------------------------------------------------------------------------
%	DEDICATION
%----------------------------------------------------------------------------------------
%
\setstretch{1.3} % Return the line spacing back to 1.3
\pagestyle{empty} % Page style needs to be empty for this page
\dedicatory{Dedicated to Science} % Dedication text
\addtocontents{toc}{\vspace{2em}} % Add a gap in the Contents, for aesthetics

%----------------------------------------------------------------------------------------
%	THESIS CONTENT - CHAPTERS
%----------------------------------------------------------------------------------------

\mainmatter % Begin numeric (1,2,3...) page numbering

\pagestyle{fancy} % Return the page headers back to the "fancy" style

% Include the chapters of the thesis as separate files from the Chapters folder
% Uncomment the lines as you write the chapters

% Chapter Template

\chapter{Introduction} % Main chapter title

\label{Chapter1} % Change X to a consecutive number; for referencing this chapter elsewhere, use \ref{ChapterX}

\lhead{Chapter 1. \emph{Introduction}} % Change X to a consecutive number; this is for the header on each page - perhaps a shortened title

A lot of surveillance application today deals with capturing natural terrains, reconstructing the surface using optimization techniques and creating detailed terrain maps for study and survey. The capturing technology has progressed a lot over the past decade and nowadays a mixture of visual, radio, infrared, laser and radar sensors are used to capture terrain information \cite{terrainmap2001}. However, for vast terrains and when very small features are the focus of studies, it is easy to lose them while reconstruction or even while capturing. Here the question of the reliability of capturing devices becomes significant, and thus the need for a technique which is successful in achieving that is necessary. Often visual data is more reliable in case of small features and thus one of the primary objective should be to create a visual map of the terrain which reliably covers the entirety of the surface.

A major problem with that is to capture points on the surface which are otherwise occluded by the non-smooth features i.e. uneven landforms. Ideally, to capture all points on the terrain, an idealistic orthographic projection of the whole surface needs to be constructed, which means taking sensor data at every point on the surface. This approach, although will result in ideal reconstruction, but is highly impractical, implausible and incomprehensible due to resource limitations. Thus the need for an algorithm arises which, given the surface description, calculates the best(optimal) points from which the data should be captured for a quasi-orthographic projection to be reconstructed. Research in the aforementioned problem is almost absent in the academia and this study tries to make a beginning. This thesis provides some novel techniques to achieve that goal and provides the comparison among those techniques.
%----------------------------------------------------------------------------------------
%	SECTION 1

%----------------------------------------------------------------------------------------
\section{Motivation}
Terrain plays a fundamental role in modulating Earth surface and capturing environmental data. Terrain mapping using visual data is one of the most important technological problems in modern science. Be it Google Maps, Navigation, Geological surveys, agriculture, disaster management, Astronomical studies and mapping of extra-terrestrial surfaces, Archaeological studies, Sociological studies or even deciding national policies, it is imperative to accept the importance of studying and generating visual representation of topographical data. Also, the assessment and management of natural resources depends on the accuracy and coverage of surface mapping, which can be achieved by incorporating appropriate dependencies on terrain. Thus representations of terrain plays a central role in environmental modelling and landscape visualization.

The integral parts of Digital Terrain Modelling includes collection of topographical data, digital elevation model (DEM) interpolation and filtering methods. In terrain modelling, accurate representation of surface shape is a common requirement, and is usually facilitated by development of locally-adaptive, process-based DEM interpolation techniques \cite{terrainrepresent1999}. Here, traditional contour data sources and remotely-sensed data sources also play a significant role. The main tasks in digital terrain modelling is summarized as follows:
\begin{enumerate}
    \item Elevation Data Capture: 
     Digital elevation data capture has improved immensely following recent developments in airborne and spaceborne remote sensing, such as laser and synthetic aperture radar systems, and the development of the Global Positioning System(GPS) for ground data survey \cite{langgps1999}. Elevation contours continue to be the principal data source for the interpolation of DEMs and themselves are useful representations of terrain. They are widely available from existing topographic maps and although often error prone, can produce high accuracy elevation maps if coupled with proper interpolation techniques.
     \item DEM Generation:
     The crux of terrain modelling is the interpolation and filtering of DEM data, but the methods are now applied to a wider variety of data sources. These include traditional data sources such as points, profiles, contours, stream-lines, and break-lines, for which specific interpolation techniques have been developed, and remotely-sensed elevation data, for which various filtering procedures are required. The task of DEM generation includes a variety of associated DEM manipulation tasks such as DEM editing, DEM resampling, and data structure conversion between regular grids and triangulated irregular networks (TINs), the two dominant forms of terrain representation.
     \item DEM Interpretation: 
     Interpretation of DEMs includes scale analyses, terrain parameters, and a variety of terrain features that can be constructed from DEMs. Scale and resolution analysis deals with the choice of scale or grid resolution which in itself is a trade-off between achieving fidelity of the true surface and practical limits on density of information. Terrain parameters, or topographic indices, are descriptions of the surface form that can be computed directly at every point on a DEM, whereas, terrain features are associated with structures such as landscape features like mountain ranges, ridges, catchments, rivers and valleys.
     \item DEM Visualization: 
     Visualization of DEMs can provide subjective assessments, such as perspective views and inter-visibility analyses for various planning and monitoring applications. Visualizations of DEMs draped with various textures can also provide valuable insight into the nature of the processes being represented. They are an essential component of many virtual environment systems.
\end{enumerate}

\begin{figure}[hb]
  \centering
  \includegraphics[width=0.5\linewidth]{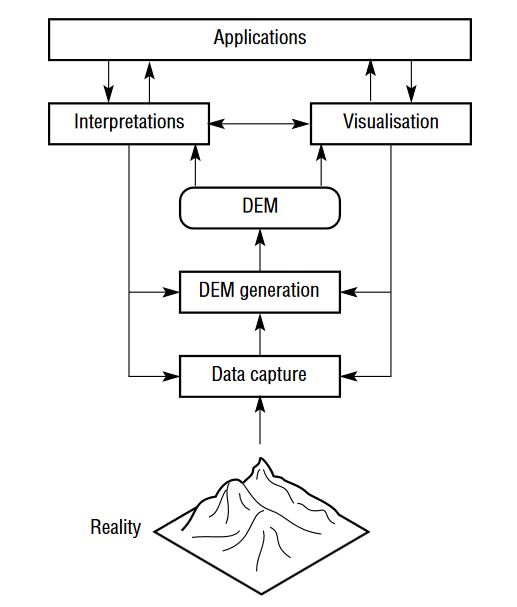}
  \caption{\textit{The main tasks in digital terrain modelling}}
  \label{fig:1}
\end{figure}

%-------------------------------------
% SUBFIGURES
% \begin{figure}
% \centering
% \begin{subfigure}{.5\textwidth}
%   \centering
%   \includegraphics[width=.4\linewidth]{image1}
%   \caption{A subfigure}
%   \label{fig:sub1}
% \end{subfigure}%
% \begin{subfigure}{.5\textwidth}
%   \centering
%   \includegraphics[width=.4\linewidth]{image1}
%   \caption{A subfigure}
%   \label{fig:sub2}
% \end{subfigure}
% \caption{A figure with two subfigures}
% \label{fig:test}
% \end{figure}
%---------------------------------------
DEM generation procedures are guided by both nature of the data source and also the application of the generated data or the targeted surface \cite{terrainrepresent1999}. Sometimes, the accurate representations of the surface shape and features is more vital as compared to the absolute elevation accuracy. There are three main classes of source elevation data:

\begin{enumerate}
    \item Surface-specific point elevation data- This included high and low points, saddle points and points on streams and ridges that make up the skeleton of the terrain. They are an ideal data source for most interpolation techniques, including triangulation methods and specially adapted gridding methods.These data are usually obtained by ground surveys or manually assisted photogrammetric stereo models.
    \begin{figure}[hb]
      \centering
      \includegraphics[width=0.5\linewidth]{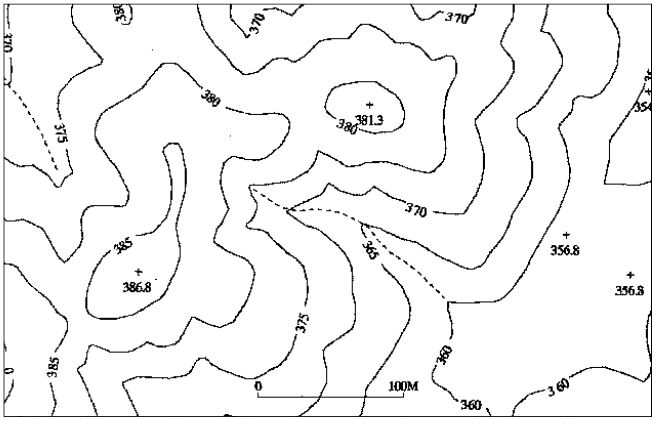}
      \caption{\textit{Contour, stream and point elevation data\\ (Source: \textit{Representation of Terrain} by M F Hutchinson and J C Gallant)}}
      \label{fig:2}
    \end{figure}
    \item Contour and stream-line data- These are the most common source of data for large terrains and the only available data source for many existing topographic maps which have been digitized. 
    \item Remotely-sensed elevation data- Gridded DEMs are calculated directly by stereoscopic interpretation of data collected by airborne and satellite sensors. The traditional source is often aerial photography or drone cameras. These are the fastest method of data collection.
    \begin{figure}[hb]
      \centering
      \includegraphics[width=0.5\linewidth]{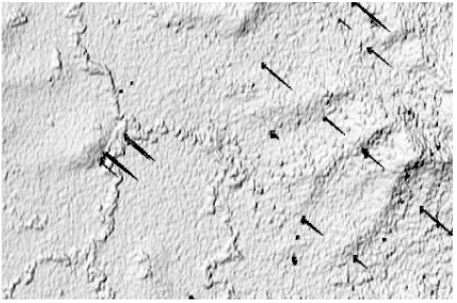}
      \caption{\textit{Shaded relief view of a 10-m-resolution DEM obtained from airborne SAR in an area with low relief (Source: \textit{Representation of Terrain} by M F Hutchinson and J C Gallant)}}
      \label{fig:3}
    \end{figure}
    
\end{enumerate}

The main aim of this thesis is to aid in the collection of data, or decide the optimal locations for the capturing aerial data. The previously generated contour maps, which are often approximate with lost surface features, can be used to decide the optimal capture locations for aerial surveys, photographic or otherwise, which can then be used to generate near orthographic terrain maps with accurate surface features retained. The goal is to achieve this with minimum number of captures, i.e. minimum data collection and processing.

%----------------------------------------------------------------------------------------
%	SECTION 2

%----------------------------------------------------------------------------------------

\section{Objectives of the Thesis}
In this thesis, the primary objective is the calculation of optimal points for visual data capture. The major focus is on the following objectives as stated below:

\begin{itemize}
    \item Detailed study of Orthographic Imaging and its application on surface topographies.
    
    \item Generation of surface topographies and surface curvatures from elevation maps. Derivation and analysis of imaging surfaces.
    
    \item Given the surface description ($S$), find the $\epsilon$-orthographic boundary region around any point on the surface and demarcate it on the 3D surface as well as on a 2D pixel representation.
    
    \item Given the surface description ($S$), the number of capture points ($N$), the imaging parameters and the physical constraints, find the position of the optimal capture points and the $\epsilon$-orthographic bounds around those points such that the visual data captured at those locations, satisfying the imaging parameters, that can maximize the coverage and generate a quasi-orthographic view of the whole surface when combined together.
    
    The surface description can be in terms of mathematical equation of a 3D surface along with the boundary constraints, can be the description of the curvature of the surface or it can be the digital elevation map created from previous captures.
    
    \item Given the surface description ($S$), the imaging parameters and the physical constraints, develop a method for finding the capture points and $\epsilon$-orthographic boundary surrounding those points such that the surface can be covered with the minimum number of capture points. 
    
    The objective is to find the positions on the surface such that the minimum number of captures are needed for quasi-orthographic reconstruction, with minimum loss of surface feature information.
    
    \item Compare the different optimization objectives and algorithms for the previous problems in terms of efficiency parameters and computational complexities. 
    
    The user or the team responsible for the mapping application is given the independence of choosing the algorithm best suited for their purpose and study based on resource availability, time and the nature of the surface. 
    
\end{itemize}

%-----------------------------------
%	SUBSECTION 1
%-----------------------------------
%\section{Problem Definition}

%-----------------------------------
%	SUBSECTION 2
%-----------------------------------

%----------------------------------------------------------------------------------------
%	SECTION 2
%----------------------------------------------------------------------------------------

\section{Contributions of the Thesis}

The key contribution of the thesis is the development of novel techniques for calculating optimal capture points on a surface to generate a quasi-orthographic projection with minimum loss of surface feature information while generating terrain maps. With the aforementioned objectives in mind, this thesis has achieved the following:

\begin{itemize}
    \item Studied and implemented a technique for topographical surface generation from digital elevation maps and calculated its surface curvatures. 
    \item Formulated and derived a method for computing imaging surface and illustrated the same for mathematical and topographical surfaces.
    \item Derived and formulated the $\epsilon$-orthographic boundary of a photographic capture at a point on a surface and showed its dependency on imaging parameters. Demonstrated the computation of the said boundary for mathematical surfaces.
    \item Developed a novel technique for finding an optimal set of $N$ capture points for a bounded surface based on $\epsilon$-orthographic bounds and maximum coverage. Both local and global solutions have been studied for different optimization objectives.
    \item Developed and compared sequential filling and batch filling algorithms for computing the minimum number of capture locations, based on local $\epsilon$-orthographic bounded regions, for complete coverage of a bounded surface with optimizing on maximum coverage and minimum overlap.
    \item Demonstrated the existence of a pareto-front for a set of optimization objectives
    \item Proposed a heuristic technique for choosing capture points based on normal-clusters
    \item Compared the different algorithms and cost functions based on computational complexities and efficiency measures.
\end{itemize}

\section{Organization of the Thesis}
The remaining part of the thesis has been organized in the following manner: 

Chapter \ref{Chapter2}: This chapter introduces and explains the concept of Orthographic projections and thus by extension, Orthographic Imaging. This also contains the basics of Imaging system and concerned parameters and also the review of some works related to Orthography.

Chapter \ref{Chapter3}: This chapter deals with the problem formulation and the mathematical model of the current thesis. It contains the derivation of the $\epsilon$-orthographic boundary, its implementation and limitations related to computation. This chapter also introduces the concept of surface curvatures and how it can be used to approximate orthographic boundaries. 

Chapter \ref{Chapter4}: In this chapter, the different algorithms for computation of optimal points have been developed based on the constraints. Both local and global optimization techniques have been explored while implementation, and the pros and cons of the techniques have been studied.

Chapter \ref{Chapter5}: In this chapter, the results of the previously introduced techniques have been compared in terms of efficiency measures and computational complexities.

Chapter \ref{Chapter6}: It concludes the thesis with possible directions for further improvements and future work.

\chapter{Orthography} % Main chapter title

\label{Chapter2} % Change X to a consecutive number; for referencing this chapter elsewhere, use \ref{ChapterX}

\lhead{Chapter 2 \emph{Orthography}} % Change X to a consecutive number; this is for the header on each page - perhaps a shortened title

%----------------------------------------------------------------------------------------
%	SECTION 1
%----------------------------------------------------------------------------------------
%\section{Orthographic Projection}
In geometry, \textit{projection} is defined as a correspondence or relation between points in a 3D figure to the points on a surface. In planar projections, a series of points are projected onto a plane such that the line joining the focus and the point on the figure, meet the projection plane. The corresponding figure created by the projection is said to be in perspective and the image is said to be a projection of the original figure. If the projected rays are orthogonal to the projection plane and thus parallel, then the projection is said to be \textit{orthogonal}. Thus orthogonal or \textit{Orthographic Projection} is a transformation or two dimensional representation of a three-dimensional object.

\begin{figure}[hb]
  \centering
  \includegraphics[width=0.5\linewidth]{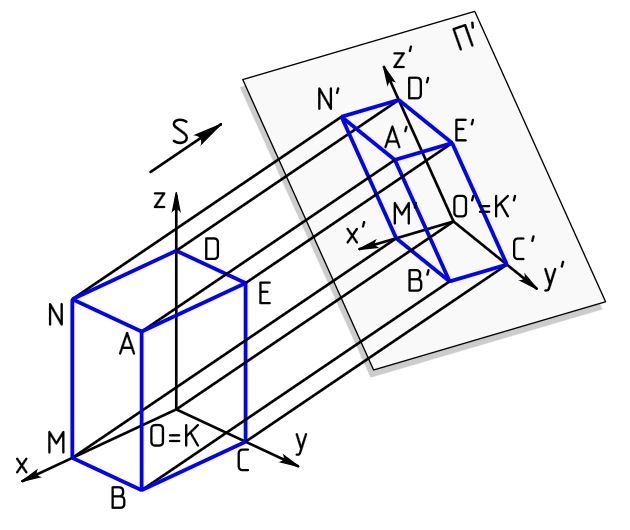}
  \caption{\textit{Axonometric orthographic projection with view plane not parallel to principal plane. }}
  \label{fig:4}
\end{figure}

In Orthographic projections sometimes the principal axes or planes of the object can be parallel to the projection plane, referred to as \textit{multiview projections}, or it can be non-parallel, referred to as \textit{axonometric projections}. Imaging, although usually perspective, is also a projection of the object on the sensing plane and thus image is a 2D planar projection of a 3D scene.  

\section{Geometry of Orthographic Projection}
Like any other geometric transformation, orthographic projection can be represented by a transformation matrix. For example, if an object is to be projected onto the $z$-plane, only the $x$ and $y$ co-ordinates are retained. So, the orthographic projection onto $z = 0$ is given by the following transformation matrix:
\begin{equation}
    P=\left[\begin{array}{lll}{1} & {0} & {0} \\ {0} & {1} & {0} \\ {0} & {0} & {0}\end{array}\right]   
\end{equation}
So, for each point $v = (v_{x}, v_{y}, v_{z})$ in the $xyz$-space, the transformation $P$ will project it to $Pv$,
\begin{equation}
    P v=\left[\begin{array}{lll}{1} & {0} & {0} \\ {0} & {1} & {0} \\ {0} & {0} & {0}\end{array}\right]\left[\begin{array}{l}{v_{x}} \\ {v_{y}} \\ {v_{z}}\end{array}\right]=\left[\begin{array}{c}{v_{x}} \\ {v_{y}} \\ {0}\end{array}\right]
\end{equation}

Homogeneous co-ordinates are often more useful for studying geometric transformations as they facilitate common vector operations such as translation, rotation, scaling and perspective projection to be represented as a matrix and thus the projected co-ordinates can be calculated by vector multiplication. Consequently, a chain of such transformations can be represented by multiplication of matrices. Let us take the example of perspective projection. If the centre of projection is the origin and the projection plane is $z = 1$. For a given point $(x, y, z)$, the point where the line and the plane intersects is $(x/z, y/z, 1)$. Here $z$ being superfluous, the projected point can be represented as $(x/z, y/z)$. In homogeneous co-ordinates, the point $(x, y, z)$ is represented by $(xw, yw, zw, w)$ and the point it maps on the plane is represented  by $(xw, yw, zw)$. So the projection in matrix form is given as 
\begin{equation}
    T = \left(\begin{array}{llll}{1} & {0} & {0} & {0} \\ {0} & {1} & {0} & {0} \\ {0} & {0} & {1} & {0}\end{array}\right)
\end{equation}
Matrices representing other geometric transformations can be combined with this by matrix multiplication. As a result, any perspective projection of space can be represented as a single matrix.\cite{ortho2005}

\begin{figure}
\centering
\subfigure[Orthographic projection onto $xy$-plane ]{\label{fig:sub1}\includegraphics[width=0.45\linewidth]{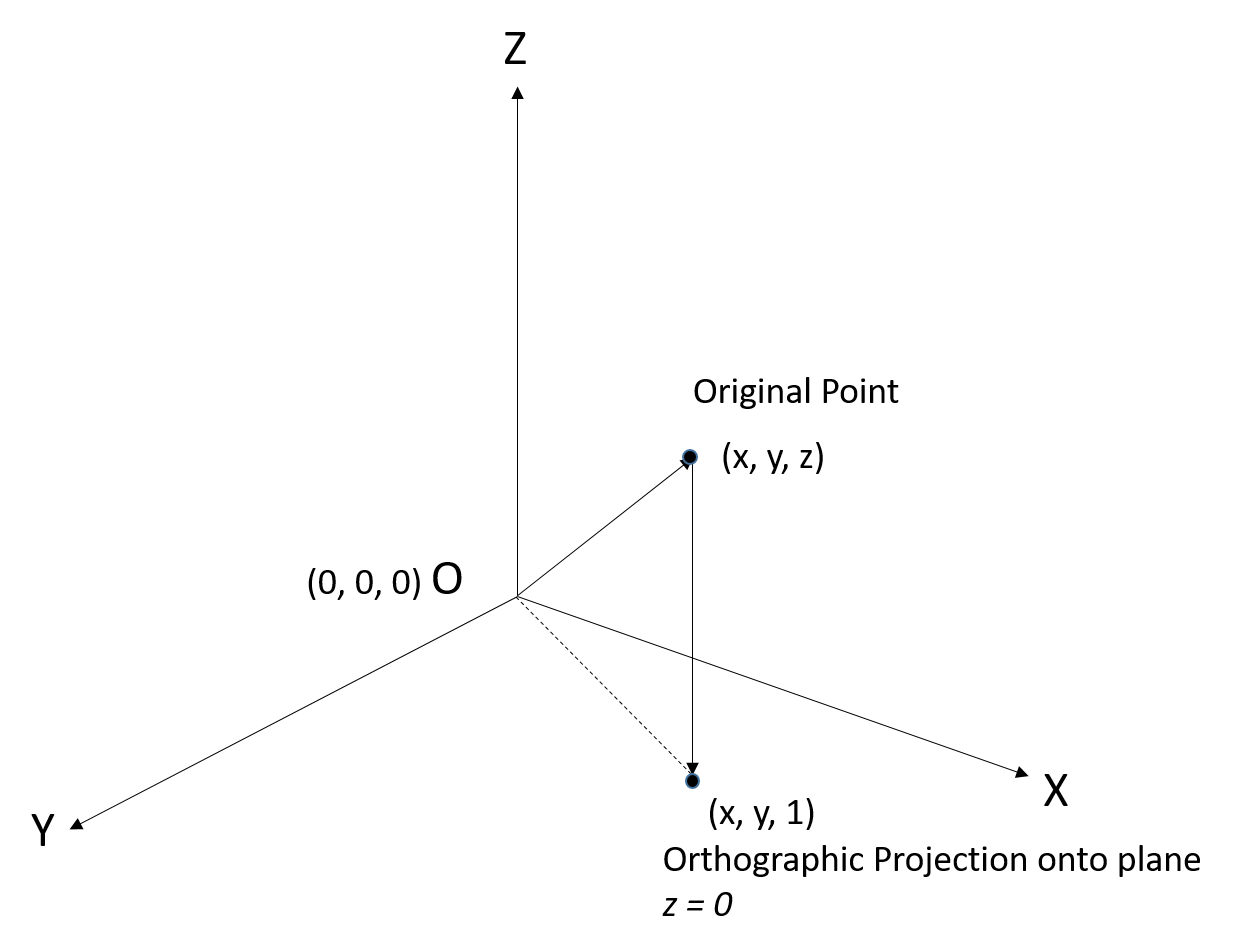}}
  %\centering
  %\caption{\textit{Orthographic projection onto $xy$-plane }}
\hfill
\subfigure[Perspective projection onto $z=1$ plane]{\label{fig:sub2}\includegraphics[width=0.45\linewidth]{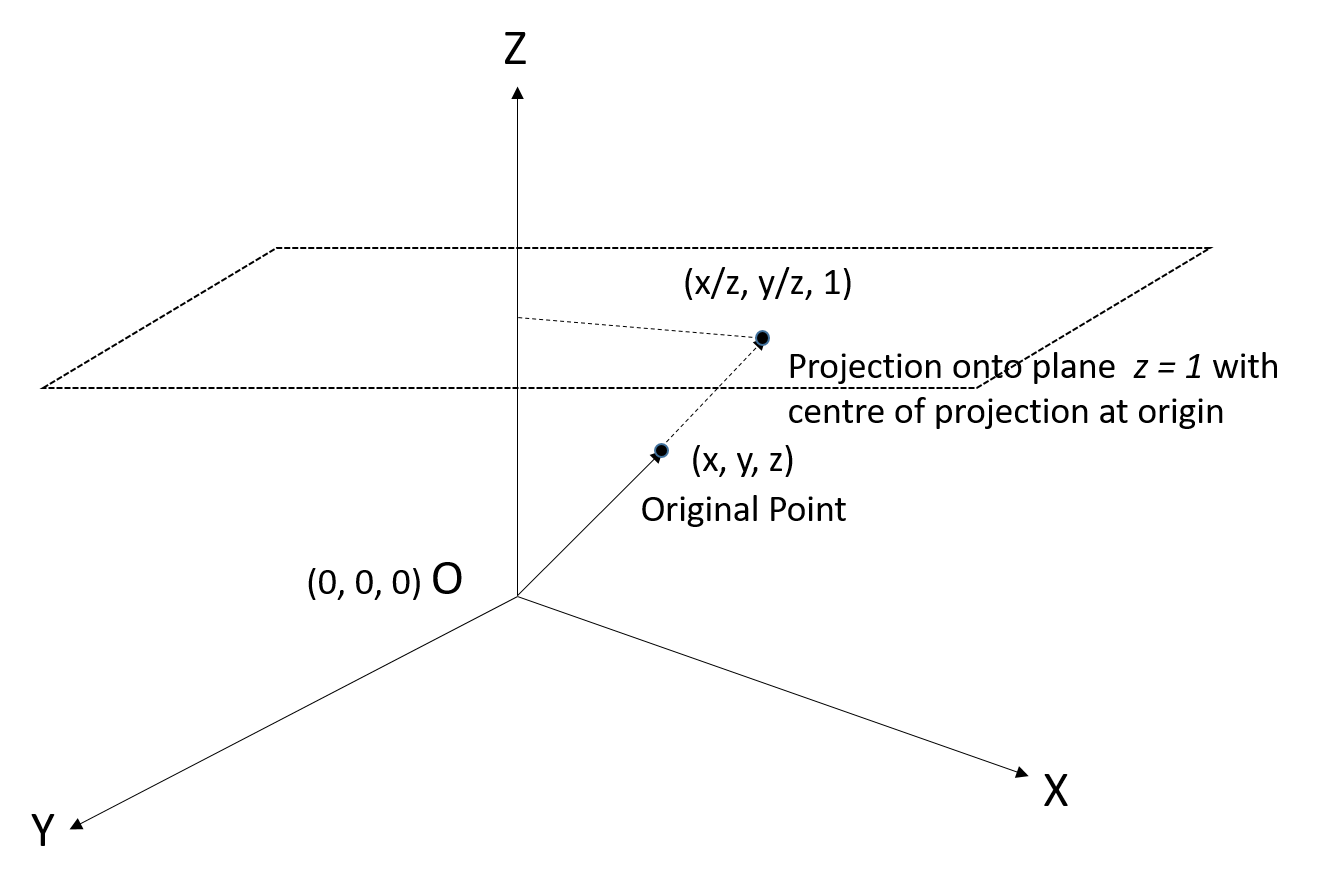}}
  %\centering
  %\caption{Perspective projection onto $z=1$ plane}
\caption{Examples of point projections}
\label{fig:5}
\end{figure}

Now, the transformation $P$ in homogeneous co-ordinates can be represented as
\begin{equation}
    P=\left[\begin{array}{llll}{1} & {0} & {0} & {0} \\ {0} & {1} & {0} & {0} \\ {0} & {0} & {0} & {0} \\ {0} & {0} & {0} & {1}\end{array}\right]
\end{equation}
For each homogeneous vector $v = (v_{x}, v_{y}, v_{z}, 1)$, the transformed vector $Pv$ is given as
\begin{equation}
    P v=\left[\begin{array}{llll}{1} & {0} & {0} & {0} \\ {0} & {1} & {0} & {0} \\ {0} & {0} & {0} & {0} \\ {0} & {0} & {0} & {1}\end{array}\right]\left[\begin{array}{c}{v_{x}} \\ {v_{y}} \\ {v_{z}} \\ {1}\end{array}\right]=\left[\begin{array}{c}{v_{x}} \\ {v_{y}} \\ {0} \\ {1}\end{array}\right]
\end{equation}

But the projection matrix $P$ defined above is not invertible since its determinant $|P| = 0$, i.e. there is no way to retrieve the lost dimension. A problem with this kind of orthographic projection is it projects both points with positive and negative $z$-values onto the projection plane. Therefore, it is useful to restrict the $z$-values (and the $x$- and $y$-values) to a certain interval, $n$ (near plane) to $f$ (far plane). In computer graphics [cite], orthographic projection can be defined by a 6-tuple, \textit{(l, r, b, t, n, f)} denoting left, right, bottom, top, near and far, which defines the clipping planes. The planes form a box with minimum corner at \textit{(left, bottom, -near)} and the maximum corner at \textit{(right, top, -far)}.\cite{orthography2018}

The box is translated to the origin as its center and scaled to a unit cube. Thus the minimum corner is at $(-1, -1, -1)$ and the maximum corner is at $(1, 1, 1)$. $S$ is the scaling matrix and $T$ is the translation matrix. So, the orthographic projection matrix is given by
\begin{equation}
    P=S T=\left[\begin{array}{cccc}{\frac{2}{r-l}} & {0} & {0} & {0} \\ {0} & {\frac{2}{t-b}} & {0} & {0} \\ {0} & {0} & {\frac{2}{f-n}} & {0} \\ {0} & {0} & {0} & {1}\end{array}\right]\left[\begin{array}{cccc}{1} & {0} & {0} & {-\frac{l+r}{2}} \\ {0} & {1} & {0} & {-\frac{t+b}{2}} \\ {0} & {0} & {-1} & {-\frac{f+n}{2}} \\ {0} & {0} & {0} & {1}\end{array}\right]
    = \left[\begin{array}{cccc}{\frac{2}{\text{r-l}}} & {0} & {0} & {-\frac{r+l}{r-l}} \\ {0} & {\frac{2}{t-b}} & {0} & {-\frac{t+b}{t-b}} \\ {0} & {0} & {\frac{-2}{f-n}} & {-\frac{f+n}{f-n}} \\ {0} & {0} & {0} & {1}\end{array}\right]
\end{equation}.

The inverse of the projection matrix, $P^{-1}$, is given as,

\begin{equation}
    P^{-1}=\left[\begin{array}{cccc}{\frac{r-l}{2}} & {0} & {0} & {\frac{l+r}{2}} \\ {0} & {\frac{t-b}{2}} & {0} & {\frac{t+b}{2}} \\ {0} & {0} & {\frac{f-n}{-2}} & {-\frac{f+n}{2}} \\ {0} & {0} & {0} & {1}\end{array}\right]
\end{equation}

Although orthographic projection does represent the 3D nature of a natural object, but it does not reflect the true representation of the object while being photographed by a camera or viewed by an observer. In particular, parallel lengths at all points in an orthographically projected image are of the same scale regardless of whether they are far away or near to the virtual viewer. As a result, the length is not foreshortened and the perception of depth is lost. 

%----------------------------------------------------------------------------------------
%	SECTION 2
%----------------------------------------------------------------------------------------
\section{Imaging System and Parameters}
In machine vision, it is very crucial to study the parameters of an imaging system. An imaging system typically consist of a camera or capturing device and a target object which is to be captured. The physical parameters of such a system is guided by the optics and the sensors of the device.\cite{imgparam} 
\begin{figure}[hbt!]
  \centering
  \includegraphics[width=0.8\linewidth]{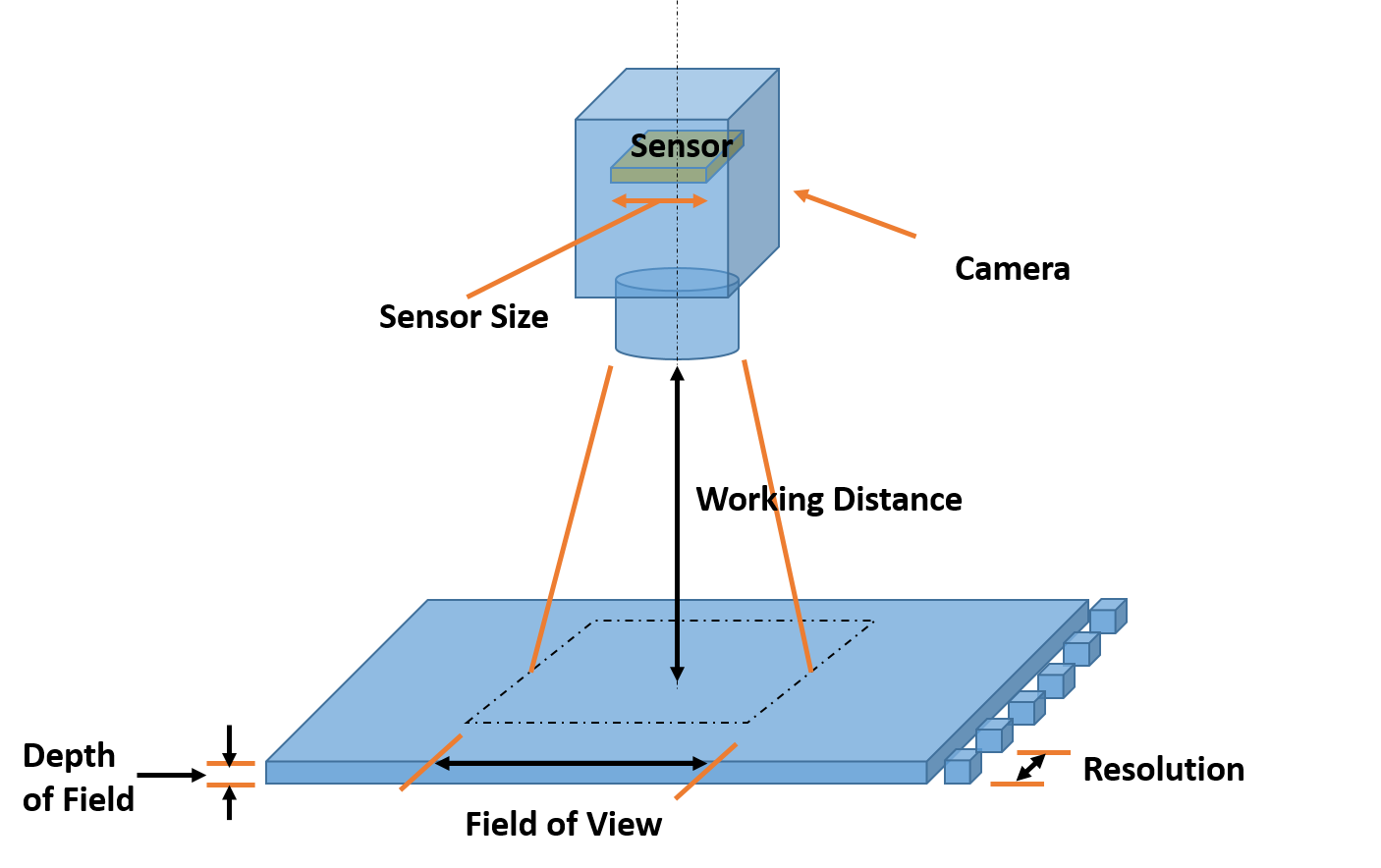}
  \caption{\textit{Illustration of the Fundamental Parameters of an Imaging System.}}
  \label{fig:6}
\end{figure}
\begin{itemize}
    \item \textbf{Field of View (FOV):} The effective viewable area of the object under inspection that fills the camera sensor is called the \textit{Field of View}. The FOV is can be expressed as physical(length) or angular(degrees) quantities.
    \item \textbf{Working Distance (WD):} The distance between the front of the camera lens and the object. The relation between the \textit{Physical FOV}, the \textit{Angular FOV} and the \textit{WD} is shown below.
    \begin{figure}[hb]
      \centering
      \includegraphics[width=0.35\linewidth]{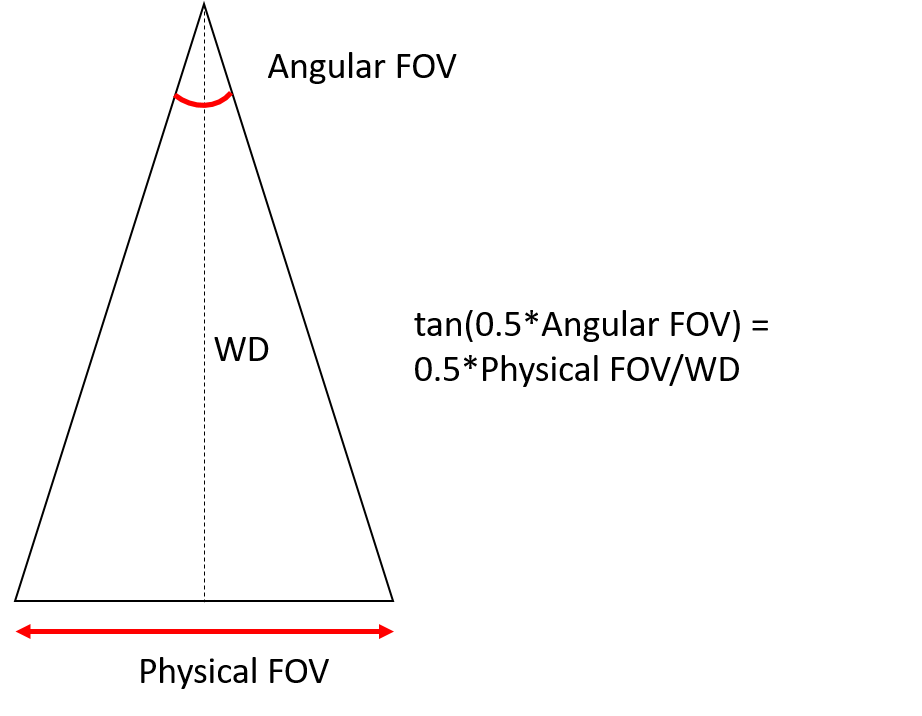}
      \caption{\textit{Relation between FOV and WD.}}
      \label{fig:7}
    \end{figure}
    \item \textbf{Resolution:} The minimum feature size of the object that can be distinguished by the imaging system.
    \item \textbf{Depth of Field (DOF):} The maximum depth of the object that can be maintained entirely within the focus of the camera. DOF is also the maximum allowable object movement while maintaining the best focus.
    \item \textbf{Sensor Size:} The size of the active area of the camera sensor specified in the horizontal direction. This is important in determining the lens magnification required to obtain a desired FOV.
    \item \textbf{Primary Magnification (PMAG):} The ratio between the sensor size and the FOV. Horizontal values are usually considered.
    $$ PMAG = \frac{Sensor\,Size [mm]}{Field\,of\,View[mm]}$$
    \begin{figure}[hb]
      \centering
      \includegraphics[width=0.5\linewidth]{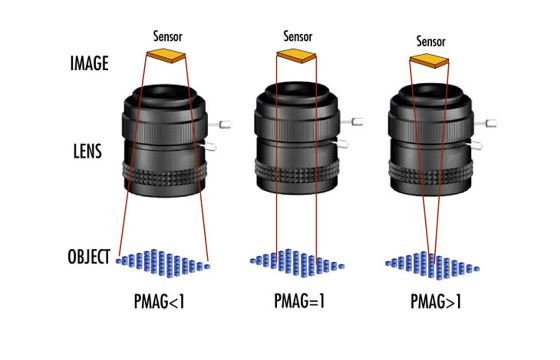}
      \caption{\textit{PMAG vs FOV. (source: Edmund Optics)}}
      \label{fig:8}
    \end{figure}
\end{itemize}
\begin{figure}[hb]
  \centering
  \includegraphics[width=0.8\linewidth]{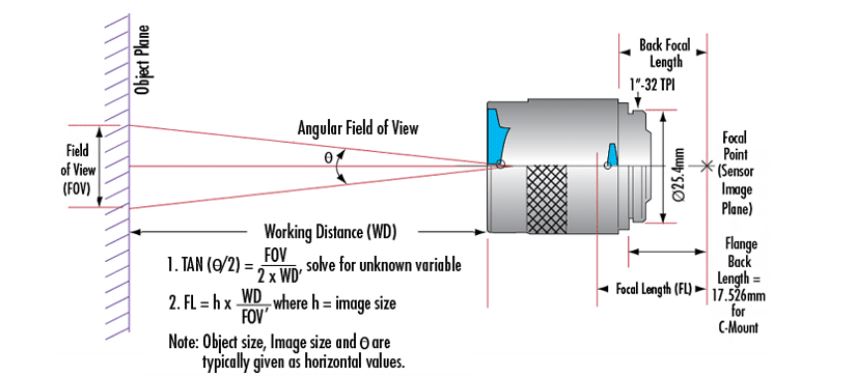}
  \caption{\textit{Parameter Diagram of Fixed  Focal Lenses.(source: Edmund Optics)}}
  \label{fig:9}
\end{figure}

%----------------------------------------------------------------------------------------
%	SECTION 3
%----------------------------------------------------------------------------------------
\section{Orthographic Imaging}
\subsection{Perspective Distortion}
Perspective transformation is the projection of the three-dimensional object onto the image plane. As an effect, distant objects appear smaller than nearer objects. In contrast to orthographic projection, the lines which are parallel in nature, appear to intersect in the projected image, i.e. they converge to a \textit{vanishing point}. Photographic lenses and the human eye work in the same way, therefore perspective projection looks most realistic. Perspective projection is usually categorized into one-point, two-point and three-point perspective, depending on the orientation of the projection plane towards the axes of the depicted object.\cite{graphics1997} 

\begin{figure}[hb]
  \centering
  \includegraphics[width=0.5\linewidth]{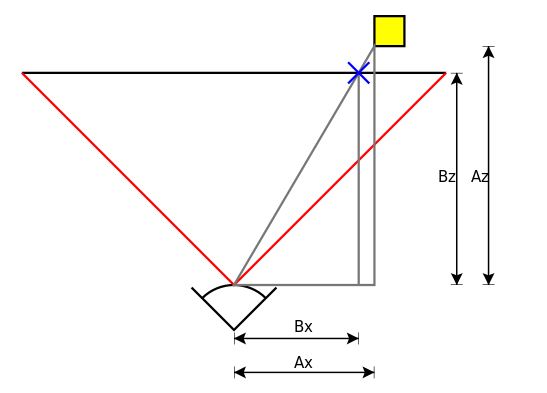}
  \caption{\textit{Calculation of image co-ordinates.}}
  \label{fig:10}
\end{figure}

The image co-ordinates ($B_{x},B_{y},B_{z}$) on the image plane or screen can be determined from the corresponding object co-ordinates ($A_{x},A_{y},A_{z}$) as shown below. Here, $B_{z}$ is the focal length- the axial distance from the camera center to the image plane, and $A_{z}$ is the object distance. 
\begin{equation}
    \frac{B_{x}}{A_{x}} = \frac{B_{z}}{A_{z}}
\end{equation}
\begin{equation}
    or, B_{x} = A_{x}\frac{B_{z}}{A_{z}}
\end{equation}
The same works for the image $y$-coordinate, by substituting $y$ for $x$ in the above diagram and equation. We can model the perspective projection by a projection matrix ($P$) as demonstrated below. Here ($x,y,z$) is the projection point co-ordinates, origin is taken as the \textit{ centre of projection (COP)} and $d$ is the distance of the projection plane from COP.

\begin{equation}
    \left[\begin{array}{llll}{1} & {0} & {0} & {0} \\ {0} & {1} & {0} & {0} \\ {0} & {0} & {-1 / d} & {0}\end{array}\right]\left[\begin{array}{l}{x} \\ {y} \\ {z} \\ {1}\end{array}\right]= \left[\begin{array}{c}{x} \\ {y} \\ {-z / d}\end{array}\right] \Rightarrow\left(-d \frac{x}{z},-d \frac{y}{z}\right)\\
\end{equation}
\begin{equation}
    P = \left[\begin{array}{llll}{1} & {0} & {0} & {0} \\ {0} & {1} & {0} & {0} \\ {0} & {0} & {-1 / d} & {0}\end{array}\right]
\end{equation}

In photography and cinematography, a \textit{normal lens} reproduces a FOV that appears natural to a human observer. In contrast, a typical lens with longer or shorter focal lengths introduces noticeable distortion due to depth compression and expansion. \textit{Perspective Distortion} is the warping or transformation of an object and its surrounding area that differs significantly from what the object would look like with a normal focal length, due to the relative scale of nearby and distant features. It is determined by the relative distances at which the image is captured and viewed, and is due to the angle of view of the image (as captured) being either wider or narrower than the angle of view at which the image is viewed, hence the apparent relative distances differing from what is expected. 

Perspective distortion takes two forms: extension distortion and compression distortion, also called wide-angle distortion and long-lens or telephoto distortion\cite{galerphoto}, when talking about images with the same field size. Extension or wide-angle distortion can be seen in images shot from close using a wide-angle lens (with an angle of view wider than a normal lens). Objects close to the lens appear larger relative to more distant objects, and distant objects appear smaller and hence farther away – thus distances are extended. Compression, long-lens, or telephoto distortion can be seen in images shot from a distance using a long focus lens or the more common telephoto sub-type (with an angle of view narrower than a normal lens). Distant objects look approximately the same size – closer objects are abnormally small, and more distant objects are abnormally large. Therefore, the viewer cannot discern relative distances between distant objects – distances are compressed.  
\\

\begin{figure}[hb]
  \centering
  \includegraphics[width=0.8\linewidth]{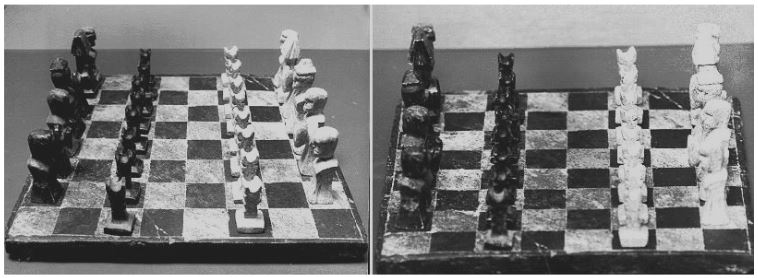}
  \caption{\textit{Wide-angle (left) showing the perspective effect in imaging vs. Tele-perspective (right). (source: Optical Measurement Techniques with Telecentric Lenses,
  by Dr. Karl Lenhardt and Bad Kreuznach)}}
  \label{fig:11}
\end{figure}

The goal of orthographic imaging is to capture the image such that the effect of perspective is eliminated. This results in an image where the perception of relative distances between the objects and the depth is lost and the objects appear to be at the same distance from the lens. A lens designed to provide an orthographic projection is known as an \textit{object-space telecentric lens}.

\subsection{Telecentric lens}
A \textit{telecentric lens}\cite{telelens2} is a compound lens that has its entrance or exit pupil at infinity. When the entrance pupil is at infinity, it produces an orthographic view of the subject. This is the case when the chief rays (oblique rays from the subject that pass through the center of the aperture) are parallel to the optical axis in front of the system. The simplest way to make such a telecentric lens is to put the aperture stop at one of the lens' focal points.  

An entrance pupil at infinity makes the lens \textit{object-space telecentric}. By entrance pupil at infinity, it means that if one looks in the front of the lens, the image of the aperture is very far away. These lenses are widely used in machine vision for orthographic imaging or otherwise because the image magnification is independent of the object's distance or position in the FOV. An exit pupil at infinity makes the lens \textit{image-space telecentric}. If both pupils are at infinity, the lens is \textit{bi-centric}. 

\begin{figure}[hb]
  \centering
  \includegraphics[width=0.7\linewidth]{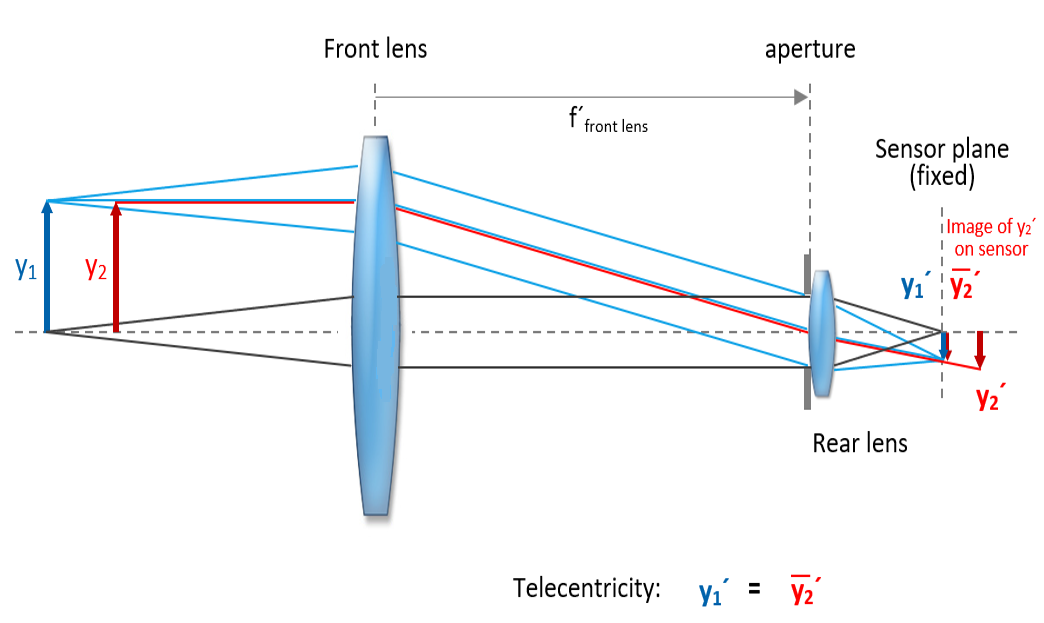}
  \caption{\textit{Objects of same size $y_{1}$ and $y_{2}$ with different working distances projected to same size on the image plane. (source: Telecentric Lenses, Vision-Doctor.com)}}
  \label{fig:12}
\end{figure}

Non-telecentric lenses exhibit varying degree of magnification at different distances from the lens. Most lenses are entocentric in nature, i.e. objects further away have lower magnification. For pericentric lenses, however, objects further away have higher magnification. The variation of magnification with distance causes several problems for machine vision and other applications:
\begin{itemize}
    \item The apparent size of objects changes with distance from the camera.
    \item Some features or objects may be hidden by objects that are closer to the lens.
    \item The apparent shape of objects varies with distance from the center of the field of view (FOV). Objects appearing close to the edges are viewed from an angle, while objects near the centre of the FOV are viewed frontally. Hence the image gets distorted towards the periphery of the FOV. (For example, circles become oval.)
\end{itemize}
In contrast, object-space telecentric lenses provide the same magnification at all distances, thus generating an orthographic projection. However, an object too close or too far away may still be out of focus, but the size and shape are unaltered.  

\subsection{Orthophotography}
An \textit{orthophoto} or \textit{orthophotograph} is an aerial image or a satellite image of a terrain or surface which is geometrically corrected or \textit{orthorectified} such that the photo or image is essentially a orthographic projection of the terrain. An orthophotograph can be used to measure distances accurately because it is an almost accurate depiction of the Earth's surface being adjusted for topographic relief, perspective distortion and camera tilt.\cite{orthograph1}
\\
\begin{figure}[hbt!]
  \centering
  \includegraphics[width=0.6\linewidth]{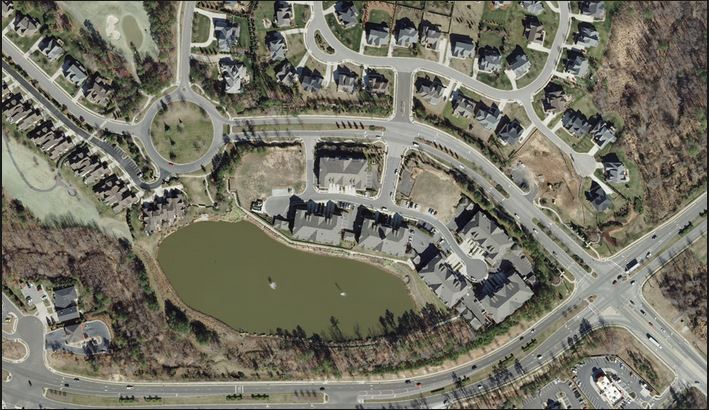}
  \caption{\textit{Digital orthophotograph of sub-urban Raleigh, North Carolina.\\
  (Source: Digital Orthoimagery, Surdex.net)}}
  \label{fig:13}
\end{figure}

Orthophotographs are commonly used in \textit{geographic information systems (GIS)} as it is 'map-accurate'. A \textit{digital elevation model (DEM)} is often required to create accurate orthophotos as distortions in the image due to the varying distance between the camera sensor and different points on the ground need to be corrected.

Consequently, an \textit{orthophotomosaic} is an image made by merging orthophotos generated from both aerial and satellite photographs which have been transformed for perspective correction, so that they appear to have been taken vertically from an infinite distance\cite{orthograph2}. Google Earth and Google Maps use orthophotomosaics as a staple. The orthophotomosaics can be incorporated with additional cartographical information to generate an \textit{orthomap} or \textit{image map}.
\\
\begin{figure}[hbt!]
  \centering
  \includegraphics[width=0.8\linewidth]{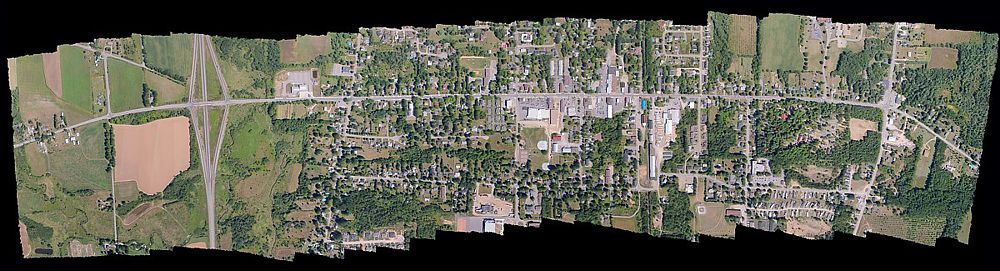}
  \caption{\textit{Orthophotomosaic map generated from GoPro aerial images.\\ 
  (GoPro Hero2 Aerial Imaging and Mapping Project, Paul Illsley).}}
  \label{fig:14}
\end{figure}
%----------------------------------------------------------------------------------------
%	SECTION 4
%----------------------------------------------------------------------------------------

\chapter{Modelling Orthographic Approximation}

\label{Chapter3}

\lhead{Chapter 3. \emph{Modelling Orthographic Approximation}}
This chapter deals with the problem formulation, the model is defined mathematically and the derivations are provided along with necessary assumptions. Section \ref{elev_map} describes a method of generating a topographical surface from a digital elevation map in \textsc{Matlab} for future analysis and optimization. Section \ref{img_surf} deals with the derivation of an imaging surface at a distance $d$ for a given surface $S$. An analysis of the variation of imaging surfaces with $d$ is also provided for one-dimensional functions. In section \ref{epsi_ortho}, the basic assumptions of $\epsilon$-orthography is given and the boundary derivation based on the assumptions is provided. Also, its implementation in \textsc{Matlab} and the computational limitations is pointed out. In section \ref{epsi_appx}, an approximation of $\epsilon$-orthography is provided so that it can be incorporated and implemented in optimization problem for the whole surface under consideration. 

\section{Generating Surface Topography from Digital Terrain Elevation Maps}\label{elev_map}
A digital terrain elevation map (DTEM) is a digital image of a terrain or a topography map where the pixel intensity at a point gives the relative elevation of the point. There are various ways of allocating pixel values (Gray-scale or RGB colormap) in a DTEM. The images that were used for the purpose of this thesis are Gray-scale DTEMs, where the elevation of a point or location in the digital map can range from $0$ to $255$, the white intensity pixels denoting the highest elevation points and the black intensity pixels denoting the lowest intensity points. The following figure is an example of a digital terrain elevation map.

\begin{figure}[hbt!]
  \centering
  \includegraphics[width=0.4\linewidth]{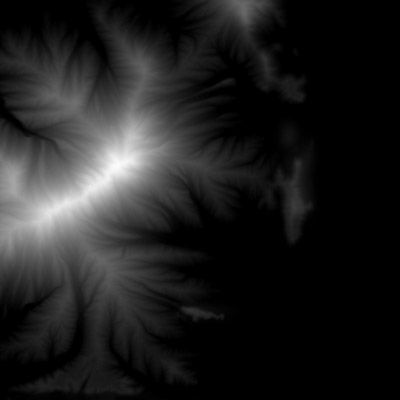}
  \caption{\textit{A gray-scale digital elevation map denoting a terrain. (Source: Creating Heightfields and Details on Terrain RAW Files, wiki.secondlife.com)}}
  \label{fig:15}
\end{figure}
The following algorithm is used for generating topographical map from a gray-scale digital elevation map. 
\begin{algorithm}
\caption{Terrain Generation from DTEM}\label{alg:dtem2terrain}
\begin{algorithmic}[1]
\State Read the Gray-scale image $I$. (If not Gray-scale, convert it from RGB to Gray-scale)
\State Smooth $I$ using a mean-smoothing filter of window size $nxn$, where $n$ is an odd integer in the range $[17,23]$.
\State $M$ = number of rows of $I$.
\State $N$ = number of columns of $I$.
\State $x$ is an array of length $N$, $x = 0:N-1$.
\State $y$ is an array of length $M$, $y = 0:M-1$.
\State Create a meshgrid $[X,Y]$ using $x$ and $y$.
\State Convert $I$ from \textit{uint8} to \textit{double precision}.
\State Create a \textit{mesh plot} of $I$ using grid $[X,Y]$. 
\end{algorithmic}
\end{algorithm}
 
% \begin{figure}[hbt!]
% \centering
% \subfigure[Mesh plot of the surface]{\label{fig:sub3}\includegraphics[width=0.48\linewidth]{Pictures/3/topography2.jpg}}
%   %\centering
%   %\caption{\textit{Orthographic projection onto $xy$-plane }}
% \hfill
% \subfigure[Contours plot of surface ]{\label{fig:sub2}\includegraphics[width=0.48\linewidth]{Pictures/3/topography1.jpg}}
%   %\centering
%   %\caption{Perspective projection onto $z=1$ plane}
% \caption{Topography map generated in \textsc{Matlab} for the DTEM in \ref{fig:15}}
% \label{fig:16}
% \end{figure}

\begin{figure}[hbt!]
  \centering
  \includegraphics[width=0.5\linewidth]{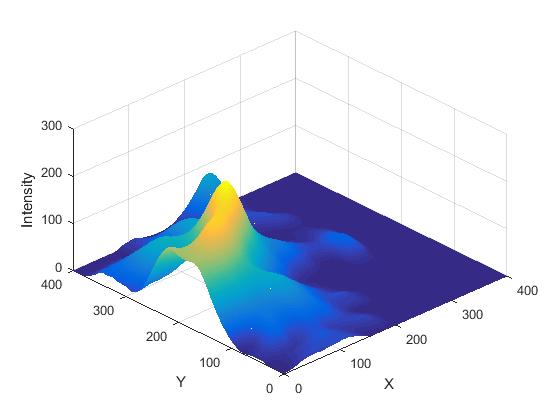}
  \caption{\textit{Mesh plot of the topographical surface generated in \textsc{Matlab} for the DTEM in figure \ref{fig:15}}}
  \label{fig:16}
\end{figure}

\begin{figure}[hbt!]
  \centering
  \includegraphics[width=0.5\linewidth]{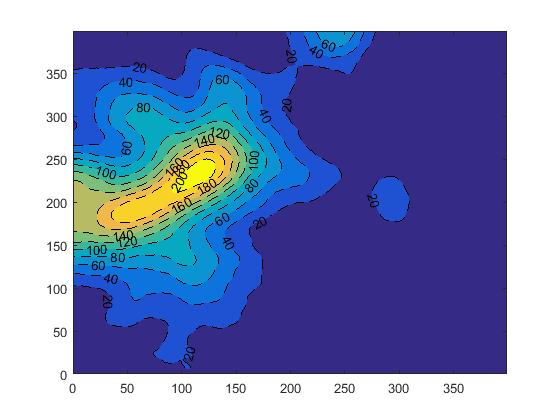}
  \caption{\textit{Contour plot of the topographical surface in figure \ref{fig:16} generated in \textsc{Matlab}}}
  \label{fig:17}
\end{figure}

The obtained image of the height-map is first smoothed out because rough surfaces with abrupt changes creates difficulty in further processing. The double precision matrix $I$ is used as a topographical surface for later processing. The surface generated in \textsc{Matlab} for the DTEM in figure \ref{fig:15} is shown in figure \ref{fig:16}.

\section{Imaging Surface} \label{img_surf}
As mentioned in Chapter \ref{Chapter2}, \textit{Working Distance} is the distance between the lens of the camera and the object. In the case of photographing a surface, let the working distance be denoted as $d$, i.e., it is assumed that to capture a point $P(x,y,z)$ on the surface, the camera needs to be placed at a height $d$ along the normal to the surface at $P$. 
\subsection{Derivation}
Let us consider a surface $S$ given by bi-variate function $f(x,y)$; so for any point $P(x,y,z)$ on the surface, 
\begin{equation}
z = f(x,y). 
\end{equation}
Then the surface normal at point $P(x,y,z)$ is given as
\begin{equation}
\Vec{n} = \left[\frac{\partial f(x, y)}{\partial x}, \frac{\partial f(x, y)}{\partial y},-1\right].
\end{equation}
If variables $p$ and $q$ are defined as 
\begin{equation}
    p = \frac{\partial f(x, y)}{\partial x} \ \ and \ \ 
    q = \frac{\partial f(x, y)}{\partial y},
\end{equation}
then the surface normal can be written as $[p, q, -1]$. The quantity $(p, q)$ is called the \textit{gradient} of $f(x, y)$ and \textit{gradient space} is the two-dimensional space of all such points $(p, q)$. The unit normal vector at $P$ is 
\begin{equation}
    \hat{n} = \frac{\Vec{n}}{|{\Vec{n}}|}.
\end{equation}
Here $|{\Vec{n}}| = \sqrt{p^2 + q^2 + 1}$ and so,
\begin{equation}
\begin{aligned}
    \hat{n} & = \frac{\Vec{n}}{\sqrt{p^2 + q^2 + 1}}\\
    & = \left[\frac{p}{\sqrt{p^2 + q^2 + 1}}, \frac{q}{\sqrt{p^2 + q^2 + 1}}, \frac{-1}{\sqrt{p^2 + q^2 + 1}}\right].
\end{aligned}
\end{equation}
Now, using equation 3.5, if $P(x,y,z)$ is a point on the surface, then the corresponding imaging point, $P'(x',y',z')$ at a height $d$ from point $P$ and along the unit surface normal $\hat{n}$ is given by
\begin{equation}
\begin{aligned}
    \Vec{P'} & = \Vec{P} + d\cdot\hat{n}\\
             & = \Vec{P} + \frac{d\cdot\Vec{n}}{\sqrt{p^2 + q^2 + 1}}\\
             & = \Vec{P} + \left[\frac{d\cdot p}{\sqrt{p^2 + q^2 + 1}}, \frac{d\cdot q}{\sqrt{p^2 + q^2 + 1}}, \frac{-d}{\sqrt{p^2 + q^2 + 1}}\right].
\end{aligned}
\end{equation}
Therefore, using equation 3.6, the co-ordinates of point $P'(x',y',z')$ can be evaluated as 
\begin{equation}
    \begin{aligned}
    x' & = x + \frac{d\cdot p}{\sqrt{p^2 + q^2 + 1}},\\
    y' & = y + \frac{d\cdot q}{\sqrt{p^2 + q^2 + 1}} \ and\\
    z' & = z - \frac{d}{\sqrt{p^2 + q^2 + 1}}.
    \end{aligned}
\end{equation}

Using the expressions for $x'$, $y'$ and $z'$ in equation 3.7, the imaging surface $S'$ at a distance $d$ from $S$ can be parametrized in terms of $x$ and $y$.

\begin{equation}
    \Vec{S'} = \left[
    \begin{array}{c}
         {x + \frac{d\cdot p}{\sqrt{p^2 + q^2 + 1}}} \\
         {y + \frac{d\cdot q}{\sqrt{p^2 + q^2 + 1}}} \\
         {f(x,y) - \frac{d}{\sqrt{p^2 + q^2 + 1}}}
    \end{array} \right]
\end{equation}
Here $p = \frac{\partial f(x, y)}{\partial x}$ and $q = \frac{\partial f(x, y)}{\partial y}$ as defined in equation 3.3.

If at any point on $S'$ is below the surface $S$, then those points are inaccessible and hence cannot be used as imaging points, i.e. if $z' \ < \  f(x',y')$, $P'(x',y',z')$ cannot be an imaging point for $P(x,y,z)$ at height $d$.

\begin{figure}[hbt!]
\centering
\subfigure[Surface plot]{\label{fig:sub3}\includegraphics[width=0.4\linewidth]{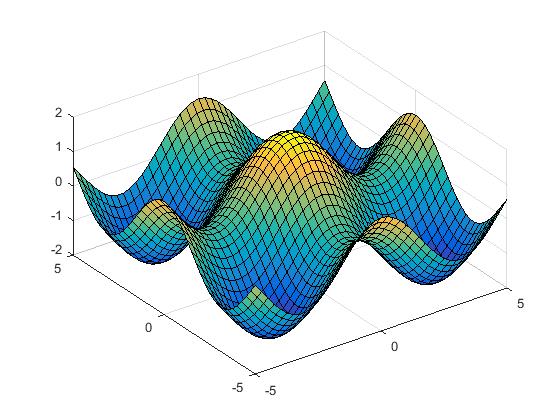}}
\hfill
\subfigure[Contour plot]{\label{fig:sub4}\includegraphics[width=0.4\linewidth]{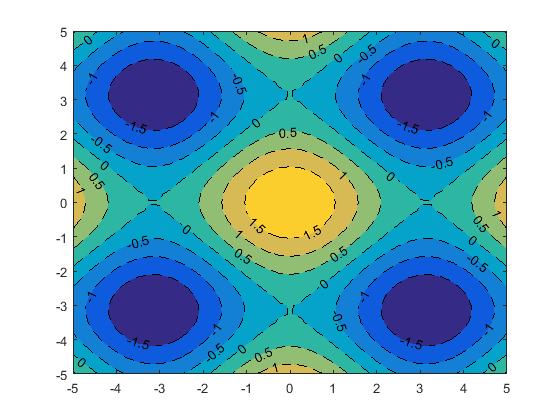}}

\subfigure[Surface normal plot]{\label{fig:sub4}\includegraphics[width=0.45\linewidth]{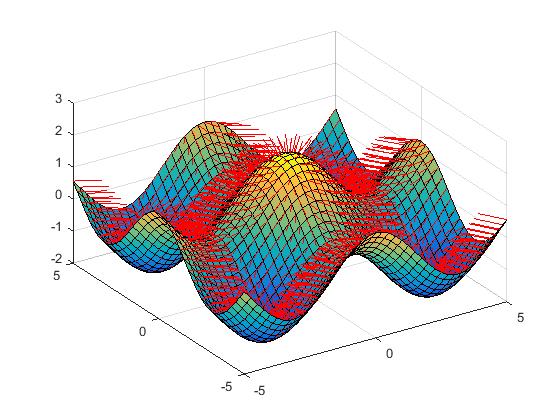}}
\caption{The surface $S$ given by $f(x,y) = cos(x) + cos(y)$ plotted in \textsc{Matlab}}.
\label{fig:18}
\end{figure}

\begin{figure}[H]
\centering
\subfigure[$d = 0.1$]{\label{fig:sub5}\includegraphics[width=0.4\linewidth]{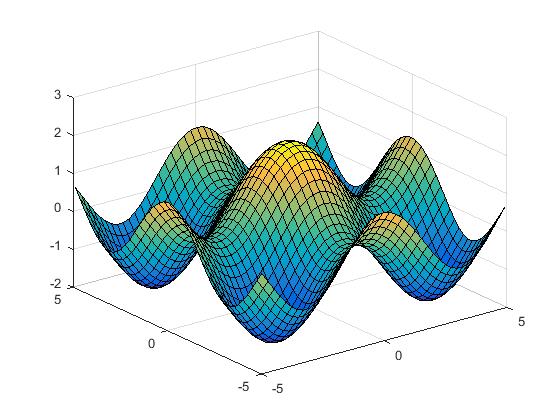}}
\hfill
\subfigure[$d = 0.1$]{\label{fig:sub6}\includegraphics[width=0.4\linewidth]{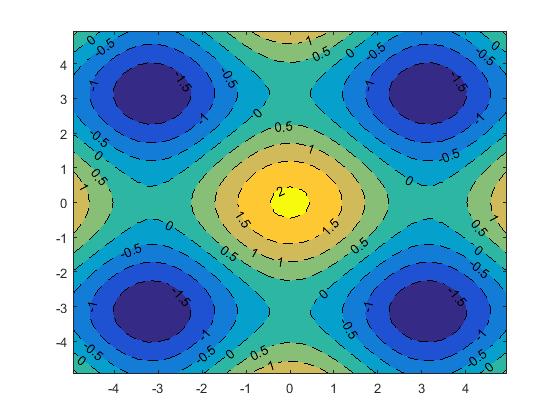}}

\subfigure[$d = 0.5$]{\label{fig:sub11}\includegraphics[width=0.4\linewidth]{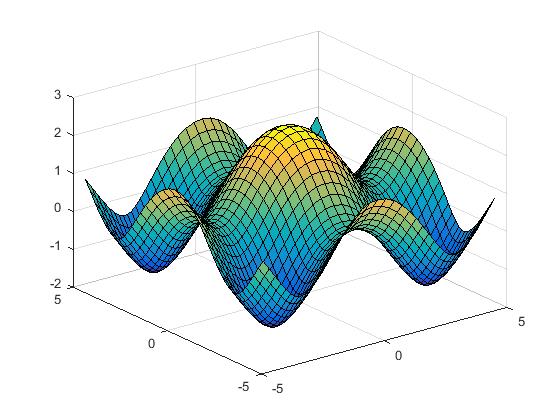}}
\hfill
\subfigure[$d = 0.5$]{\label{fig:sub12}\includegraphics[width=0.4\linewidth]{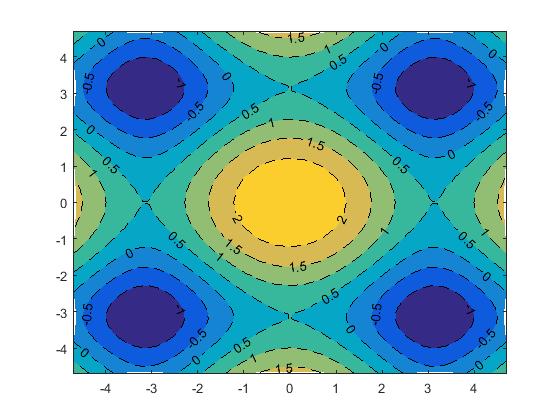}}

\subfigure[$d = 1$]{\label{fig:sub7}\includegraphics[width=0.4\linewidth]{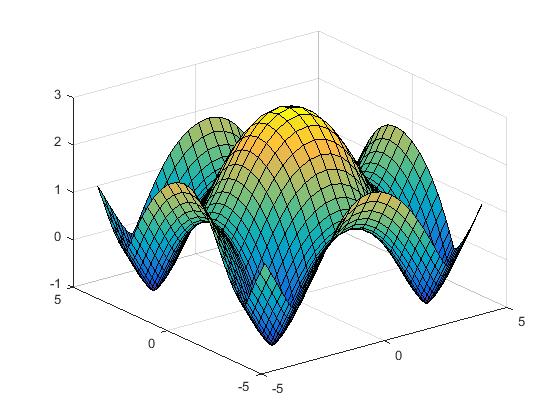}}
\hfill
\subfigure[$d = 1$]{\label{fig:sub8}\includegraphics[width=0.4\linewidth]{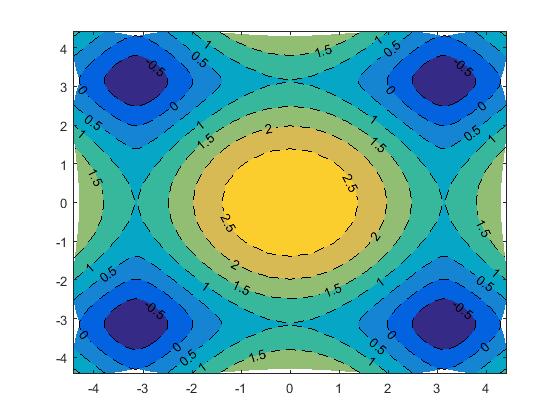}}

\subfigure[$d = 2$]{\label{fig:sub9}\includegraphics[width=0.4\linewidth]{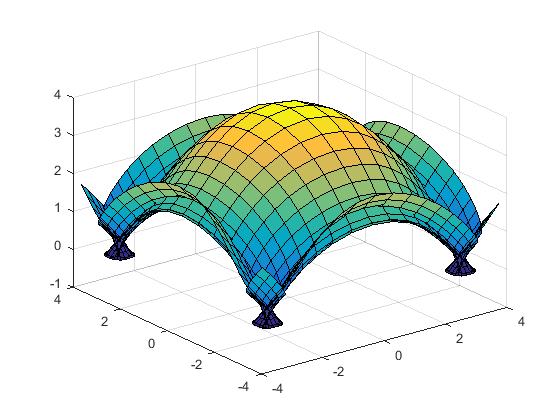}}
\hfill
\subfigure[$d = 2$]{\label{fig:sub10}\includegraphics[width=0.4\linewidth]{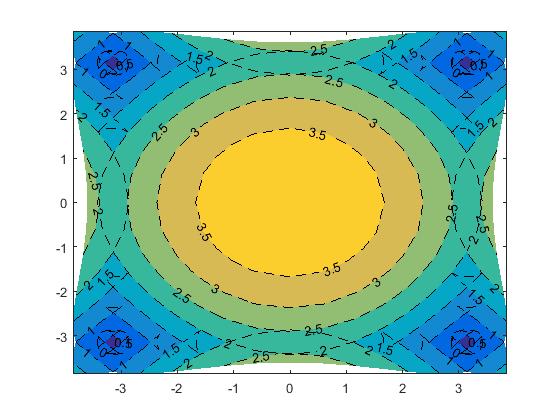}}

\caption{Imaging surfaces $S'$ and corresponding contour plots for different values of $d$.}
\label{fig:19}
\end{figure}

The following figures, plotted in \textsc{MATLAB}, demonstrate the imaging surfaces for the surface $S$ given by $f(x,y) = cos(x) + cos(y)$ in the range $(-5\leq x \leq 5)$ and $(-5 \leq y \leq 5)$ (fig \ref{fig:18}), calculated and plotted at different values of $d$ (fig \ref{fig:19}).

\pagebreak

In case the surface $S$ cannot be expressed mathematically as a bi-variate function and is a double precision matrix ($I$) as shown in section \ref{elev_map}, then instead of calculating mathematical gradients ($\frac{\partial f(x, y)}{\partial x}$ and $\frac{\partial f(x, y)}{\partial y}$) for finding surface normals, the numerical gradients can be calculated as an approximation. Considering a topographical matrix $I$(of size $M \times N$) to have unit-spaced data, the gradient matrices $G_{x}$ and $G_{y}$ are calculated as follows:

The horizontal gradient values of interior points are the central differences,
\begin{equation}
    G_{x}(i,j) = \frac{I(i,j+1) - I(i,j-1)}{2}, \ \ \  j = 2,...,N-1 \ \  and \ \ i = 1,...,M 
\end{equation}
and the horizontal gradient values along the edges of the matrix I are calculated as single-side differences.
\begin{equation}
\begin{aligned}
    G_{x}(i,1)  & = I(i,2) - I(i,1), \\
    G_{x}(i,N)  & = I(i,N) - I(i,N-1), \ \ \ i = 1,...,M
\end{aligned}
\end{equation}

Similarly, the vertical gradient values of interior points and along the edges of the matrix are calculated as
\begin{equation}
    G_{y}(i,j) = \frac{I(i+1,j) - I(i-1,j)}{2}, \ \ \  i = 2,...,M-1 \ \ and \ \ j = 1,...,N, 
\end{equation}
\begin{equation}
\begin{aligned}
    G_{y}(1,j)  & = I(2,j) - I(1,j), \\
    G_{y}(M,j)  & = I(M,j) - I(M-1,j), \ \ \ j = 1,...,N.
\end{aligned}
\end{equation}

Using equations 3.9 - 3.12 and putting $p = G_{x}$ and $q = G_{y}$ in equation 3.8, the points on surface $S'$ can be calculated and hence the imaging surface at distance $d$ can be generated. An example is shown in figure \ref{fig:20} below. 

\begin{figure}[hbt!]
  \centering
  \includegraphics[width=0.5\linewidth]{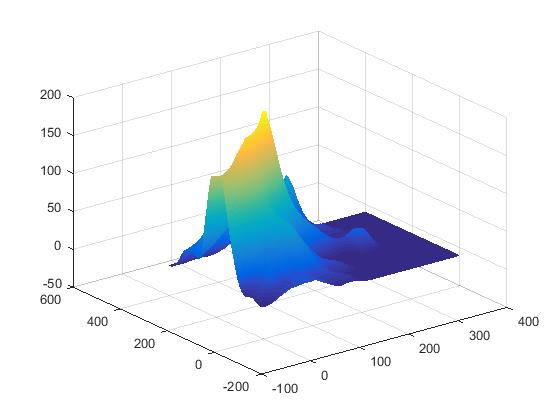}
  \caption{\textit{The imaging surface $S'$ at $d = 10$ plotted in \textsc{Matlab} for the surface in fig \ref{fig:16}}}
  \label{fig:20}
\end{figure}

\subsection{Analysis in 1D} \label{imgsurf}
Visualizing the variation of the \textit{imaging surface} with imaging height $d$ is difficult for bi-variate functions. The purpose of this section is to create an analogy of imaging surface for uni-variate functions, which is essentially finding the imaging curve, and derive the analogous vector equation of 3.8. Through this analysis, some comments are provided for the virtual bounds of $d$.

Let us consider a function $f(x)$. It is visualized by the curve $C$ which contains points $P(x,y)$ such that
\begin{equation}
    y = f(x).
\end{equation}
So the curve $C$ can be parametrized by vector $\Vec{P}$ as
\begin{equation}
    \Vec{P}(x) = \left[ \begin{array}{c}{x}\\{y}
    \end{array}\right] = \left[ \begin{array}{c}{x}\\{f(x)}
    \end{array}\right].
\end{equation}
The tangent at $P(x,y)$ can be parametrized as 
\begin{equation}
    \Vec{T}(x) = \frac{d\Vec{P}}{dx} = \left[ \begin{array}{c}{1}\\{f'(x)}
    \end{array}\right] \\
\end{equation}
and as tangent and normal are perpendicular, if $\Vec{N}(x)$ is the normal at that point, then
\begin{equation}
\begin{split}
    \Vec{N}(x)\cdot\Vec{P}(x) = 0 
    \\
    So, \ \ \Vec{N}(x) = \left[ \begin{array}{c}{-f'(x)}\\{1}
    \end{array}\right].
\end{split}
\end{equation}

Therefore, the unit normal vector at $P$ is given by
\begin{equation}
\begin{aligned}
    \hat{N}(x) & = \frac{\Vec{N}(x)}{|\Vec{N}(x)|} = \frac{\Vec{N}(x)}{\sqrt(1 + (f'(x))^{2})} \\
    & = \frac{1}{\sqrt{1 + (f'(x))^{2}}}\left[ \begin{array}{c}{-f'(x)}\\{1}
    \end{array}\right].
\end{aligned}    
\end{equation}
Now, if $P'(x',y')$ is the point on the imaging curve $C'$ located at a distance $d$ along the unit normal $\hat{N}$ at $P(x,y)$, then 
using equations 3.14 and 3.17, 

$\Vec{P'}$ can be parametrized in $x$ as
\begin{equation}
\begin{aligned}
    \Vec{P'} & = \Vec{P} + d\cdot\hat{N} \\
             & = \left[ \begin{array}{c}{x}\\{f(x)}
    \end{array}\right] + \frac{d}{\sqrt{1 + (f'(x))^{2}}}\left[ \begin{array}{c}{-f'(x)}\\{1}
    \end{array}\right] \\
            & = \left[ \begin{array}{c}{x - \frac{d\cdot f'(x)}{\sqrt{1 + (f'(x))^{2}}}}\\{f(x) + \frac{d}{\sqrt{1 + (f'(x))^{2}}}}
    \end{array}\right].
\end{aligned}
\end{equation}
the coordinates($x'$ and $y'$) of imaging curve $C'$ can be parametrized in terms of $x$ as

\begin{equation}
    \begin{aligned}
        x' & = x - \frac{d\cdot f'(x)}{\sqrt{1 + (f'(x))^{2}}} \ \ and \\
        y' & = f(x) + \frac{d}{\sqrt{1 + (f'(x))^{2}}}.
    \end{aligned}
\end{equation}

For non-smooth functions, numerical gradients can be used instead for aforementioned calculations at non-differentiabile points. Now, it is obvious that if a point $P'(x',y')$ on curve $C'$ is such that it satisfies $y'(x) < f(x')$, then it lies below the curve $C$($f(x)$) and hence it cannot be accepted as a valid imaging point. Thus, if $d$ is kept unchanged for all the points, only those values of $d$'s are valid for which such invalid imaging points are not generated. In other terms for a $d$ to be valid, curves $C$ and $C'$ should not intersect at any point. This gives a mathematical bound for imaging height $d$ -
\begin{itemize}
    \item $d > 0$
    \item $d < D$ such that $ \forall \ d \geq D, \ \exists$ some $x$ in $dom(f)$,  s.t. $y'(x) < f(x')$, where $x'$ and $y'$ are as given in equation 3.19.
\end{itemize}
The mathematical upper bound $D$ depends on the curvature or nature of the function $f(x)$ and also the imaging range, i.e., the range of values of $x$ that is to be imaged. D can be calculated numerically by solving the following equation and applying \textit{bisection} as stated in algorithm \ref{alg:bisecD}.
\begin{equation}
\begin{aligned}
    & y'(x) = f(x') \\ 
    or, \ & f(x') =  f(x) + \frac{d}{\sqrt{1 + (f'(x))^{2}}}
\end{aligned}
\end{equation}
For $d < D$, the above equation will have no solution and for $d \geq D$, the above equation will have one or more solution(s). 

\begin{algorithm}
\caption{Finding upper bound $D$ by Bisection }\label{alg:bisecD}
\begin{algorithmic}[1]
\State Set lower bound of $d$, $L = 0$.
\State Set a very large upper bound $U$ such that by replacing $d = U$ in equation 3.20, it has a solution.
\State Set $D = (L+U)/2$.
\State Replace $d = D$ in equation 3.20 and look for a solution.
    \If{solution exists}
        \State Set $U = D$
    \Else{
        \State Set $L = D$
    }

\State Repeat steps 5-8 until convergence criterion is met.
\EndIf
\end{algorithmic}
\end{algorithm}
For some smooth functions, there may not be any upper limit on $d$ (i.e. $D = \infty$). For those functions, equation 3.20 does not have a solution for any $d > 0$, i.e. $y'(x) > f(x')$ for all positive values of $d$ and all $x$ in $dom(f)$. However, the practical upper bound depends on the limitations of resolution of the capturing device and also the concerned application. 

\begin{figure}[hbt!]
  \centering
  \includegraphics[width=0.7\linewidth]{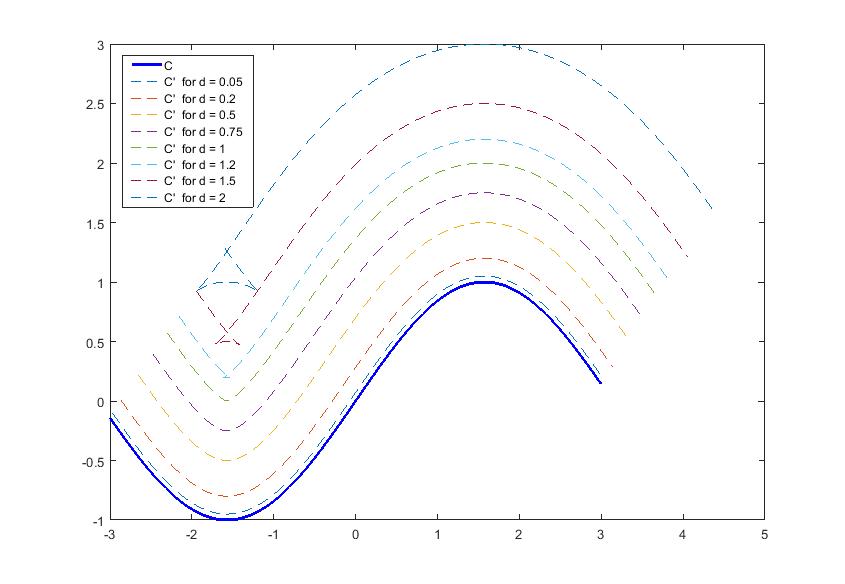}
  \caption{\textit{The imaging curves($C'$) of $f(x) = sin(x)$ (curve drawn in blue) have been plotted for various values of $d$. As $d$ increases, $C'$ moves further away from $C$, thus the upper bound on $d$, $D = \infty$  }}
  \label{fig:21}
\end{figure}

\begin{figure}[hbt!]
  \centering
  \includegraphics[width=0.7\linewidth]{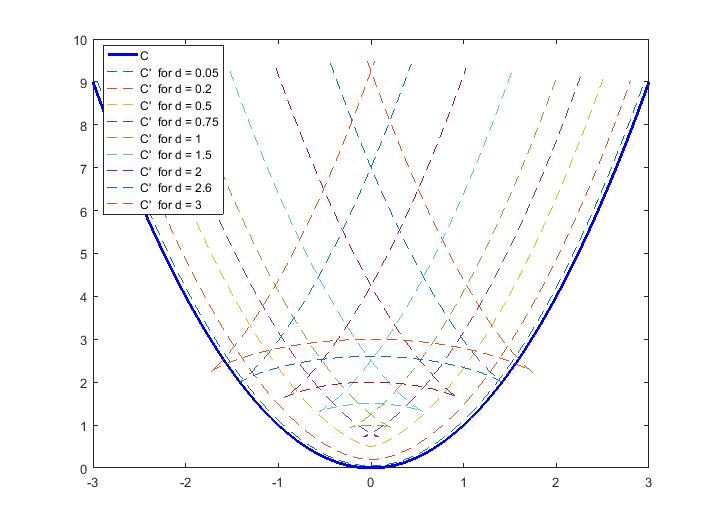}
  \caption{\textit{The imaging curves($C'$) of $f(x) = x^2$ (curve drawn in blue) have been plotted for various values of $d$. As $d$ increases, $C'$ moves further away from $C$. For lower values of $d$, $C'$s do not intersect $C$ and therefore are valid imaging curves. But for bigger values of $d$, they intersect $C$ and hence are invalid. Here the upper bound on $d$, $D \approx 2.6$}.}
  \label{fig:22}
\end{figure}

\begin{figure}[hbt!]
  \centering
  \includegraphics[width=0.7\linewidth]{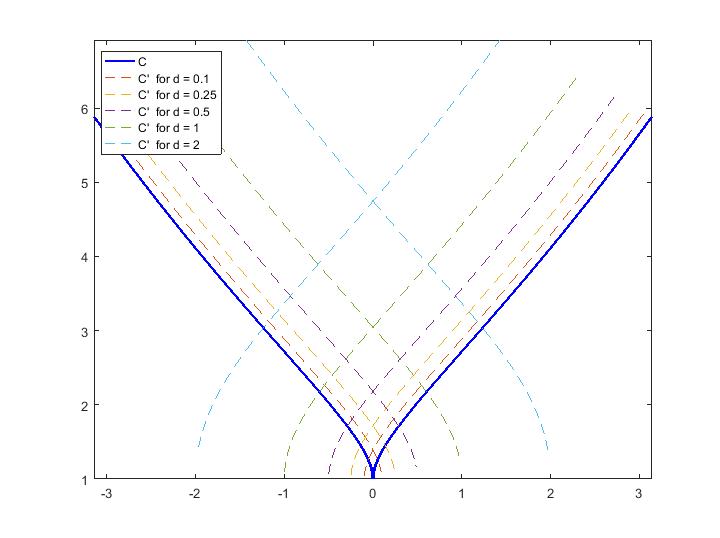}
  \caption{\textit{The imaging curves($C'$) of $f(x) = \exp{\sqrt{|x|}}$ (curve drawn in blue) have been plotted for various values of $d$. Here $C$ is non-smooth and it is non-differentiable at $x = 0$. In this case, $C'$ generated for any $d>0$ is invalid, as it always intersects $C$, i.e. there are always points around $0$ which generate invalid imaging points. Also, as $d$ increases, the number of invalid imaging points increases.}}
  \label{fig:23}
\end{figure}

The aforementioned equations have been implemented in \textsc{Matlab} and the function curves($C$) have been plotted along with the imaging curves($C'$) for different values of $d$ (figures \ref{fig:21}, \ref{fig:22}, \ref{fig:23}).

\pagebreak

\begin{figure}[htb!]
\centering
\subfigure[$m = 0.5$, valid for $0 < d < \infty$]{\label{fig:sub13}\includegraphics[width=0.45\linewidth]{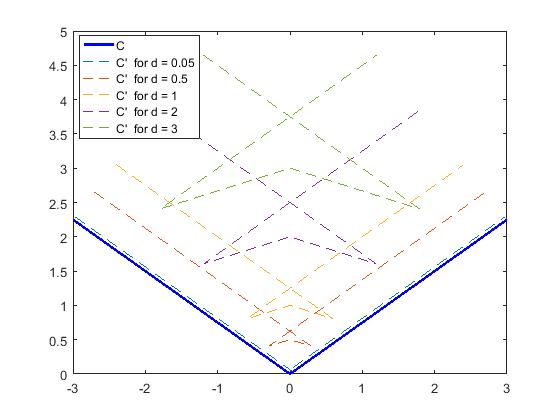}}
\hfill
\subfigure[$m = 1$, valid for $0 < d < \infty$]{\label{fig:sub14}\includegraphics[width=0.45\linewidth]{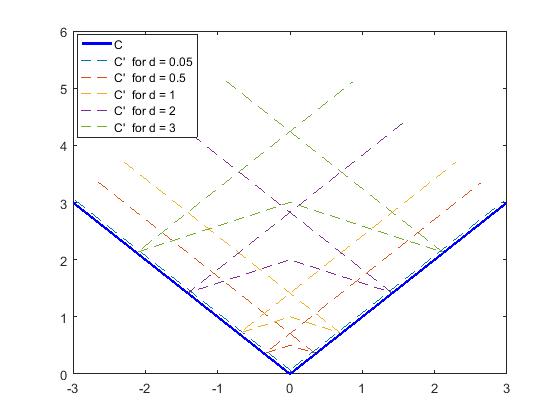}}

\subfigure[$m = 1.5$, not valid for any $d$]{\label{fig:sub15}\includegraphics[width=0.45\linewidth]{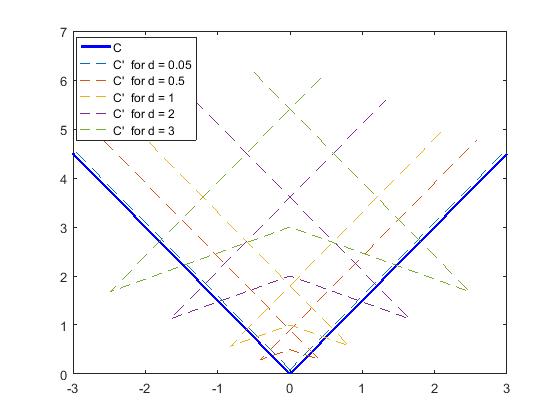}}

\caption{Imaging curves $C'$ of $g(m,x) = |mx|$ is plotted different values of $d$.}
\label{fig:24}
\end{figure}

It is interesting to note that all non-smooth curves do not generate invalid imaging curves. Let us consider a class of non-smooth functions defined on the real line ($-\infty < x < \infty$)
\begin{equation}
    g(m,x) = |mx|, \ \ \ m > 0.
\end{equation}
Any function in this class is not differentiable at $x = 0$. However, for values of $m$ in the range $(0,1]$, the function $f(x) = g(m,x)$ will always generate valid imaging curves ($D = \infty$). But for any value of $m > 1$, the imaging curves will always intersect the function curve and thus will never be valid. This is illustrated in the figure \ref{fig:24}.

\section{$\epsilon$-Orthography} \label{epsi_ortho}
As discussed previously, in an ideal orthographic imaging, every point($P$) on a surface must be imaged separately from a point($P'$) along the surface normal at $P$, which is practically impossible. A practical approximation of orthography is to consider a very small($\epsilon$) angular \textit{field of view(FOV)} and the points on the surface within this $\epsilon$-FOV to be roughly orthographic. Here $\epsilon$ is a very small angle ($\sim \ 10^{\circ} - 20^{\circ}$).   
\subsection{Assumptions and Circular Case}
It has been established in \ref{img_surf} that for orthographic imaging of a surface point $P$, the imaging point must be along the surface normal at $P$ and at a height $d$. A reverse exercise would involve choosing a point above a surface and try to  find surface points $P$ for which it can act as an imaging point $P'$. Also, the $\epsilon$ approximation as stated before is considered. The following discussion is for curves but can be easily extended to surfaces.

For a circle $C$, that radius to a point on the circumference is always orthogonal to the tangent at that point. Consequently, the centre $O$ of the circle satisfies the properties of a valid imaging point for any point $P$ on the circle. So the imaging point at $O$ can be used to capture a length of $2 \pi R$ or the entire circumference as shown in figure below. The total number of captures required to cover the entire circle is $\ceil{\frac{2\pi}{\epsilon}}$.

\begin{figure}[hbt!]
  \centering
  \includegraphics[width=0.4\linewidth]{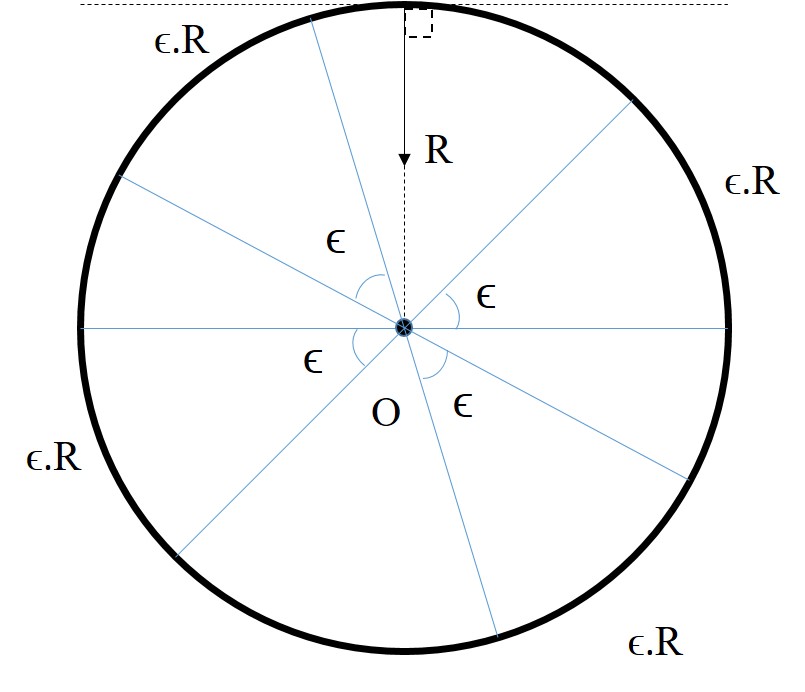}
  \caption{\textit{The centre $O$ is the imaging point and the imaging height $d = R$. Normal at any point on the curve passes through $O$ and hence all the points on $C$ can be orthographically imaged from $O$ without any approximation. The $\epsilon$-FOV, shown in blue, is exaggerated for illustration.}}
  \label{fig:25}
\end{figure}

Now, if the imaging point is shifted by $x$ from $O$ to an eccentric point $Q$, the symmetry is broken. In such a case, the normal from only two diagonally opposite points on the circumference passes through $Q$. These two points can be found by drawing a line through $O$ and $Q$ and finding the intersection of $\overline{AB}$ with $C$. Let those points be $P_{1}$ and $P_{2}$. Considering $\epsilon$-orthographic approximation, the two arcs subtended by $\epsilon$ containing $P_{1}$ and $P_{2}$ are orthographically imaged from $Q$. Therefore

\begin{equation}
\begin{aligned}
& \angle{AQB} = \angle{CQD} = \epsilon, \ \angle{AQO} = \epsilon/2 \\
& \text{In $\triangle{AOQ}$, let $\angle{OAQ} = \theta /2$. So, $\angle{AOP_{1}} = (\epsilon + \theta)/2 $.} \\ 
& \text{By symmetry, $\angle{BOP_{1}} =  (\epsilon + \theta)/2 $. Hence $\angle{AOB} = \epsilon + \theta$}. \\
\end{aligned}
\end{equation}
\begin{figure}[hbt!]
  \centering
  \includegraphics[width=0.5\linewidth]{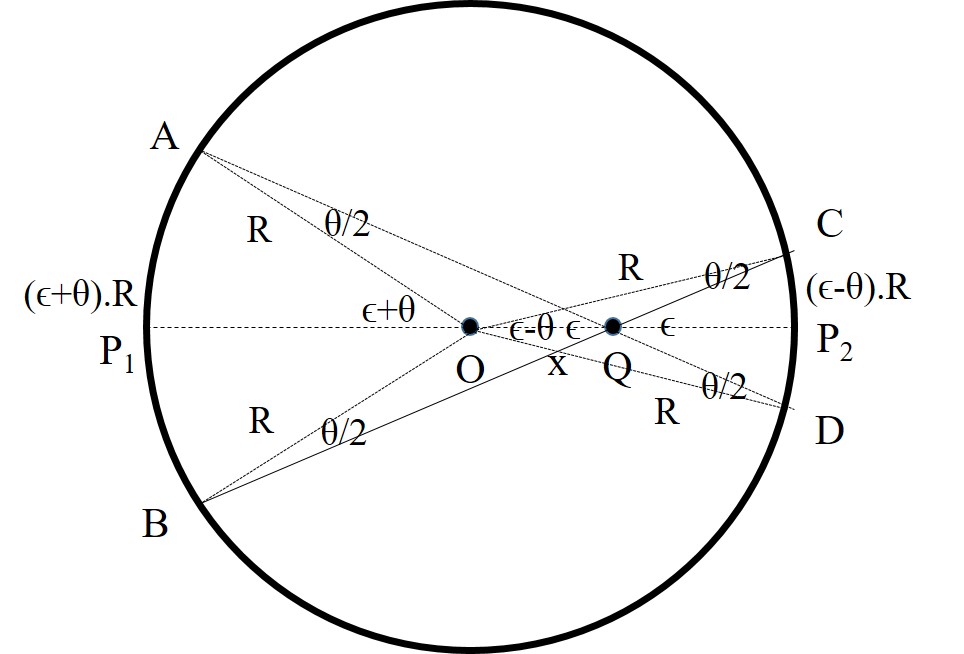}
  \caption{\textit{The eccentric point case is illustrated. The total length of the circle $C$ that can be imaged from $Q$ is independent of the location of $Q$. Here $0 < x \leq R$.}}
  \label{fig:26}
\end{figure}
\begin{equation}
\begin{aligned}
& \text{Now, $\triangle{AOD}$ is isosceles as $\overline{OA} = \overline{OD} = R$. So, $\angle{ODA} = \angle{OAD} = \angle{OAQ} = \theta/2$.} \\
& \text{As  $\angle{DQP_{2}} = \epsilon/2$, therefore $\angle{DOP_{2}} = (\epsilon - \theta)/2$}. \\
& \text{By symmetry, $\angle{COP_{2}} = (\epsilon - \theta)/2$. So, $\angle{COD} = \epsilon - \theta$}. \\
& \text{Therefore, arc} AB = (\epsilon + \theta)\cdot R \ \text{and arc} CD = (\epsilon - \theta)\cdot R. \\
& \text{So total length of curve $C$ that can be imaged from $Q$ is arc$AB + $  arc$CD = 2\epsilon R$}.
\end{aligned}
\end{equation}

\subsection{Derivation of $\epsilon$-Orthographic Bound for Curves}
Let us consider a curve $C$ given by a univariate function $f(x)$. By equations 3.15 and 3.16, the tangent($\Vec{T}$) and normal($\Vec{N}$) vectors at point $P(x,f(x))$ are given as
\begin{equation}
    \Vec{T}(x) =  \left[ \begin{array}{c}{1}\\{f'(x)}
    \end{array}\right] \ \ \ \ 
    \Vec{N}(x) = \left[ \begin{array}{c}{-f'(x)}\\{1}
    \end{array}\right].
\end{equation}

Let point $P'(x',f(x'))$ be situated at a small distance $\Delta x$ to the left of $x$. Let $p = f'(x)$ and $p' = f'(x'). $If $\Delta x$ is very small then $f(x')$ and $f'(x')$ can be approximated as
\begin{equation}
\begin{aligned}
    f(x') & = f(x - \Delta x) \approx f(x) - \Delta x\cdot f'(x) \\
    f'(x') & = f'(x - \Delta x) \approx f'(x) - \Delta x\cdot f''(x),
\end{aligned}
\end{equation}
therefore,
\begin{equation}
\begin{aligned}
    p' & = p + \Delta p \\
    \Delta p & \approx - \Delta x\cdot f''(x)
\end{aligned}
\end{equation}

\begin{figure}[hbt!]
  \centering
  \includegraphics[width=0.5\linewidth]{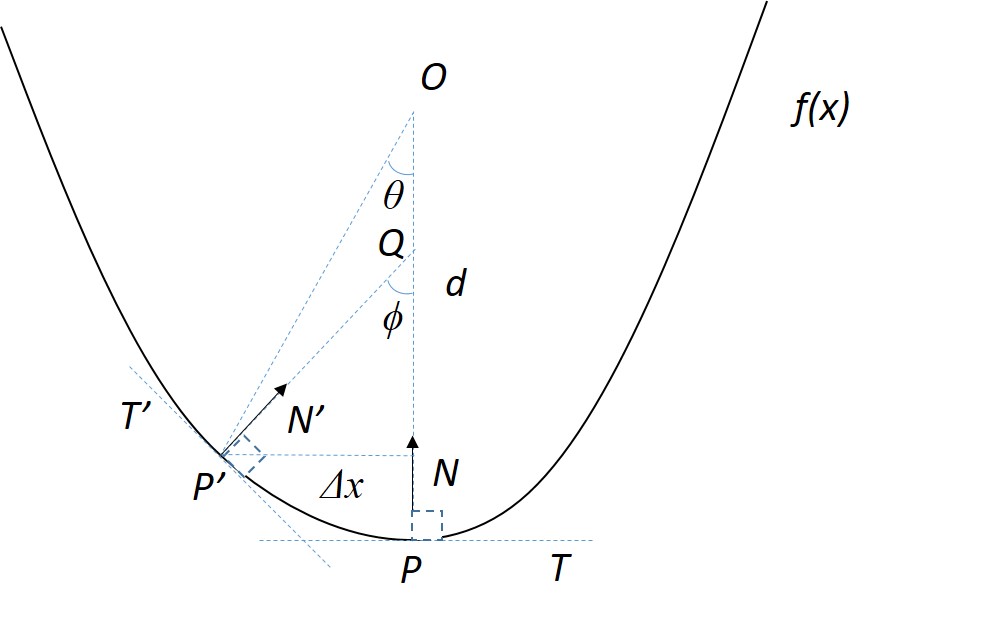}
  \caption{\textit{Illustrating the variables for derivation.}}
  \label{fig:27}
\end{figure}

Tangent $\Vec{T'}$ and normal $\Vec{N'}$ are constructed at $P'$. The normals $\Vec{N}$ and $\Vec{N'}$ intersect at $Q$ at an angle $\phi$. Therefore,
\begin{equation}
    \begin{aligned}
    cos(\phi) & = \frac{\Vec{N}\cdot\Vec{N'}}{|\Vec{N}|\cdot|\Vec{N'}|} \\
              & = \frac{1}{\sqrt{p^{2} + 1}\sqrt{p'^{2} + 1}}\cdot\left[ \begin{array}{c}{-p}\\{1}
    \end{array}\right] \cdot \left[ \begin{array}{c}{-p'}\\{1}
    \end{array}\right] \\
              & = \frac{pp' + 1}{\sqrt{p^{2} + 1}\sqrt{p'^{2} + 1}}. 
    \end{aligned}
\end{equation}
So, 
\begin{equation}
    \phi = cos^{-1}\Big(\frac{pp' + 1}{\sqrt{p^{2} + 1}\sqrt{p'^{2} + 1}}\Big).
\end{equation}
Also, if the line joining $P'$ and the imaging point $O$ intersect $\overline{OP}$ at angle $\theta$, as $\Delta x$ is much smaller compared to $d$,
\begin{equation}
\begin{aligned}
    & tan(\theta)  = \frac{\Delta x}{d} \\
    or, \ & \theta  = tan^{-1}\big(\frac{\Delta x}{d}\big).
\end{aligned}
\end{equation}
$\epsilon$-orthographic bounds are dependent on both the FOV and the curvature at the concerned point.  For a point $P'$ on $C$ to lie within the $\epsilon$-orthographic region for capturing point $O$ at a height $d$ from the point $P$, it must satisfy-
\begin{itemize}
    \item $\theta \leq \epsilon$, so that the point $P'$ lies within the $\epsilon$-FOV.
    \item $\phi \leq \epsilon$. This condition is required because as curvature of $C$ increases around $P$, although a point close to it may remain within the $\epsilon$-FOV bound, the high curvature causes very small region around $P$ to be approximately orthographic. With reference to figure \ref{fig:27}, if curvature at $P$ increases, $OP$ and $OP'$ may differ so much that both points cannot be considered in the same orthographic image.
\end{itemize}
The algorithm for computing the orthographic bounds for a smooth curve $f(x)$ at a central point $P_{0}(x_{0},f(x_{0}))$, for $\epsilon$ angular FOV, resolution $dx$ and imaging height $d$ is stated.

\begin{algorithm}
\caption{Finding Orthographic Bounds of a Curve }\label{alg:ortholeft}
\begin{algorithmic}[1]
\Procedure{Left Orthographic Bound}{$x_{0},\epsilon, d, dx$}
    \State Find $p_{0} = f'(x_{0})$
    \State Set $x = x_{0} - dx$
    \State Find $p = f'(x)$ and calculate $\phi$ using equation 3.28
    \State Set $\Delta x = |x - x_{0}|$
    \State Calculate $\theta$ using equation 3.29
    \While{$\phi \leq \epsilon$ and $\theta \leq \epsilon$}
        \State Set $x = x - dx$
        \State Find $p = f'(x)$ and calculate $\phi'$ using equation 3.28
        \State $\phi = \phi'$'
        \State Set $\Delta x = |x - x_{0}|$
        \State Calculate $\theta'$ using equation 3.29
        \State $\theta = \theta'$
    \EndWhile
    \State \textbf{return} $x = x + dx$
\EndProcedure
\Procedure{Right Orthographic Bound}{$x_{0},\epsilon, d, dx$}
    \State Find $p_{0} = f'(x_{0})$
    \State Set $x = x_{0} + dx$
    \State Find $p = f'(x)$ and calculate $\phi$ using equation 3.28
    \State Set $\Delta x = |x - x_{0}|$
    \State Calculate $\theta$ using equation 3.29
    \While{$\phi \leq \epsilon$ and $\theta \leq \epsilon$}
        \State Set $x = x + dx$
        \State Find $p = f'(x)$ and calculate $\phi'$ using equation 3.28
        \State $\phi = \phi'$'
        \State Set $\Delta x = |x - x_{0}|$
        \State Calculate $\theta'$ using equation 3.29
        \State $\theta = \theta'$
    \EndWhile
    \State \textbf{return} $x = x - dx$
\EndProcedure
\end{algorithmic}
\end{algorithm}
\begin{figure}[hbt!]
  \centering
  \includegraphics[width=0.6\linewidth]{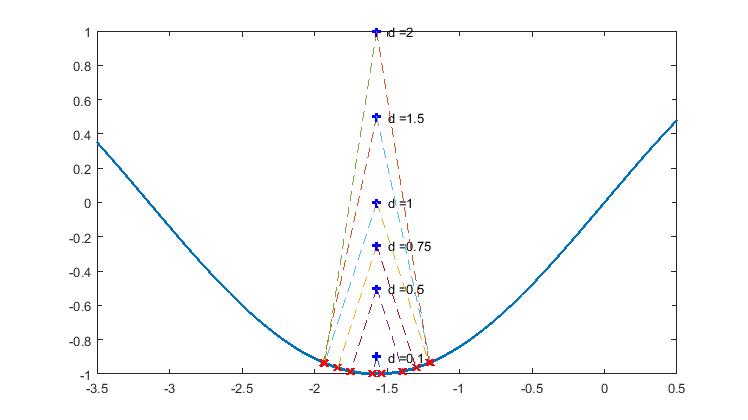}
  \caption{\textit{The orthographic bounds for a convex curve for increasing $d$.}}
  \label{fig:28}
\end{figure}

\begin{figure}[hbt!]
  \centering
  \includegraphics[width=0.6\linewidth]{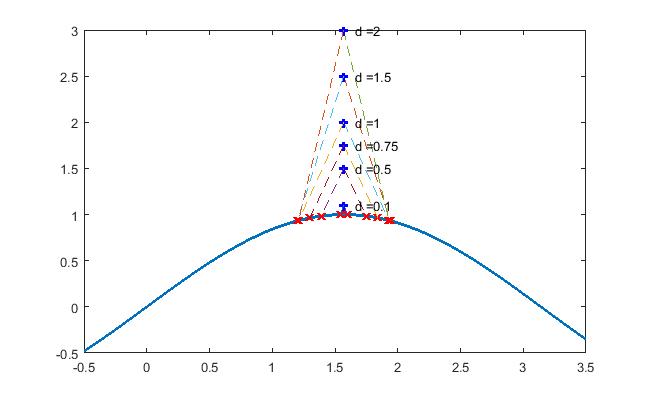}
  \caption{\textit{The orthographic bounds for a concave curve for increasing $d$.}}
  \label{fig:29}
\end{figure}
Algorithm \ref{alg:ortholeft} has been implemented in \textsc{Matlab} and the bounds have been calculated and plotted for a convex curve and a concave curve as shown in figures \ref{fig:28} and \ref{fig:29}. It is to be noted that although $d$ keeps increasing, the bounds do not spread after a point. If the bounds were a function of $\epsilon$ only, then with increase in $d$, they would have spread apart indefinitely which is not a true characteristic of orthography. This shows that $\epsilon$-orthography not only depends on the FOV but also the curvature.

\subsection{Derivation of $\epsilon$-Orthographic Boundary for Surfaces}\label{subsection:deriveortho}
The derivation is very similar to that for the curves. A surface $S$ is given by a bi-variate function, $z = f(x,y)$. Given a central point $P(x,y,z)$ on the surface, an imaging height of $d$ and useful FOV $\epsilon$, the goal is to find the orthographic boundary surrounding $P$, or in other words, the area around $P$ which can be considered as an approximate orthographic image. 

From equation 3.2 and 3.3, the surface normal at point $P(x,y,z)$ is $[p, q, -1]$, where $p = \frac{\partial f(x,y)}{\partial x}$ and $q = \frac{\partial f(x,y)}{\partial y}$. The Hessian matrix of $f(x,y)$ is 
\begin{equation}
    H=\left[\begin{array}{cc}{\frac{\partial^{2} f(x, y)}{\partial x^{2}}} & {\frac{\partial^{2} f(x, y)}{\partial x \partial y}} \\ {\frac{\partial^{2} f(x, y)}{\partial y \partial x}} & {\frac{\partial^{2} f(x, y)}{\partial y^{2}}}\end{array}\right].
\end{equation}
Let us take a point $P'(x',y',z')$ very close to $P$ such that
\begin{equation}
    \begin{aligned}
    x' & = x + \Delta x \\
    y' & = y + \Delta y.
    \end{aligned}
\end{equation}
As $\Delta x$ and $\Delta y$ are very small quantities, the change in the  surface normal vector is also small. So the surface normal at $P'$, $\Vec{N'} =  [p', q', -1]$ and it can be approximated as
\begin{equation}
    \begin{aligned}
    & \left[\begin{array}{cc}{\Delta p}\\{\Delta q}\end{array}\right] =  H\cdot \left[\begin{array}{cc}{\Delta x}\\{\Delta y}\end{array}\right] \\
    & p' = p + \Delta p \\
    & q' = q + \Delta q \\
    \end{aligned}
\end{equation}

Similarly as equation 3.27 and 3.29, $\phi$ and $\theta$ are calculated as 
\begin{equation}
    \begin{aligned}
    cos(\phi) & = \hat{N}.\hat{N'} \\
              & = \frac{\Vec{N}\cdot\Vec{N'}}{|\Vec{N}||\Vec{N'}|} \\
              & = \frac{1}{\sqrt{p^{2} + q^{2} + 1}\sqrt{p'^{2} + q'^{2} + 1}}\left[ \begin{array}{c}{p}\\{q}\\{-1}
    \end{array}\right] \cdot \left[ \begin{array}{c}{p'}\\{q'}\\{-1}
    \end{array}\right] \\
              & = \frac{pp' + qq' + 1}{\sqrt{p^{2} + q^{2} + 1}\sqrt{p'^{2} + q'^{2} + 1}}. 
    \end{aligned}
\end{equation}

So, 
\begin{equation}
    \phi = cos^{-1}\Big(\frac{pp' + qq' + 1}{\sqrt{p^{2} + q^{2} + 1}\sqrt{p'^{2} + q'^{2} + 1}}\Big)
\end{equation}

and
\begin{equation}
\begin{aligned}
    & tan(\theta)  = \frac{\sqrt{\Delta x^{2} + \Delta y^{2}}}{d} \\
    or, \ & \theta  = tan^{-1}\Big(\frac{\sqrt{\Delta x^{2} + \Delta y^{2}}}{d}\Big).
\end{aligned}
\end{equation}

Now, if $P'$ belongs to the orthographic region around point $P$ for an imaging height $d$, then both $\theta \leq \epsilon$ and $\phi \leq \epsilon$.

\subsection{Implementation}\label{subsection:impleortho}
The following algorithm utilizes equations 3.31 - 3.35 for numerically computing the $\epsilon$-orthographic boundary for a smooth surface $S$ ($z = f(x,y)$) at a central point $P_{0}(x_{0},y_{0},z_{0})$, for $\epsilon$ angular FOV, resolutions $dx$ and $dy$, and imaging height $d$.

\begin{algorithm}
\caption{Finding Orthographic Boundary of a Surface }\label{alg:orthoboundary}
\begin{algorithmic}[1]
\State Find surface normal components $p_{0}$ and $q_{0}$ at $P_{0}$.
\State Create an empty point set $P$ for storing the eligible points inside the orthographic region.
\State Append $P_{0}$ to $P$
\State Compute the number of $x$ or $y$ co-ordinates in the grid. $n_{x} = (r_{H} - r{L})/dx = R_{x}/dx$ and $n_{y} = R_{y}/dy$.
\State Set $s = max(n_{x},n_{y})$
\State Set $buff = 0$
\For{$n = 1:s$}
    \State Vector $out = PairGen(n)$
    \State Set $count = 0$
    \For{$j = 1:length(out)$}
        \State $x_{1} = x_{0} + dx\cdot out(1,j)$
        \State $y_{1} = y_{0} + dy\cdot out(2,j)$ 
        \State Compute surface normal vector at $P_{1}(x_{1},y_{1},z_{1})$
        \State Calculate $\phi$ using equation 3.34
        \State Calculate $\theta$ using  equation 3.35
        \If{$\theta \leq \epsilon \ \ and \ \ \phi \leq \epsilon$}
            \State Append $P_{1}(x_{1},y_{1},z_{1})$ to 
            \State $count = count + 1$
        \EndIf
    \EndFor
    \If{$count = 0$}
        \State $buff = buff + 1$
    \EndIf
    \If{$buff > 3$}
        \State Break  from loop
    \EndIf
\EndFor
\end{algorithmic}
\end{algorithm}

\begin{figure}[htb!]
\centering
\subfigure[$d = 2, \ (x_{0},y_{0}) = (0,0)$]{\label{fig:sub5}\includegraphics[width=0.4\linewidth]{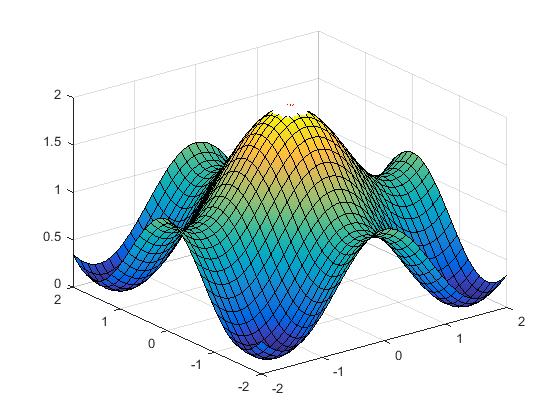}}
\hfill
\subfigure[$d = 2, \ (x_{0},y_{0}) = (0,0)$]{\label{fig:sub6}\includegraphics[width=0.4\linewidth]{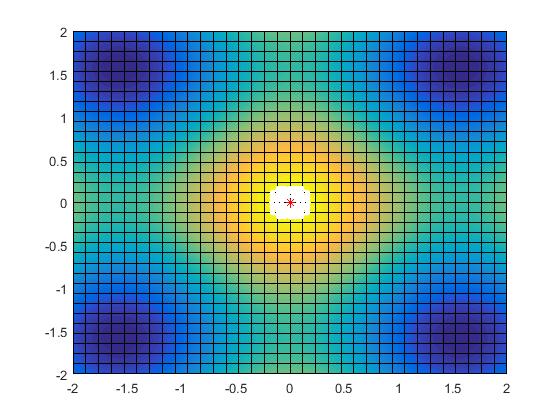}}

\subfigure[$d = 2, \ (x_{0},y_{0}) = (0,-1)$]{\label{fig:sub11}\includegraphics[width=0.4\linewidth]{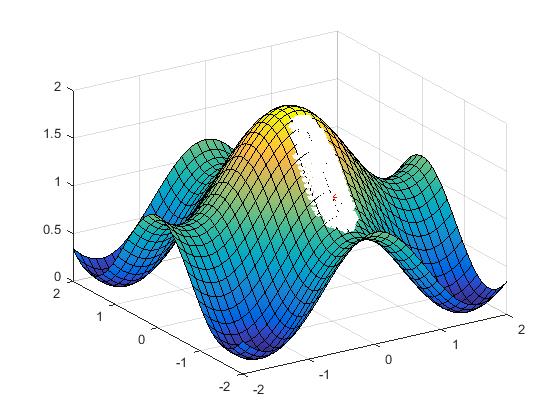}}
\hfill
\subfigure[$d = 2, \ (x_{0},y_{0}) = (0,-1)$]{\label{fig:sub12}\includegraphics[width=0.4\linewidth]{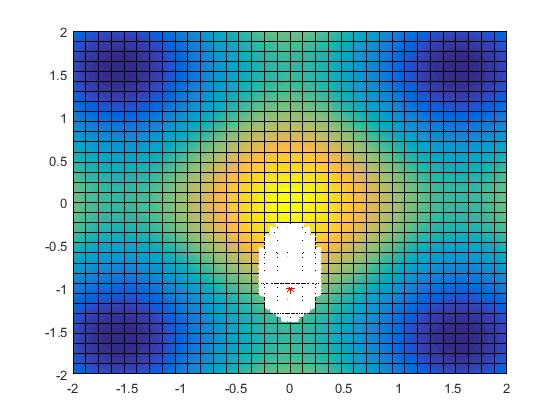}}

\subfigure[$d = 2, \ (x_{0},y_{0}) = (-1,-1)$]{\label{fig:sub7}\includegraphics[width=0.4\linewidth]{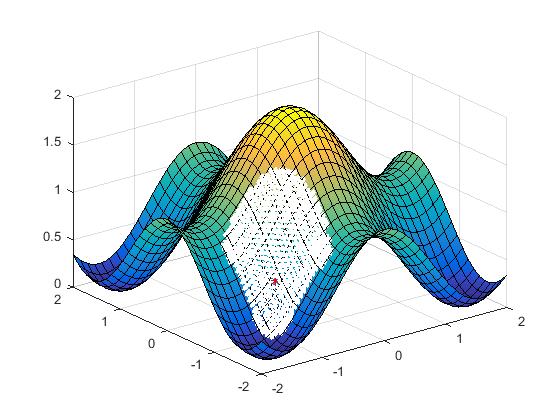}}
\hfill
\subfigure[$d = 2, \ (x_{0},y_{0}) = (-1,-1)$]{\label{fig:sub8}\includegraphics[width=0.4\linewidth]{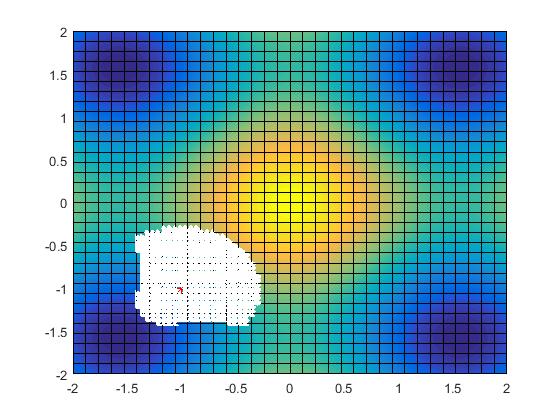}}

\caption{Orthographic regions drawn on curve $f(x,y) = cos^{2}(x) + cos^{2}(y)$ shown in white. The figures on right show the boundary shape. The central point $(x_{0},y_{0})$ is plotted in red. (Here $\epsilon = 10^{\circ}$)}
\label{fig:30}
\end{figure}

The function \textit{PairGen} generates a vector of all pairs of integers $n_{1}$ and $n_{2}$ such that $|n_{1}| + |n_{2}| = n$, i.e. all co-ordinates located at absolute distance $n$. Algorithm \ref{alg:orthoboundary} has been implemented in \textsc{Matlab} to generate $\epsilon$-orthographic regions on smooth curves (figure \ref{fig:30}). This algorithm is also valid for non-smooth curves, for which numerical gradients can be calculated at non-differentiable points. 

Also, in all the aforementioned derivations, the gradient components at $P'$ have been calculated by approximations for fast computation. Where computational capability is not an issue, the actual gradients can be calculated by differentiating the function $f(x,y)$. The following figures illustrate the implemented algorithm.

\subsection{Special Surfaces}
\textbf{\textit{Conjecture:}} Points on surfaces of constant Gaussian curvature (\ref{curvatures}) form $\epsilon$-orthographic regions of same area for constant imaging height $d$. The upper bound on $d$ depends on the nature(parameters) of such surfaces.

Surfaces of constant curvatures can be classified into the following three classes-
\begin{enumerate}
    \item \textit{\textbf{Zero Curvature Surfaces }}- A surface with Gaussian curvature($\kappa$) equal to zero at all points is a plane. For a plane, which is inherently orthographic, the calculated region is thus of same shape as the \textit{FOV}. As the \textit{FOV} is considered circular in all our calculations, the orthographic region is thus circular for a planar surface, the radius of which depends on the imaging height as given by \textit{equation} 3.41. Thus the problem of finding optimal capture points is reduced to a \textit{Circle Packing} problem.
    
    \item \textit{\textbf{Positive Curvature Surfaces }}- A surface with equal positive Gaussian curvature($\kappa$) at all points is sphere. The equation of a sphere is given by
    \begin{align*}
        z^{2} = a^{2} - x^{2} - y^{2},
    \end{align*}
    where $a$ is the radius of the sphere.\\
    Using the definition of $\epsilon$-orthography, it can be shown that for a sphere, the orthographic regions are also circular and of constant radii, dependent on the imaging height $d$. This property is demonstrated by plotting in \textsc{MATLAB} using algorithm  \ref{alg:orthoboundary} as shown in fig \ref{fig:64}. This is due to the fact that the two principal curvatures ($\kappa_1$ and $\kappa_2$, [\ref{curvatures}]) at any point on the sphere are equal and constant. However, unlike a plane, sphere is not inherently orthographic but behaves like one. Because of this property, similar to the planar case, the problem of finding optimal capture points can be reduced to the problem of \textit{circle packing} on a hemisphere. It can be conjectured that, with appropriate transformations, the problem can be reduced to \textit{circle packing} within a circle.
    \begin{figure}[htb!]
    \centering
    \subfigure[3D view of the surface]{\label{fig:sub64a}\includegraphics[width=0.4\linewidth]{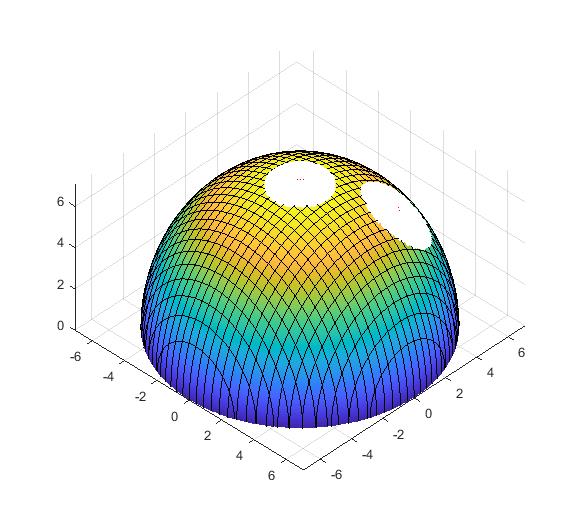}}
    \hfill
    \subfigure[2D top view of the surface]{\label{fig:sub64b}\includegraphics[width=0.4\linewidth]{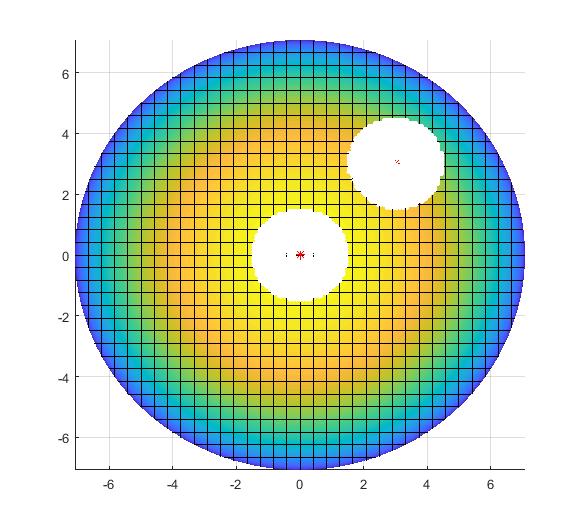}}
    \caption{$\epsilon$-Orthographic regions plotted on a sphere- a surface of constant positive Gaussian curvature.}
    \label{fig:64}
    \end{figure}
    
    \item \textit{\textbf{Negative Curvature Surfaces }}- A surface with equal negative Gaussian curvature($\kappa$) at all points is a pseudosphere. The equation of a pseudosphere is given by
    \begin{align*}
        z^{2} = \left[ a\cdot sech^{-1}\bigg(\sqrt{\frac{x^2 + y^2}{a}}\bigg) - \sqrt{a^2 - x^2 - y^2} \right]^2
    \end{align*}
    where $a$ is the radius of the pseudosphere.\\
    Unlike the other two cases, for a pseudosphere, the orthographic boundaries are not circular, and the limit on imaging height $d$ is dependent on the radius $a$. A pseudosphere is plotted in \textsc{MATLAB} along with four orthographic regions, calculated using algorithm \ref{alg:orthoboundary}, as illustrated in figure \ref{fig:65}. Among the two principal curvatures ($\kappa_1$ and $\kappa_2$) calculated at any point on the surface, one is positive and the other is negative. Interestingly, as we move along the surface, from the flat region to the narrow region, the magnitude of the positive principal curvature increases and the negative principal decreases such that the product of the two (Gaussian curvature) remains the same. As a consequence of this property, it has been empirically observed that the size of the $\epsilon$-orthographic regions remain the same, although the shape may vary (figure \ref{fig:65}).
    \begin{figure}[htb!]
    \centering
    \subfigure[3D view of the surface]{\label{fig:sub65a}\includegraphics[width=0.4\linewidth]{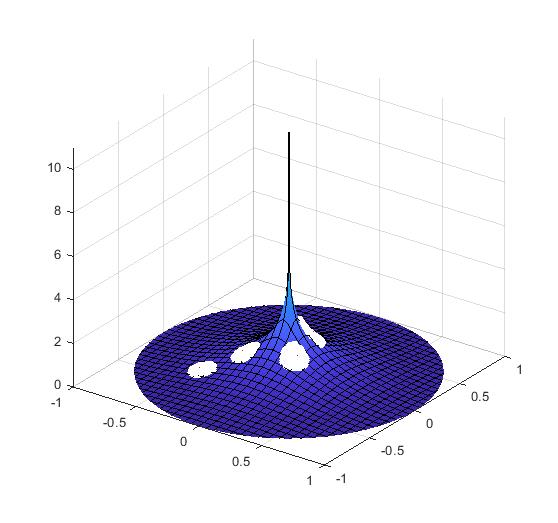}}
    \hfill
    \subfigure[2D top view of the surface]{\label{fig:sub65b}\includegraphics[width=0.4\linewidth]{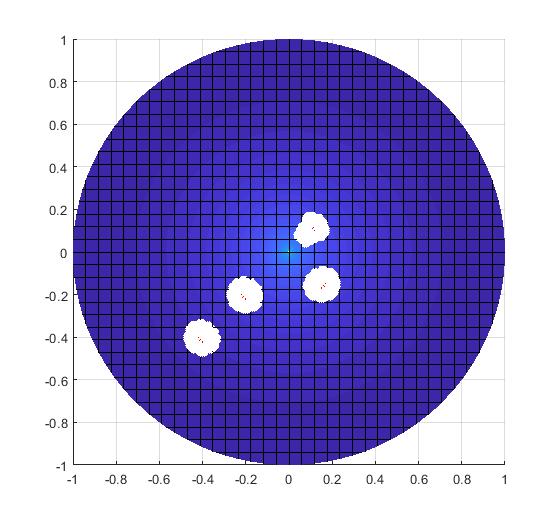}}
    \caption{$\epsilon$-Orthographic regions plotted on a pseudosphere- a surface of constant negative Gaussian curvature (large regions are plotted for demonstration).}
    \label{fig:65}
    \end{figure}
\end{enumerate}

\subsection{Limitations}
\begin{itemize}
    \item The calculation of orthographic regions is computationally expensive, specially for higher resolutions. Consequently, finding the overlap between two regions is also expensive. 
    \item The entire orthographic region needs to be calculated to find the boundary. 
    \item For non-smooth surfaces, gradients cannot be calculated at non-differentiable points.
    \item For natural surfaces and very fast varying curvatures,  orthographic boundaries are difficult to calculate.
    \item The exact boundaries cannot be used for optimization problem to calculate optimal capture points, where the boundaries need to be computed at multiple points simultaneously, which is repetitive and slow. 
\end{itemize}

\section{Approximation of $\epsilon$-Orthography}\label{epsi_appx}
As pointed out in the limitations, calculating the exact orthographic region and hence the boundary is not practical because it is computationally expensive and thus time consuming. However, instead of considering exact boundaries, they can be approximated to some regular shapes for faster boundary computation as well as calculating overlap between regions.
\subsection{Curvatures} 
\begin{figure}[hbt!]
  \centering
  \includegraphics[width=0.45\linewidth]{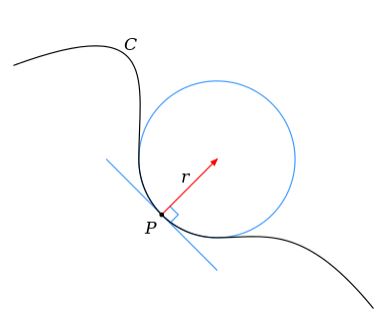}
  \caption{\textit{Illustration of osculating circle and radius of curvature.}}
  \label{fig:31}
\end{figure}

The curvature of a smooth curve at a point is usually defined as the curvature of its osculating circle at that point. If $C$ is a plane curve, then the curvature of $C$ at a point is the measure of how sensitive its tangent line is to moving the point to other nearby points. Geometrically, the curvature of a straight line is considered to be constantly zero and. Also, a circle of small radius should have a large curvature and of large radius should have a small curvature. Thus curvature of a circle is defined to be the reciprocal of its radius.
\cite{kcurvature}
\begin{equation}
    \kappa = \frac{1}{R}.
\end{equation}
Given a point $P$ on $C$, there is a circle or line which most closely approximates the curve near $P$, which is the osculating circle at $P$. So, the curvature of $C$ at $P$ is then defined to be the curvature of that circle or line. Consequently, the radius of curvature is the reciprocal of the curvature. (Figure \ref{fig:31})

For a plane curve $C$, parametrically expressed in \textit{Cartesian co-ordinates} as $\gamma(t) = (x(t),y(t))$, the curvature $\kappa$  given as
\begin{equation}
    \kappa=\frac{\left|x^{\prime} y^{\prime \prime}-y^{\prime} x^{\prime \prime}\right|}{\left(x^{\prime 2}+y^{\prime 2}\right)^{\frac{3}{2}}},
\end{equation}
and the signed curvature $k$ is given as
\begin{equation}
    k=\frac{x^{\prime} y^{\prime \prime}-y^{\prime} x^{\prime \prime}}{\left(x^{\prime 2}+y^{\prime 2}\right)^{\frac{3}{2}}},
\end{equation}
where $x^{\prime} = \frac{dx}{dt}$ and $y^{\prime} = \frac{dy}{dt}$. 
\subsection{Curvature of Surfaces} \label{curvatures}
\begin{figure}[hbt!]
  \centering
  \includegraphics[width=0.55\linewidth]{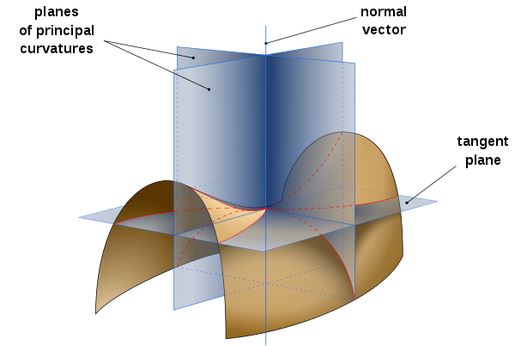}
  \caption{\textit{Saddle surface with normal planes in directions of principal curvatures. (source: Wikipedia)}}
  \label{fig:32}
\end{figure}
Consider a point $P$ on the surface $S$. All curves lying on the surface and passing through $P$ with the same tangent vector $\Vec{T}$ will have the same normal curvature, which is the same as the curvature of the curve obtained by intersecting the surface with the \textit{tangent plane} and the plane containing the surface normal vector $\Vec{N}$. Taking all possible tangent vectors, the maximum and minimum values of the normal curvature at a point are called the principal curvatures, $k_{1}$ and $k_{2}$, and the directions of the corresponding tangent vectors are called \text{principal normal directions}. Curvatures are evaluated along surface normal sections. 

An \textit{intrinsic} measure of surface curvature at a point $P$ on the surface is the \textit{Gaussian Curvature} ($K$). Whereas, an \textit{extrinsic} measure of surface curvature is the \textit{Mean Curvature} ($H$). They are given as
\begin{equation}
\begin{aligned}
    K  & =  k_{1}\times k_{2},   \\
    H  & =  \frac{k_{1} + k_{2}}{2},
\end{aligned}
\end{equation}
where $k_{1}$ and $k_{2}$ are principal curvatures. For a surface given by $z = f(x,y)$, the Gaussian and Mean curvatures can be expressed in terms of $x$ and $y$ as
\begin{equation}
    \begin{aligned}
    f_{x} & = \frac{\partial f}{\partial x} \ \ \ \ f_{y} = \frac{\partial f}{\partial y} \\
    f_{xx} & = \frac{\partial^{2} f}{\partial x^{2}} \ \ \ \ f_{xy} = \frac{\partial^{2} f}{\partial x \partial y} \ \ \ \    f_{yy}  = \frac{\partial^{2} f}{\partial y^{2}} \\
    K & = \frac{f_{xx}\cdot f_{yy} - {f_{xy}}^{2}}{(1 + {f_{x}}^{2} + {f_{y}}^{2})^{2}} \\
    H & = \frac{(1 + {f_{x}}^{2})\cdot f_{yy} + (1 + {f_{y}}^{2})\cdot f_{xx} - 2f_{x}\cdot f_{y}\cdot f_{xy}}{(1 + {f_{x}}^{2} + {f_{y}}^{2})^{\frac{3}{2}}}.
    \end{aligned}
\end{equation}
For Gaussian curvature, both convex minima and concave maxima have positive curvatures, and saddle points have negative curvature. Whereas, for Mean curvature, convex portions have positive and concave portions have negative curvature, and saddle points have curvatures close to zero. This is illustrated in the following figure \ref{fig:33}. 
\begin{figure}[hbt!]
  \centering
  \includegraphics[width=1\linewidth]{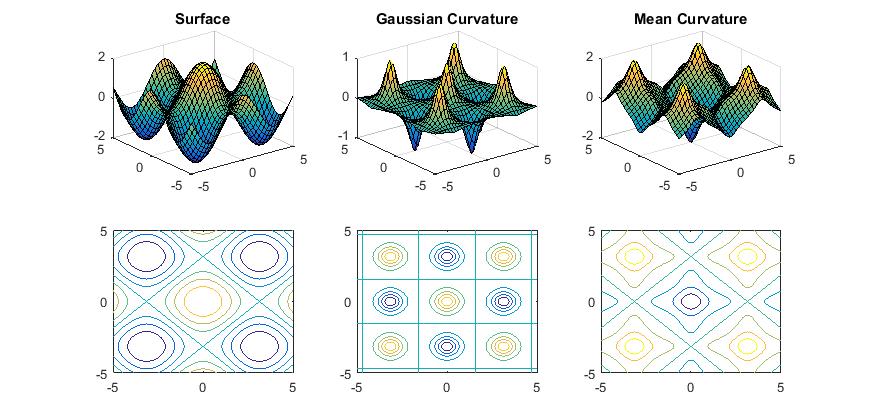}
  \caption{\textit{Surface plots and contour plots for $f(x,y) = cos(x) + cos(y)$.}}
  \label{fig:33}
\end{figure}

\subsection{Boundary Approximation}
As stated before, the approximation of orthographic boundaries of the regions calculated in sections \ref{subsection:deriveortho} and \ref{subsection:impleortho} is necessary for faster computation and calculation of overlaps between regions. The following approaches can be explored for approximating the boundary.
\subsubsection{Polygonal Approximation}
In this approach, instead of calculating all the points on the boundary, $N$ points are calculated in $N$ different directions from the central point. By setting $\Delta x$ and $\Delta y$ in equation 3.35 according to the direction of calculation, and by calculating $\theta$ and $\phi$ using equations 3.34 and 3.35, and checking at each step to see whether they maintain the $\epsilon$ constraint, the boundary point in the concerned direction can be evaluated numerically. If we consider a $N$-polygon, then the directions in which the boundary points must be calculated, should be at equal angles to each other at the central point $(x_{0}, y_{0})$, i.e. the directions should be at an angle $\theta = \frac{360^{\circ}}{N}$ from each other. 
\begin{figure}[hbt!]
  \centering
  \includegraphics[width=0.5\linewidth]{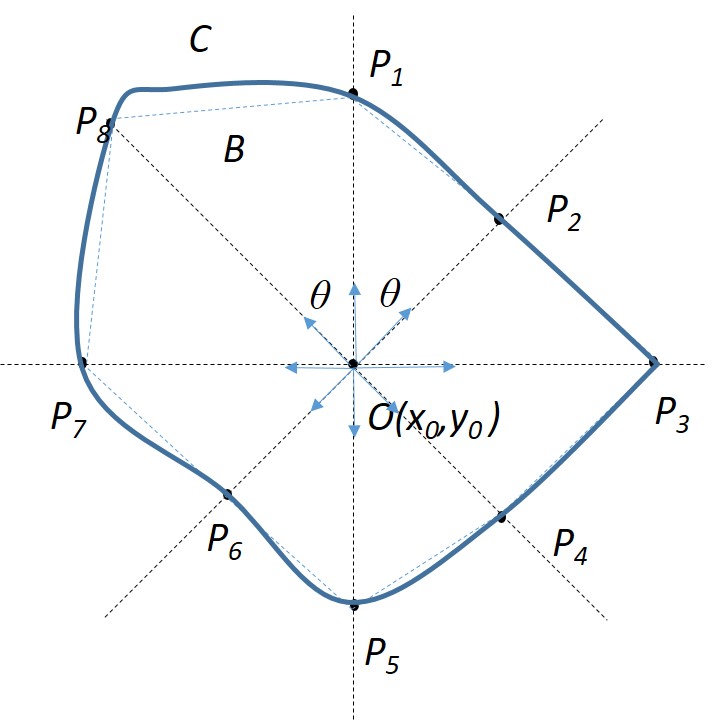}
  \caption{\textit{Boundary points detected for $N = 8$. Here $\theta = 45^{\circ}$. Actual boundary $C$ shown in bold. }}
  \label{fig:34}
\end{figure}
Figure \ref{fig:34} shows an example of polygonal aprroximation of the boundary. Here the boundary points $P_{i}$'s are evaluated in 8 directions centered at $O(x_{0},y_{0})$. Obviously, larger the number of directions taken, better will be the approximate boundary. This approach may lead to both over-estimation and under-estimation of boundary depending on the convexity of the boundary curve.
\subsubsection{Elliptical Approximation}
This is an extension or further approximation of the \textit{polygonal approximation} approach but here only even sided polygons are considered. In this case, the orthographic boundary is approximated as an ellipse. 

\begin{itemize}
    \item Boundary points are calculated in $N$ different equiangularly spaced directions. Hence, we get $N$ boundary points $P_{i}, \ \ i = 1,2,...,N$. 
    \item Now, the distances between the diagonally opposite boundary points is calculated, and thus we have $N/2$ diagonals ($d_{i}$).
    \item The maximum and minimum diagonals are considered, $d_{max} = max(d_{i})$ and $d_{min} = min(d_{i})$.
    \item The boundary is approximated as an ellipse with the major axis as $d_{max}$ and the minor axis as $d_{min}$ and the major axis is aligned along the longest diagonal.
\end{itemize}
\begin{figure}[hbt!]
  \centering
  \includegraphics[width=0.5\linewidth]{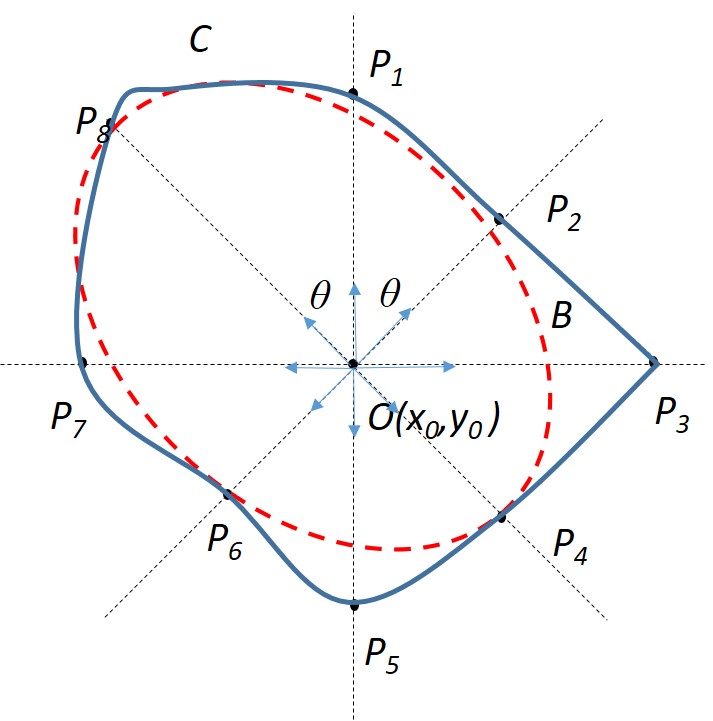}
  \caption{\textit{Actual boundary $C$ and approximated elliptical boundary $B$ for the polygonal approximation of fig \ref{fig:34}.}}
  \label{fig:35}
\end{figure}
In the illustrated figure \ref{fig:35}, the maximum length diagonal is $\overline{P_{4}P_{8}}$ and the minimum length diagonal is $\overline{P_{2}P_{6}}$. The major axis of the constructed ellipse($B$) is $\overline{P_{4}P_{8}}$ and the minor axis is of same length as  $\overline{P_{2}P_{6}}$. It is to be noted that central point $O(x_{0},y_{0})$ is not the centre of the ellipse.
\subsubsection{Circular Approximation - I}
This is a further simplification of the \textit{elliptical approximation}. In this case, the orthographic boundary is approximated as a circle.
\begin{itemize}
    \item Boundary points are calculated in $N$ different equiangularly spaced directions as discussed in \textit{polygonal} case. Hence, we get $N$ boundary points $P_{i}, \ \ i = 1,2,...,N$. 
    \item Now, the distances of the boundary points from the central point $(x_{0},y_{0})$ is calculated, and thus we have $N$ distances ($d_{i}$).
    \item The average of all the distance lengths is calculated, $d_{avg} = \Sigma d_{i}$.
    \item The boundary is approximated as a circle centered at $(x_0,y_0)$ and of radius $R = d_{avg}$.
\end{itemize}
\begin{figure}[hbt!]
  \centering
  \includegraphics[width=0.5\linewidth]{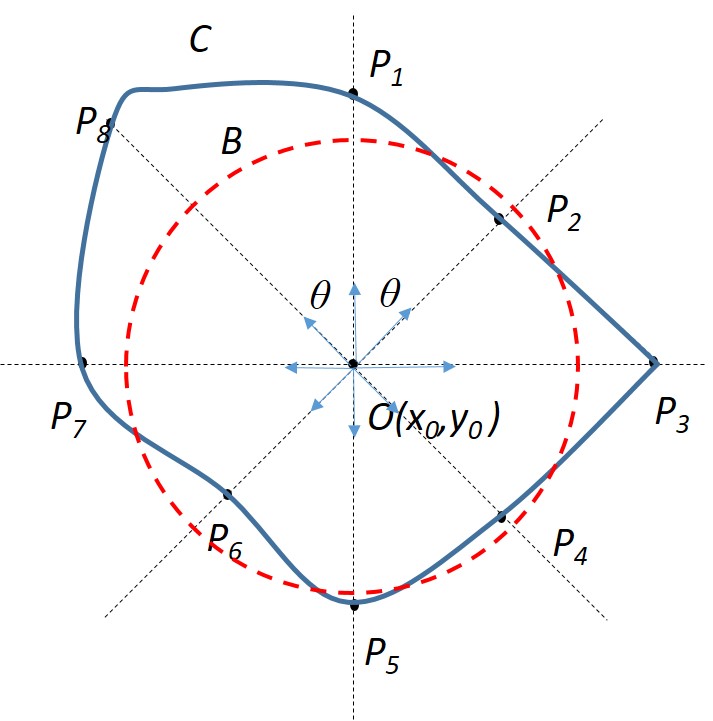}
  \caption{\textit{Actual boundary $C$ shown in blue and approximated circular boundary $B$ shown in red for the polygonal approximation of fig \ref{fig:34}.}}
  \label{fig:36}
\end{figure}
In the illustrated figure \ref{fig:36}, $d_{i} = \overline{OP_i}$, and the average of all 8 $\overline{OP_i}$'s is calculated. The boundary circle $B$ is constructed with centre at $O(x_0,y_0)$ and radius equal to the average of $OP_i$'s. 
\subsubsection{Circular Approximation - II}
This approach is the simplest and most intuitive approach among the ones discussed. It is based on the observation that on a smooth surface, the orthographic regions are small in places of high absolute curvature and relatively larger in places of low absolute curvature values. Considering the surfaces shown in figures \ref{fig:30} and \ref{fig:33}, it can be observed that at points where the orthographic region is large, the \textit{absolute gaussian curvature} is small and the points at which the region is small, the curvature is large. 

If the orthographic region or boundary is estimated by a circle, an inverse relation between the radius of the boundary and the curvature of the central point can be formulated. Also, for a planar surface or a zero curvature surface, the orthographic region is circular with radius
\begin{equation}
    R = d\cdot tan(\epsilon),
\end{equation}
where $d$ is the imaging distance and $\epsilon$ is the useful FOV as discussed in the derivation of $\epsilon$-orthography. For any point on a non-planar surface having an absolute curvature $|K| \geq 0$, the boundary will shrink from this circle. Therefore, if the region boundary is approximated by a circle of radius $r$, $r \leq R$.

Now, let us consider a surface $S$ and its Gaussian curvature($K$) is given by equation 3.40. Given the bounds of the surface, the maximum absolute curvature is calculated.
\begin{equation}
    K_{max} = \max_{x,y}|K(x,y)|, \ \ \ \ (x,y) \ \in \ Bound(S).
\end{equation}
Let us fix a ratio($m$) between the largest radius possible $(r_{max} = R)$ for the points of least absolute curvature and the least radius possible for the point having curvature $K_{max}$. So, $m = \frac{r_{max}}{r_{min}}$. Therefore, the radius of the approximated circular boundary can be expressed as a function of point $(x,y)$ as
\begin{equation}
    r(x,y) = R - \frac{|K(x,y)|}{K_{max}}\cdot R \cdot(1 - \frac{1}{m}).
\end{equation}
The value of $m$ can be tuned by experimental observations. For the optimization problems in the next chapter, the value of $m$ is set to 5.

\subsubsection{Comparison of Boundary Approximations}
The four different approaches for approximating the orthographic boundary can be compared in terms of accuracy of approximation and computation time.
\begin{itemize}
    \item \textbf{Approach 1} (\textit{Polygonal Approximation}): In this case the computational time depends on the number of directions $N$ or the number of boundary points used for approximation, which increases with increase in $N$. Also, calculation of overlap between two polygonal regions is very time consuming. But this approach gives the best approximation of the orthographic boundary.
    \item \textbf{Approach 2} (\textit{Elliptical Approximation}): This has greater computation time than approach 1 because, after the $N$ boundary points are computed, they need to be compared to calculate the major and minor axes of the ellipse, after which the equation of the ellipse has to be calculated. This is a more crude approximation if compared to approach 1 but better when compared to latter approaches. Also, finding the area of overlap between two ellipses is mathematically complicated and computationally expensive if done numerically.
    \item \textbf{Approach 3} (\textit{Circular Approximation - I}): This approach has almost the same computational time as approach 1 which increases with $N$. Also, higher the value of $N$, better is the approximation. This approach is much preferable for optimization applications, as finding the overlap between two circles is much easier to formulate mathematically than two ellipses.
    \item \textbf{Approach 4} (\textit{Circular Approximation - II}): This approach has the least computational time compared to the previous approaches and thus preferable for further optimization problem formulation. Also, the boundaries being circular, the area of overlap between regions can be formulated mathematically and calculated much faster than approach 1 or 2. But this is the crudest approximation to the orthogonal boundary and the error is very high for natural or fast-varying surfaces.  
\end{itemize}
Because the computation of optimal capture points requires fast calculation of orthogonal boundaries and also the calculation of overlap between regions, \textit{Approach 4} is used in the next chapter for formulating the optimization problem. \\ \\

\chapter{Optimization Problem Formulation}

\label{Chapter4}

\lhead{Chapter 4: \emph{Optimization Problem Formulation}}
In Chapter \ref{Chapter3}, methods for calculating the orthographic regions on a surface were formulated and discussed. Each point on the surface has an orthographic region associated with it. The goal of this chapter is to find the optimal set of points such that, the orthographic regions evaluated at those points cover the entire surface with least overlap among them. For most of the problem formulation discussed in this chapter, it is assumed that the height of the capturing device remains constant for all points. Unless mentioned otherwise, it has also been assumed that from a capture point above the surface, only one view or capture is taken i.e. the capturing device is oriented in a fixed direction. This is the direction opposite to the surface normal at the central point of focus on the surface, which may change for different points, but is fixed for a particular point. Section \ref{section:1} gives an intuitive method of deciding capture points based on density of surface normal vectors. In this section the constraints on imaging height and fixed view direction is relaxed. Section \ref{section:2} deals with the formulation of the optimization problem for determining the capture points for a surface. In this section, different cost functions and variable constraints are discussed for the optimization problem. In section \ref{section:3}, the local and global optimization are compared with respect to the cost functions. Section \ref{section:4} gives a method of deciding orthographic regions or grouping points on the surface based on the optimal points calculated.

\section{Selection of Capture Points Based on Normal Clusters}\label{section:1}
One of the key constraints of the problem being discussed so far is that the camera orientation at a capture point is fixed, or in other words the capturing device at a height $d$ along the surface normal at point $P$ on the surface is allowed to capture only one orthographic view, the region surrounding $P$. Now, suppose this constraint is relaxed such that captures in multiple directions are allowed at a fixed capture point, then it is of interest to identify such points above the surface which can act as a common imaging point for multiple orthographic regions. Also, in this case the constraint on the imaging height $d$ is also relaxed. Effectively, we don't mind what the height is, provided there is an opportunity to do multiple captures.

The qualification for a point($O$) above the surface($S$) to be an imaging point for the orthographic region centred at point $P$ on $S$ is that the surface normal at $P$ must pass through $O$. So, if surface normals are drawn for all the points on $S$, higher the number of normals that pass through $O$, the more favourable it is as a choice of imaging point. In other words, more the number of normals intersecting at or passing very close to a point, the better $P$ is as a capture point. So, in this context, studying normal densities and identifying normal clusters in space above a surface is worthwhile. The figure \ref{fig:37} below, plotted in \textsc{Matlab}, demonstrates normal densities for some known curves.
% \begin{figure}[hbt!]
%   \centering
%   \includegraphics[width=0.5\linewidth]{Pictures/3/fig40.jpg}
%   \caption{\textit{Actual boundary $C$ shown in blue and approximated circular boundary $B$ shown in red for the polygonal approximation of fig \ref{fig:34}.}}
%   \label{fig:36}
% \end{figure}
\begin{figure}[htb!]
\centering
\subfigure[$f(x) = sin(x), \ -\pi \leq x \leq \pi$]{\label{fig:sub5}\includegraphics[width=0.35\linewidth]{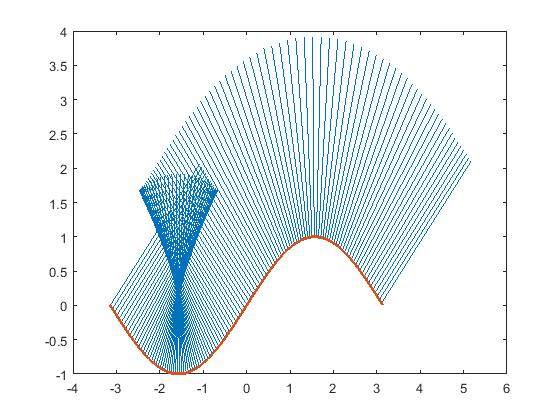}}
\hfill
\subfigure[$f(x) = x^{2}, \ -2 \leq x \leq 2$]{\label{fig:sub6}\includegraphics[width=0.35\linewidth]{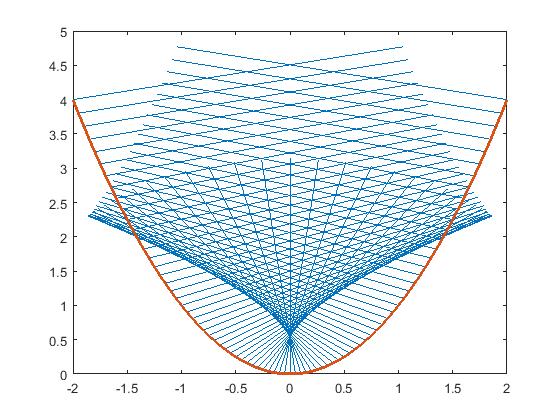}}
\hfill
\subfigure[$f(x) = x^{3}, \ -2 \leq x \leq 2$]{\label{fig:sub11}\includegraphics[width=0.35\linewidth]{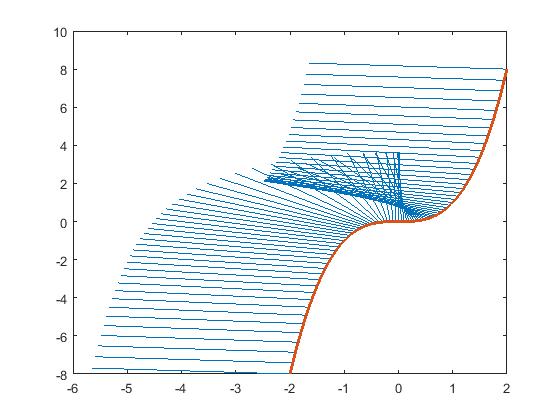}}

\caption{\textit{The curves are plotted in red and normals are plotted in blue. The number of normals intersecting at a point or the density of normal lines, proportional to the intensity of blue, can be observed.}}
\label{fig:37}
\end{figure}

\begin{figure}[htb!]
\centering
\subfigure[Normals drawn at  intervals $\Delta x = 0.05$]{\label{fig:38a}\includegraphics[width=0.45\linewidth]{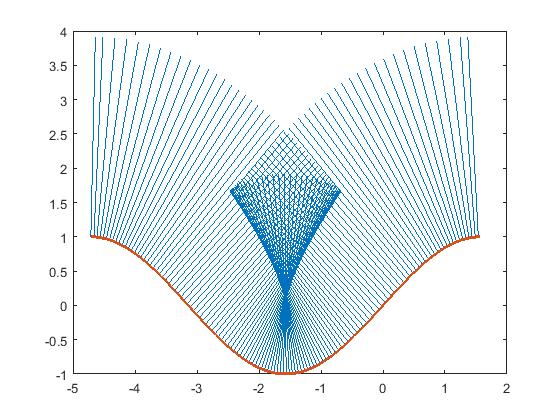}}
\hfill
\subfigure[Normals drawn at  intervals $\Delta x = \frac{\pi}{4}$]{\label{fig:38b}\includegraphics[width=0.5\linewidth]{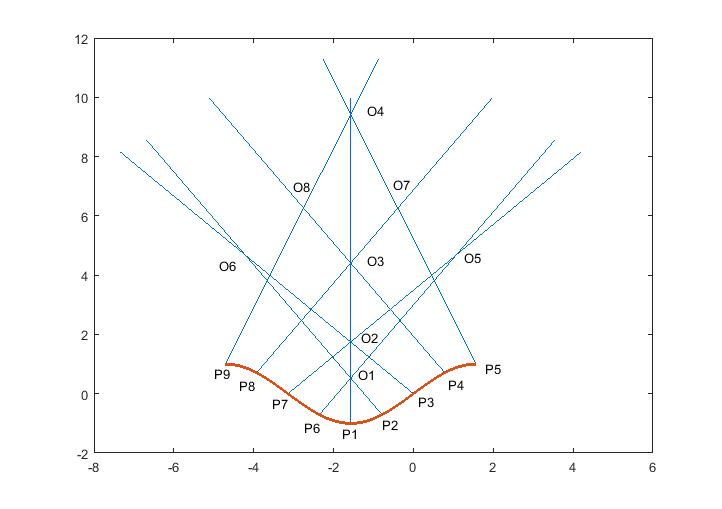}}
\caption{\textit{Normals(shown in blue) drawn for the convex part of a sine curve.}}
\label{fig:38}
\end{figure}

Although a region of high normal density or \textit{normal cluster} is favourable for choosing capture points, all the points having the same density are not equally favourable. Let us consider the example of the curve $f(x) = sin(x)$ and look at the range $-\frac{3\pi}{2} \leq x \leq \frac{\pi}{2}$, which is the convex part or a trough (figure \ref{fig:38}).  
In figure \ref{fig:38b}, all the capture points $O_{i}$'s have two or more normals passing through them and thus of comparable normal densities. Comparing with figure \ref{fig:38a}, all of these points are part of normal clusters. But, if we look at the points $P_{i}$'s on the curve, at which the normals are drawn, it is observed that for points $O_{5}$ and $O_{6}$, the orthographic regions centred at $P_{6}$, $P_{7}$ and $P_{2}$, $P_{3}$ respectively, will have huge overlap between them. Also for capture points like $O_{7}$ and $O_{8}$, the surface points are far apart and cover only small orthographic regions (as they are high curvature points). Similarly, points $O_{1}$ and $O_{4}$ cover orthographic regions of high overlap and low coverage respectively. In this case, points like $O_{2}$ and $O_{3}$ are more favourable as capture points as the corresponding surface points for them are spaced well apart for larger coverage and small overlap. This shows that not all points belonging to normal clusters and having same density are equally favourable as a choice of capture points.

\begin{figure}[htb!]
\centering
\subfigure[Surface plot for $f(x,y) = cos(x) + cos(y)$]{\label{fig:sub5}\includegraphics[width=0.45\linewidth]{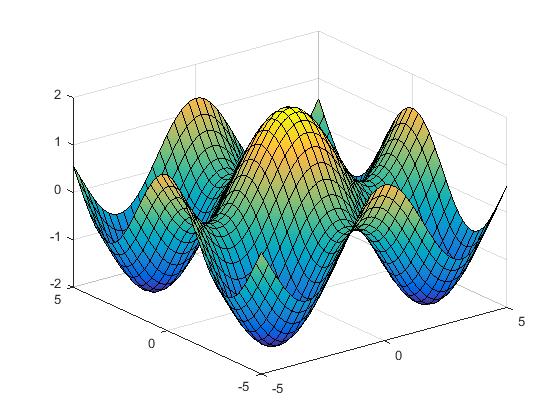}}
\hfill
\subfigure[Surface plot for $f(x,y) = xy\cdot cos(x)$]{\label{fig:sub6}\includegraphics[width=0.45\linewidth]{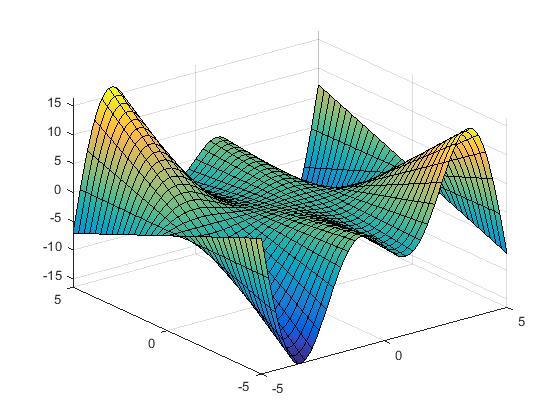}}

\subfigure[Surface normal plot for $f(x,y) = cos(x) + cos(y)$]{\label{fig:sub5}\includegraphics[width=0.45\linewidth]{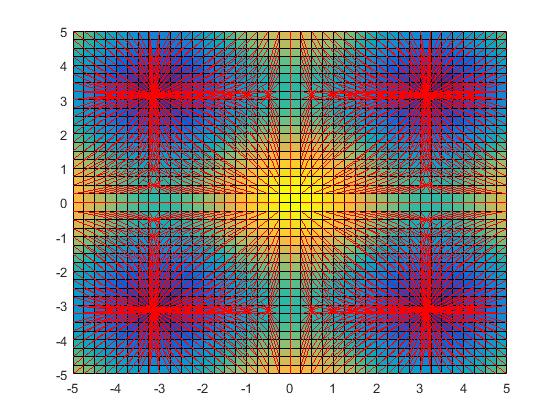}}
\hfill
\subfigure[Surface normal plot for $f(x,y) = xy\cdot cos(x)$]{\label{fig:sub6}\includegraphics[width=0.45\linewidth]{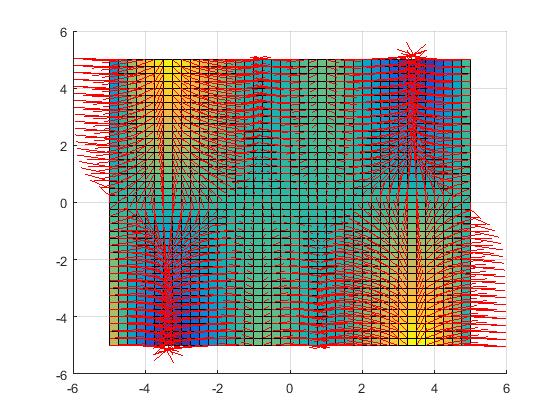}}

\subfigure[Contour plot for Mean curvature of $f(x,y) = cos(x) + cos(y)$]{\label{fig:sub5}\includegraphics[width=0.45\linewidth]{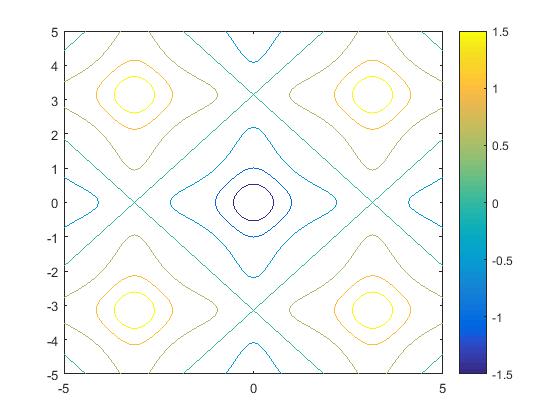}}
\hfill
\subfigure[Contour plot for Mean curvature of $f(x,y) = xy\cdot cos(x)$]{\label{fig:sub6}\includegraphics[width=0.45\linewidth]{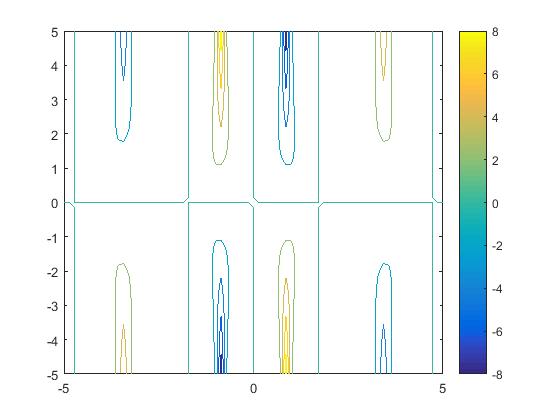}}

\caption{\textit{The comparison of normal clusters in red as observed in surface normal plots with the corresponding mean curvature of the surfaces plotted in \textsc{Matlab}.}}
\label{fig:39}
\end{figure}

From the figure \ref{fig:37}, we notice that the normal densities are higher for convex curves and lower for concave curves. Thus we can assume safely for a surface, the normal densities will be higher above a convex region and lower above a concave region. Also, as seen in section \ref{epsi_appx}, convex regions on a surface have high and concave regions have low \textit{mean curvatures}. Therefore heuristically, capture points above regions of high mean curvature must have higher density normal cluster and above regions of lower mean curvature must have low normal densities. Thus points located above high-mean curvature regions of a surface must be preferable as capture points for orthography. The correlation between mean curvature and normal clusters can be observed in figure \ref{fig:39}.

\section{Problem Formulation}\label{section:2}
The goal of the optimization problem is to find the optimal capture points such that the orthographic regions calculated for those points cover the entire surface area under consideration with minimum overlap
and that minimum number of such capture points are used.
\subsection{Assumptions}
\begin{itemize}
    \item The surface $S$ is smooth and slow-varying and that the gradients and thus surface normals can be calculated at all points
    \item Curvature function of the surface is smooth and can be evaluated at all points.
    \item The imaging height $d$ is taken to be constant for all points on the surface.
    \item The orientation of the capturing device is fixed at a capturing point $O$ and it points towards the surface point $P$, the surface normal at which passes through $O$, located at the imaging height $d$ from the surface. Multiple views at $O$ are not allowed. 
    \item The orthographic region boundaries are approximated to be circular and calculated as discussed in section 3.4.3.
\end{itemize}

\subsection{Variables and Constraints}
\begin{itemize}
    \item A surface $S(x,y,z)$ given by a smooth bi-variate function $z = f(x,y)$.
    \item The imaging distance - $d$.
    \item The useful \textit{angular FOV} - $\epsilon$.
    \item The $X$- and $Y$- bounds of the surface - $R_{high}$ and $R_{low}$.
    \begin{equation*}
        R_{low} \leq x \leq R_{high} \ \ \ \ R_{low} \leq y \leq R_{high}
    \end{equation*}
    
    \item The gaussian curvature - $K(x,y)$.
    \begin{equation*}
        K(x,y)  = \frac{\frac{\partial^{2} f}{\partial x^{2}}\cdot \frac{\partial^{2} f}{\partial y^{2}} - {\frac{\partial^{2} f}{\partial x \partial y}}^{2}}{\bigg(1 + {\big(\frac{\partial f}{\partial x}\big)}^{2} + {\big(\frac{\partial f}{\partial y}\big)}^{2}\bigg)^{2}}
    \end{equation*}
    \item The maximum absolute gaussian curvature - $K_{max}$.
    \begin{equation*}
       K_{max} = \max_{x,y}|K(x,y)|, \ \ \ \  R_{low} \leq (x,y) \leq R_{high} 
    \end{equation*}
     
    \item The maximum boundary circle radius- $R = d\cdot tan(\epsilon)$.
    \item The maximum to minimum radii ratio- $m = \frac{r_{max}}{r_{min}}$.
    \item The radius function of approximated boundary - $r(x,y)$.
    \begin{equation*}
        r(x,y) = R - \frac{|K(x,y)|}{K_{max}}\cdot R \cdot(1 - \frac{1}{m}), \ \ \ \ R_{low} \leq (x,y) \leq R_{high}  
    \end{equation*}
    \item The number of capture points - $N$.
    \item The set of capture points - $X = \{X_{1},X_{2},...,X_{N}\}$,  $X_{i} = (x_{i},y_{i})$.
    \item The function to calculate the area of overlap between two circles - $A(r_{1},r_{2},D)$. 
    \begin{equation}\label{eq:4.1}
    \begin{aligned}
        \text{Centres : } &  \ \ C_{1}(x_{1},y_{1}) \ \  \text{and} \ \  C_{2}(x_{2},y_{2}) \\
        \text{Radii : } &  \ \ r_{1} \ \ \text{and} \ \ r_{2} \\
        D  = & \ \norm{C_{1} - C_{2}} \\
        A(r_{1},r_{2},D) = & \ 0 \ \ \ \ \ \ \text{if} \ \ D \geq r_{1}+r_{2} \\
                         = & \ \pi r_{2}^{2} \ \ \ \ \ \ \text{if} \ \ D \leq |r_{1} - r_{2}| \ \ and \ \ r_{1} \geq r_{2} \\
                         = & \ \pi r_{1}^{2} \ \ \ \ \ \ \text{if} \ \ D \leq |r_{1} - r_{2}| \ \ and \ \ r_{2} \geq r_{1}  \\
                         = & \ \frac{1}{2}{r_{2}}^{2}\cdot\big(\theta - sin(\theta)\big) + \frac{1}{2}{r_{1}}^{2}\cdot\big(\phi - sin(\phi)\big) \ \ \ \ \ \ \text{otherwise}. \\
                         \text{where,  } \theta = & \ 2\cdot cos^{-1}\bigg( \frac{{r_2}^2 + D^2 - {r_1}^2}{2\cdot d\cdot r_{2}}\bigg) \ \ \text{and} \ \ \phi = 2\cdot cos^{-1}\bigg( \frac{{r_1}^2 + D^2 - {r_2}^2}{2\cdot d\cdot r_{1}}\bigg)
    \end{aligned}
    \end{equation}
    \item The positive value function - $\big[ \ . \ \big]_{+}$
    \begin{equation}
    \begin{aligned}
        \big[ x \big]_{+} & = \ x, \ \ \ \ \ \text{if} \ \ x \geq 0 \\
                          & = \ 0, \ \ \ \ \ \text{if} \ \ x < 0
    \end{aligned}    
    \end{equation}
\end{itemize}

\begin{itemize}
    \item \textbf{Optimization Variable : } $X = \{X_{i}\}$, \ \ \   $X_{i}= \left[x_{i} \ \ \ y_{i}\right]^{T}$ \ \ \ \    $ i = 1,2,...,N$
    \item \textbf{Constraints : } $lb = \left[\begin{array}{c} R_{low}\\R_{low}\end{array} \right]$, \ \ \   $ub = \left[\begin{array}{c} R_{high}\\R_{high}\end{array} \right]$
\end{itemize}

\subsection{Cost Functions}
Two primary objectives-
\begin{itemize}
    \item Minimizing overlap between orthographic regions, $\mathcal{R}_{1},\mathcal{R}_{2},...,\mathcal{R}_{N}$
    \item Maximizing the area covered by orthographic regions, $\mathcal{R}_{1},\mathcal{R}_{2},...,\mathcal{R}_{N}$  
\end{itemize}
- Area of $\mathcal{R}_{i} \ = \ \pi \cdot{\left[r(x_{i},y_{i})\right]^{2}} \ = \ \pi \cdot{\left[r(X_{i})\right]^{2}}$ 

- Overlap between $\mathcal{R}_i$ and $\mathcal{R}_j, \ \mathcal{L}\big(X_i,X_j\big) \ = \ A\big( r(X_i),r(X_j),\norm{X_i - X_j}\big)$ \ \  (ref eq \ref{eq:4.1})

- $R \ = \ r_{max} \ = \ d\cdot tan(\epsilon)$
\begin{enumerate}
    \item \textbf{Cost Function ($F_1$) : } Penalty based on apparent overlap only.
    \begin{equation}
        F_1(X) = \mathop{\mathlarger{\mathlarger{\sum}}\limits_{i = 1}^{N}  \mathlarger{\mathlarger{\sum}}\limits_{j = 1}^{N}}_{i < j}  \bigg[r(X_i) + r(X_j) - \norm{X_i - X_j}\bigg]_{+}
    \end{equation}
    \item \textbf{Cost Function ($F_2$) : } Penalty based on apparent overlap and area covered, higher weightage given to coverage.
    \begin{equation}
        F_2(X) = \mathop{\mathlarger{\mathlarger{\sum}}\limits_{i = 1}^{N}  \mathlarger{\mathlarger{\sum}}\limits_{j = 1}^{N}}_{i < j}  \bigg[r(X_i) + r(X_j) - \norm{X_i - X_j}\bigg]_{+} - \mathlarger{\mathlarger{\sum}}\limits_{i = 1}^{N} \pi\left[ \frac{r(X_i)}{R}\right]^2
    \end{equation}
    \item \textbf{Cost Function ($F_3$) : } Penalty based on apparent overlap and area covered, equal weightage given to both.
    \begin{equation}
        F_3(X) = 5\cdot\mathop{\mathlarger{\mathlarger{\sum}}\limits_{i = 1}^{N}  \mathlarger{\mathlarger{\sum}}\limits_{j = 1}^{N}}_{i < j}  \bigg[r(X_i) + r(X_j) - \norm{X_i - X_j}\bigg]_{+} - 0.5\cdot\mathlarger{\mathlarger{\sum}}\limits_{i = 1}^{N} \pi\left[ \frac{r(X_i)}{R}\right]^2
    \end{equation}
    \item \textbf{Cost Function ($F_4$) : } Penalty based on apparent overlap and area covered, variable weightage given to each.
    \begin{equation}\label{eq:4.6}
        F_4(X) = w_1\cdot\mathop{\mathlarger{\mathlarger{\sum}}\limits_{i = 1}^{N}  \mathlarger{\mathlarger{\sum}}\limits_{j = 1}^{N}}_{i < j}  \bigg[r(X_i) + r(X_j) - \norm{X_i - X_j}\bigg]_{+} - w_2\cdot\mathlarger{\mathlarger{\sum}}\limits_{i = 1}^{N} \pi\left[ \frac{r(X_i)}{R}\right]^2
    \end{equation}
    \item \textbf{Cost Function ($F_5$) : } Penalty based on left over area (\textit{Total Area - Area Covered}).
    \begin{equation}
    \begin{aligned}
        F_5(X) & = - \left[\mathlarger{\mathlarger{\sum}}\limits_{i = 1}^{N} \pi\left[ r(X_i)\right]^2 - \mathop{\mathlarger{\mathlarger{\sum}}\limits_{i = 1}^{N}  \mathlarger{\mathlarger{\sum}}\limits_{j = 1}^{N}}_{i < j}\mathcal{L}\big(X_i,X_j\big) \right] \\
        & = \mathop{\mathlarger{\mathlarger{\sum}}\limits_{i = 1}^{N}  \mathlarger{\mathlarger{\sum}}\limits_{j = 1}^{N}}_{i < j} \mathcal{L}\big(X_i,X_j\big) - \mathlarger{\mathlarger{\sum}}\limits_{i = 1}^{N} \pi\left[ r(X_i)\right]^2 \\
        & = \mathop{\mathlarger{\mathlarger{\sum}}\limits_{i = 1}^{N}  \mathlarger{\mathlarger{\sum}}\limits_{j = 1}^{N}}_{i < j} A\big( r(X_i),r(X_j),\norm{X_i - X_j}\big) - \mathlarger{\mathlarger{\sum}}\limits_{i = 1}^{N} \pi\left[ r(X_i)\right]^2
    \end{aligned}
    \end{equation}
\end{enumerate}

\section{Circle Filling Algorithms}\label{section:3}
The goal is to fill the surface with minimum number of circles such that the whole area is covered with minimum overlap among the circular regions. So, here both the number($N$) and location($X_i$) of the centres of the circles are variables. Two major algorithms were explored for this purpose- \textit{Batch Filling} and \textit{Sequential Filling}.
\subsection{Batch Filling}
The basic idea of \textit{batch filling} is to fill the surface with a fixed number of circles($N$) at a time while satisfying the primary objectives. The general structure of the algorithm is given in \ref{alg:batch fill}.
\begin{algorithm}
\caption{Batch Filling of Orthographic Circles}\label{alg:batch fill}
\begin{algorithmic}[1]
\State Set $N$ to be a small number based on the surface area to be covered and imaging height $d$. $N$ needs to be larger for large surfaces and small $d$, and smaller for small surfaces and large $d$.
\State Initialize the locations of $N$ circles at random, while satisfying the constraints- \ \ \ \ \ \ \ \    $X_0 = \{X_{i}^0 \ | \ i = 1,2,...,N\}$ \ \ \ \ $X_{i}^0 \ \in Bound(S) $.
\State Select a cost function $F_i$.
\State Optimize the cost function $F_i$ for the location variable $X = \{X_i \ | \ i=1,2,...,N\}$. 
\State Calculate the total overlap($\mathcal{L}$) between the circular regions $\mathcal{R}_1,\mathcal{R}_2,...,\mathcal{R}_N$.
\State Calculate the total area covered(excluding the overlap) by the circles.
\State Check if the total area is covered. If not, increase $N$ by $1$ and repeat Steps 2 to 7.
\State Stop when the total area is covered.
\end{algorithmic}
\end{algorithm}

\textit{Batch Filling} can be of two types- \textbf{\textit{Fixed Cost Batch Filling}} and \textbf{\textit{Variable Cost Batch Filling}}. In \textit{fixed cost} method, the cost function($F_i$) remains same through out the evaluation process, whereas in \textit{variable cost} method, the cost function can be changed depending on the number of circles. Empirically, it can be inferred that, for smaller number of circles, cost functions with penalty based on coverage should be preferred; whereas, for larger number of circles, cost functions with penalty based on overlap should be considered. With reference to equation \ref{eq:4.6}, weights $w_1$ and $w_2$ should be chosen such that for smaller values of $N$, the ratio $\frac{w_1}{w_2} >> 1$ and for larger values of $N$, $\frac{w_1}{w_2} \leq 1$. This ratio can be changed gradually and expressed as a function of $N$.

\subsection{Sequential Filling}
The basic idea of \textit{sequential filling} is to fill the surface gradually with the number of circles increasing gradually in each iteration while satisfying the primary objectives. The general structure of the algorithm is given in \ref{alg:sequential fill}.
\begin{algorithm}
\caption{Sequential Filling of Orthographic Circles}\label{alg:sequential fill}
\begin{algorithmic}[1]
\State Start with an empty set of locations, $X_p = \{\ \}$.
\State Set an integer step size, $n$, for sequential adding of circles. 
\State Initialize the locations of $n$ circles at random, while satisfying the constraints- \ \ \ \ \ \ \ \ \ \ \    $X_0 = \{X_{i}^0 \ | \ i = 1,2,...,n\}$ \ \ \ \ $X_{i}^0 \ \in Bound(S) $.
\State Select a cost function $F_i$.
\State Optimize the cost function $F_i$ for the location variable $X = \{X_i \ | \ i=1,2,...,n\}$ \ and on the set $X_p$.
\State Add the calculated locations $\{X_1,X_2,...,X_n \}$ to the set $X_p$.
\State Calculate the total overlap($\mathcal{L}$) between the circular regions $\mathcal{R}_1,\mathcal{R}_2,...,\mathcal{R}_P$, where $P$ is the total number of locations stored in $X_p$.
\State Calculate the total area covered(excluding the overlap) by the circles.
\State Check if the total area is covered. If not repeat Steps 3 to 9.
\State Stop when the total area is covered.
\end{algorithmic}
\end{algorithm}

\textit{Sequential Filling} can also be of two types- \textbf{\textit{Fixed Cost Sequential Filling}} and \textbf{\textit{Variable Cost Sequential Filling}}. Similar to batch filling, in \textit{fixed cost} method, the cost function($F_i$) remains same through out the evaluation process, whereas in \textit{variable cost} method, the cost function can be changed depending on the number of circles. The empirical inference for the variation of cost function with the number of circles as stated in \textit{variable cost batch filling} is also applicable here. Also, it can be noted that at each iteration of the algorithm, $n$ new circles are added for optimization. Here $n$ should be a small integer ($n \in \{1,2,3\}$).

\section{Surface Division}\label{section:4}
After the optimal locations have been evaluated, the surface needs to be segmented into orthographic sections such that each point on the surface is associated to one orthographic region. Suppose $N$ is the number of optimal location points evaluated by the circle filling algorithm. The set of points $X_p = \{X_1,X_2,...,X_N\}$ are the location of the centres of the corresponding orthographic circular regions $\mathcal{R} = \{\mathcal{R}_1,\mathcal{R}_2,...,\mathcal{R}_N\}$. Now, the points in the overlapping regions have to be allocated to one of the regions it lies in. For a point $X(x,y)$ lying inside the intersection of two or more regions $\mathcal{R}_i$, the point will be allocated to the region whose centre(${X_i}$) is closest to it. In algebraic terms, suppose after evaluation of the algorithm, a point 
\begin{equation}
    X(x,y) \in \{\mathcal{R}^{X}_1,\mathcal{R}^{X}_2,...,\mathcal{R}^{X}_K\}, \ \ \ \ s.t \ \ \ \{\mathcal{R}^{X}_j \ | \ j=1,2,...,K\} \subseteq \mathcal{R}. 
\end{equation}
Then $X$ is allocated to the region $\mathcal{R}^{X}_i \in \{\mathcal{R}^{X}_1,\mathcal{R}^{X}_2,...,\mathcal{R}^{X}_K\}$, s.t.
\begin{equation}
    i = \argmin_{1\leq j \leq K} \  \norm{X - X_j}
\end{equation}
where, $X_j$ is the centre of $\mathcal{R}^{X}_j$ and $\mathcal{R}^{X}_j \in \{\mathcal{R}^{X}_1,\mathcal{R}^{X}_2,...,\mathcal{R}^{X}_K\}$.

\begin{figure}[!htb]
  \centering
  \includegraphics[width=0.5\linewidth]{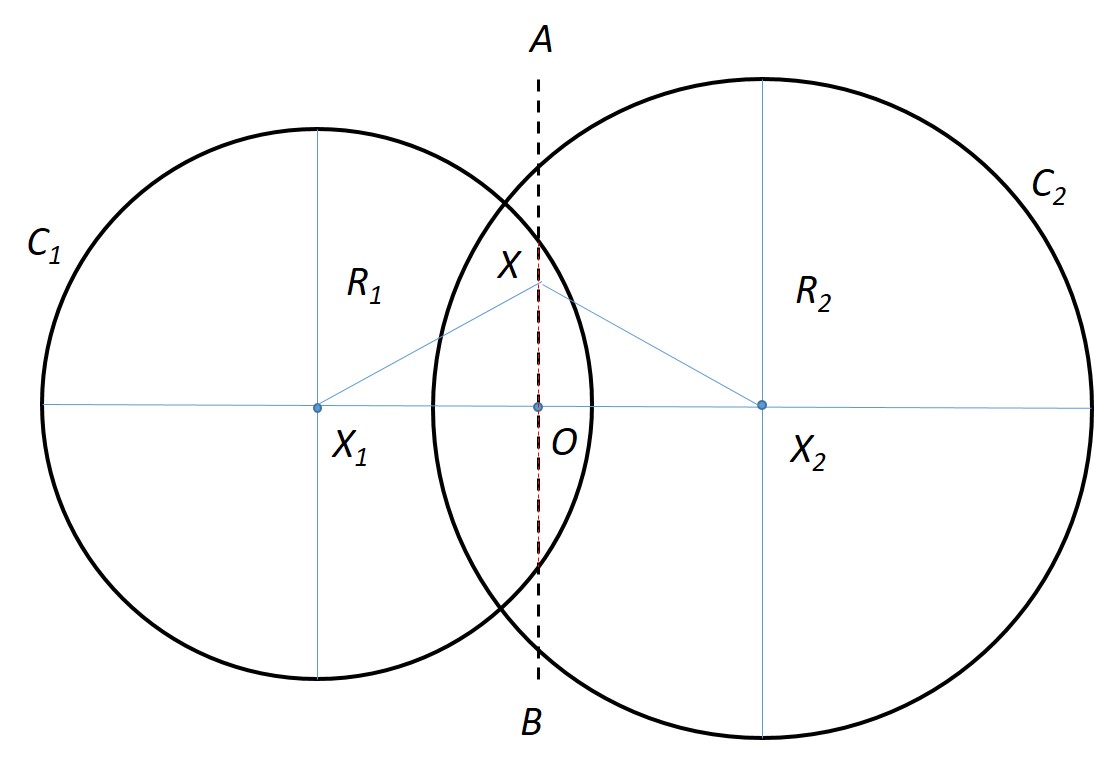}
  \caption{\textit{Illustartion of two intersecting circular regions.}}
  \label{fig:60}
\end{figure}
Consider two intersecting circles $C_1$ and $C_2$, with radii $R_1$ and $R_2$ as shown in figure \ref{fig:60}. The centers of the circles are shown as $X_1$ and $X_2$. The line segment $\overline{X_1X_2}$ is bisected at $O$ and $\overline{AB}$ is the perpendicular bisector. Now, any point $X$ on $\overline{AB}$ is equidistant from the two centres. So, any point on the left of $\overline{AB}$ is allocated to $C_1$ and any point on the right is allocated to $C_2$. Hence the decision boundary for the two circles is shown in red.   
\chapter{Results and Analysis}
\label{Chapter5}

\lhead{Chapter 5. \emph{Results and Analysis}}
In this chapter, the results obtained after implementation of the problem formulated in chapter \ref{Chapter4} have been compiled and analyzed. The \textit{Optimization Toolbox} of \textsc{Matlab} was used for evaluation of the optimization step of the algorithms. In section \ref{section:5.1}, the dependency of solutions on local and global optimization is stated, and the pros and cons of using each is evaluated. The cost functions proposed in section \ref{section:2} of the preceding chapter have been compared in section \ref{section:5.2} based on evaluation time grounds and efficiency measures. In section \ref{section:5.3}, the existence of non-uniqueness of solutions have been validated by plotting pareto-fronts among the cost functions. Section \ref{section:5.4} deals with the comparison of \textit{Circle Filling Algorithms} as proposed in section \ref{section:3} of the previous chapter. Section \ref{section:5.5} is a discussion on the final results obtained after surface division as proposed in section \ref{section:4} of the last chapter. Sections \ref{section:5.6}, \ref{section:5.7} and \ref{section:5.8} is an analysis of the effects of parameters imaging distance $d$, useful angular FOV $\epsilon$ and the surface curvature, respectively, on the results.   

\section{Local vs Global Optimization}\label{section:5.1}
Optimization is the most important step in each iteration of the circle filling algorithms described in \ref{section:3}. The solution of the optimization problem is significantly dependent on the initialization. Thus, while implementation, both local and global solvers were explored in \textsc{Matlab}. \textbf{\textit{fmincon}} was used as the local solver and \textit{interior-point} algorithm was used for obtaining solutions. For global optimization, two different global solvers were used- \textbf{\textit{GlobalSearch}} and \textbf{\textit{MultiStart}}.
\begin{itemize}
    \item In \textit{GlobalSearch}, the \text{fmincon} solver first runs the optimization based on the initial randomization. Based on the initial convergence, the solver uses a scatter-search mechanism for generating start points. As it runs local solvers at the start points generated, it analyzes start points and rejects those points that are unlikely to improve the best local minimum found so far. \textit{GlobalSearch} can be used to find a single global minimum most efficiently on a single processor. The number of start points and thus local solvers can be controlled.
    \item In \textit{MultiStart}, the start points are generated uniformly within the bounds and may also be user generated. Local solvers can run in parallel at each of the start points and all the local solutions are stored. The solution with the minimum cost is estimated as the global solution. The number of starting points and thus the local solvers can be changed.
\end{itemize}
 For comparison, the solvers were evaluated on the surface $S$ given by $z = f(x,y) = cos(x) + cos(y)$, for an imaging height $d = 3$, for angular FOV $\epsilon = 10^{\circ}$ and surface bounds $-5 \leq x \leq 5$ \ and \  $-5 \leq y \leq 5$. $F_{5}$ as discussed in section \ref{section:2} was used as the cost function, for which penalty is based on the left-over area. Total area covered and overlap among regions were used as efficiency criterion. The results generated in \textsc{Matlab} are presented in table \ref{tab:1}. 

% Please add the following required packages to your document preamble:
% \usepackage{multirow}
\begin{table}[htb!]
\begin{tabular}{ccccc}
\cline{2-5}
\multicolumn{1}{c|}{} & \multicolumn{1}{c|}{No. of circles} & \multicolumn{1}{c|}{Area Covered} & \multicolumn{1}{c|}{\% area covered} & \multicolumn{1}{c|}{Overlap} \\ \cline{2-5} 
\multicolumn{1}{l}{} & \multicolumn{1}{l}{} & \multicolumn{1}{l}{} & \multicolumn{1}{l}{} & \multicolumn{1}{l}{} \\
\multirow{4}{*}{Local Solution} & 20 & 76.03 & 54.89 & 0 \\
 & 30 & 108.46 & 78.31 & 3.38 \\
 & 40 & 126.16 & 91.09 & 10.85 \\
 & 48 & 137.35 & 99.17 & 22.25 \\
\multicolumn{1}{l}{} & \multicolumn{1}{l}{} & \multicolumn{1}{l}{} & \multicolumn{1}{l}{} & \multicolumn{1}{l}{} \\
\multirow{4}{*}{MultiStart Solution} & 20 & 76.03 & 54.89 & 0 \\
 & 30 & 110.54 & 79.81 & 3.15 \\
 & 40 & 128.86 & 93.04 & 11.58 \\
 & 48 & 138.02 & 99.65 & 21.21 \\
\multicolumn{1}{l}{} & \multicolumn{1}{l}{} & \multicolumn{1}{l}{} & \multicolumn{1}{l}{} & \multicolumn{1}{l}{} \\
\multirow{4}{*}{GlobalSearch Solution} & 20 & 76.03 & 54.89 & 0 \\
 & 30 & 110.41 & 79.72 & 3.22 \\
 & 40 & 127.23 & 91.86 & 12.53 \\
 & 48 & 137.56 & 99.32 & 22.79 \\
\multicolumn{1}{l}{} & \multicolumn{1}{l}{} & \multicolumn{1}{l}{} & \multicolumn{1}{l}{} & \multicolumn{1}{l}{} \\
\multicolumn{1}{l}{} & \multicolumn{1}{l}{} & \multicolumn{1}{l}{} & \multicolumn{1}{l}{} & \multicolumn{1}{l}{}
\end{tabular}
\caption{Comparison of local and global solutions based on area covered and overlap.}
\label{tab:1}
\end{table}
For the \textit{GlobalSearch} solver, the number of trial points were set to $50$ and for the \textit{MultiStart} solver, the number of start points used were $30$. The initialization point was selected at random within the constraints specified. The number of solutions were selected based on the convergence of the local solvers. For \textit{MultiStart}, the best solution was chosen for analysis. The \textit{local solution} given was for one random initialization and may vary. 

As it can be observed from table \ref{tab:1}, \textit{MultiStart} solver gives the better result than \textit{GlobalSearch} and both the global solvers give comparatively better results than a local solution, as expected. However,for cost function $F_5$, while a local solver can take about an hour for evaluation, a global solver can take hours. If more number of trial points are taken, the chances of obtaining a better solution increases. As a \textit{MultiStart} solver can run in parallel for different point, it is faster compared to \textit{GlobalSearch}. For obtaining the same level of efficiency in solution, it was observed that number of trial points in the range 30-50 range was sufficient in \textit{MultiStart} which took around 10-15 hours for evaluation, whereas, \textit{GlobalSearch} took around 20-24 hours for the same. Hence for this cost function, local solver is almost always a better choice than global solvers.
 
The following figures show the circle centers and the corresponding circle for the function $f(x,y) = cos(x) + cos(y)$ and for the solutions given in table \ref{tab:1}. \\
\\

\begin{figure}[!htb]
  \centering
  \includegraphics[width=0.7\linewidth]{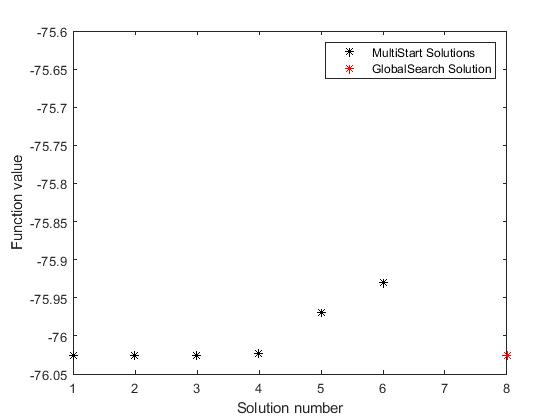}
  \caption{\textit{The \textbf{MultiStart} solutions shown in black and the  \textbf{GlobalSearch} solution shown in red for $n = 20$.}}
  \label{fig:40}
\end{figure}

\begin{figure}[!htb]
\centering
\subfigure[The solution points plotted on the surface.]{\label{fig:sub5}\includegraphics[width=0.48\linewidth]{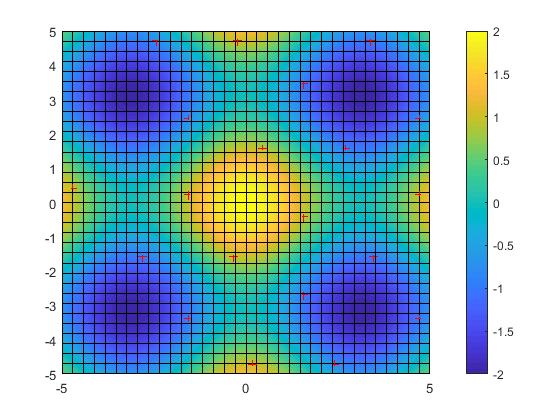}}
\hfill
\subfigure[The solution points along with the associated circles plotted on the surface contour.]{\label{fig:sub6}\includegraphics[width=0.48\linewidth]{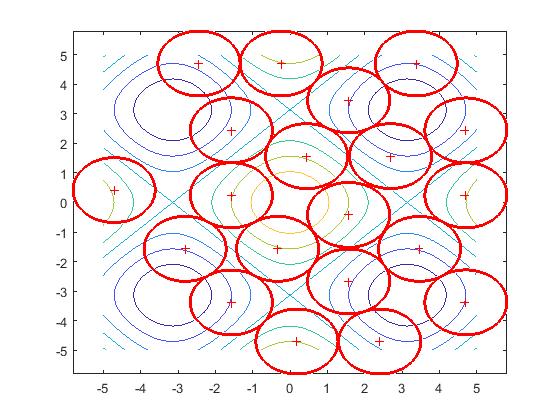}}
\caption{\textit{The circles and the centres of the best solution for the \textbf{MultiStart} solver are plotted for  $n = 20$.}}
\label{fig:41}
\end{figure}

\begin{figure}[!htb]
\centering
\subfigure[The solution points plotted on the surface.]{\label{fig:sub5}\includegraphics[width=0.48\linewidth]{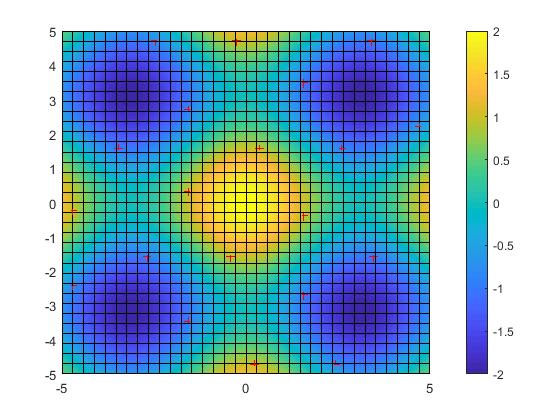}}
\hfill
\subfigure[The solution points along with the associated circles plotted on the surface contour.]{\label{fig:sub6}\includegraphics[width=0.48\linewidth]{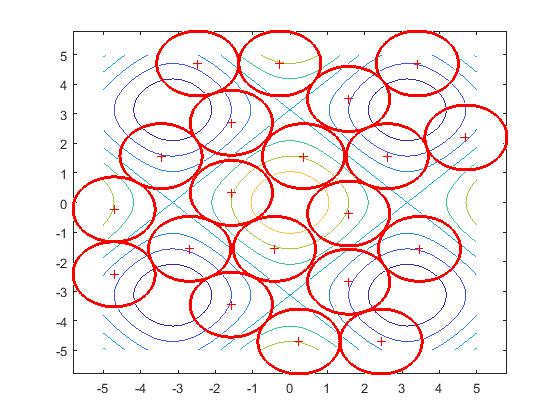}}
\caption{\textit{The circles and the centres of the solution for the \textbf{GlobalSearch} solver are plotted for  $n = 20$.}}
\label{fig:42}
\end{figure}

\begin{figure}[htb!]
  \centering
  \includegraphics[width=0.7\linewidth]{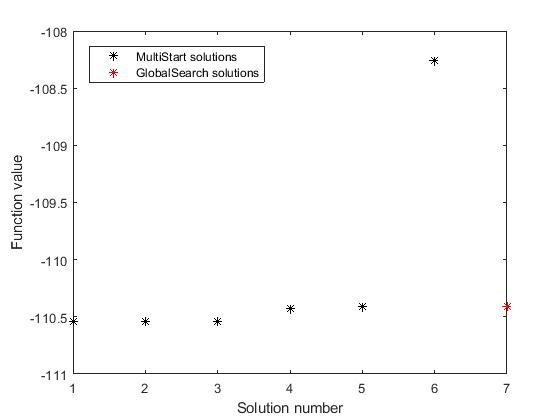}
  \caption{\textit{The \textbf{MultiStart} solutions shown in black and the  \textbf{GlobalSearch} solution shown in red for $n = 30$.}}
  \label{fig:43}
\end{figure}

\begin{figure}[!htb]
\centering
\subfigure[The solution points plotted on the surface.]{\label{fig:sub5}\includegraphics[width=0.48\linewidth]{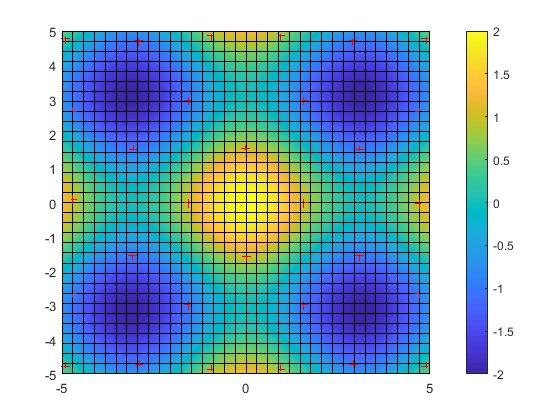}}
\hfill
\subfigure[The solution points along with the associated circles plotted on the surface contour.]{\label{fig:sub6}\includegraphics[width=0.48\linewidth]{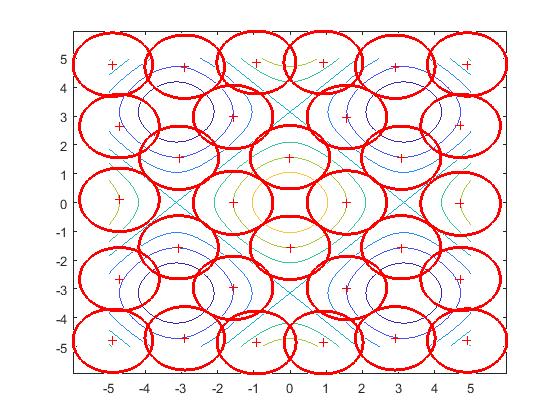}}
\caption{\textit{The circles and the centres of the best solution for the \textbf{MultiStart} solver are plotted for  $n = 30$.}}
\label{fig:44}
\end{figure}

\begin{figure}[!htb]
\centering
\subfigure[The solution points plotted on the surface.]{\label{fig:sub5}\includegraphics[width=0.48\linewidth]{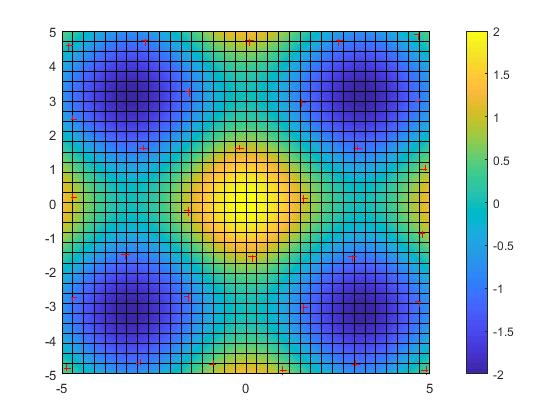}}
\hfill
\subfigure[The solution points along with the associated circles plotted on the surface contour.]{\label{fig:sub6}\includegraphics[width=0.48\linewidth]{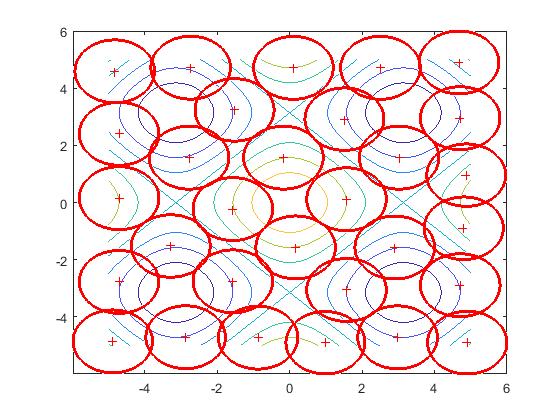}}
\caption{\textit{The circles and the centres of the solution for the \textbf{GlobalSearch} solver are plotted for  $n = 30$.}}
\label{fig:45}
\end{figure}

\begin{figure}[!htb]
  \centering
  \includegraphics[width=0.7\linewidth]{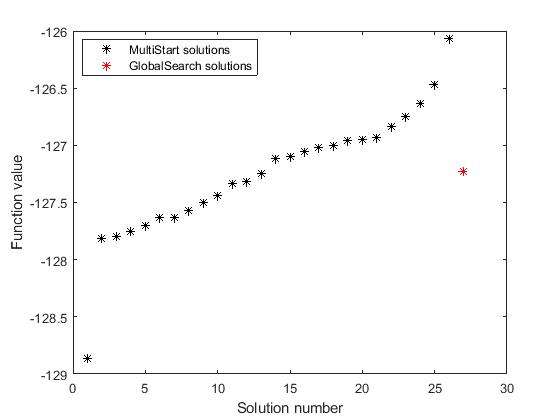}
  \caption{\textit{The \textbf{MultiStart} solutions shown in black and the  \textbf{GlobalSearch} solution shown in red for $n = 40$.}}
  \label{fig:46}
\end{figure}

\begin{figure}[!htb]
\centering
\subfigure[The solution points plotted on the surface.]{\label{fig:sub5}\includegraphics[width=0.48\linewidth]{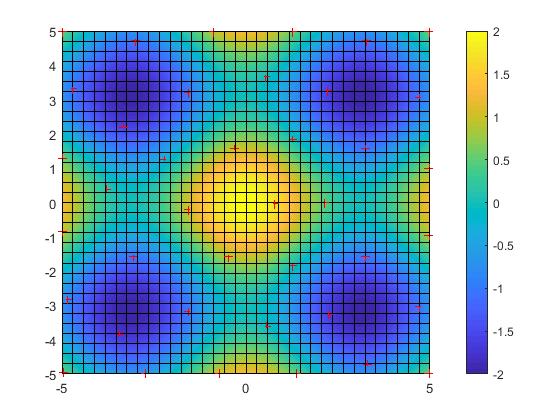}}
\hfill
\subfigure[The solution points along with the associated circles plotted on the surface contour.]{\label{fig:sub6}\includegraphics[width=0.48\linewidth]{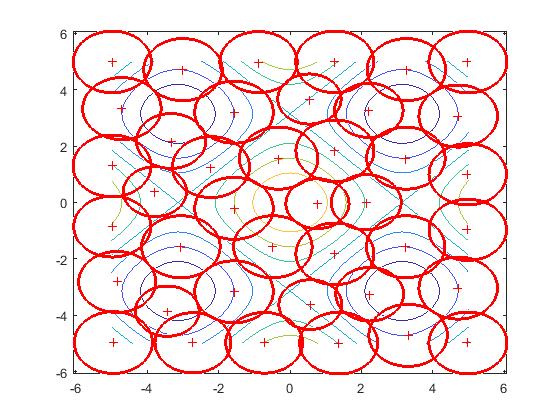}}
\caption{\textit{The circles and the centres of the best solution for the \textbf{MultiStart} solver are plotted for  $n = 40$.}}
\label{fig:47}
\end{figure}

\begin{figure}[!htb]
\centering
\subfigure[The solution points plotted on the surface.]{\label{fig:sub5}\includegraphics[width=0.48\linewidth]{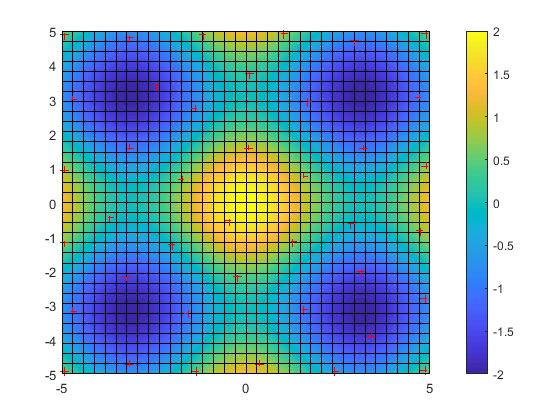}}
\hfill
\subfigure[The solution points along with the associated circles plotted on the surface contour.]{\label{fig:sub6}\includegraphics[width=0.48\linewidth]{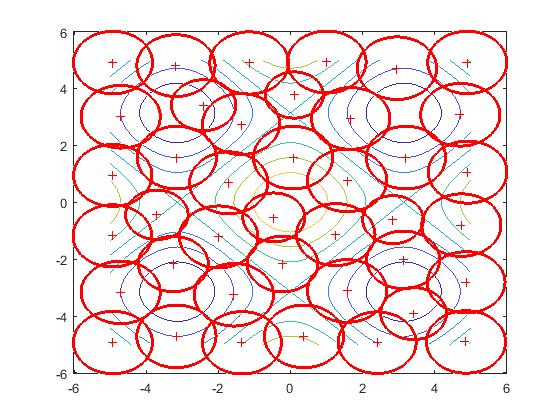}}
\caption{\textit{The circles and the centres of the solution for the \textbf{GlobalSearch} solver are plotted for  $n = 40$.}}
\label{fig:48}
\end{figure}

\begin{figure}[!htb]
  \centering
  \includegraphics[width=0.7\linewidth]{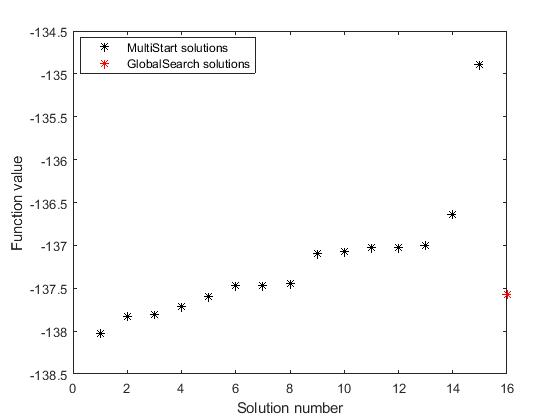}
  \caption{\textit{The \textbf{MultiStart} solutions shown in black and the  \textbf{GlobalSearch} solution shown in red for $n = 48$.}}
  \label{fig:49}
\end{figure}

\begin{figure}[!htb]
\centering
\subfigure[The solution points plotted on the surface.]{\label{fig:sub5}\includegraphics[width=0.48\linewidth]{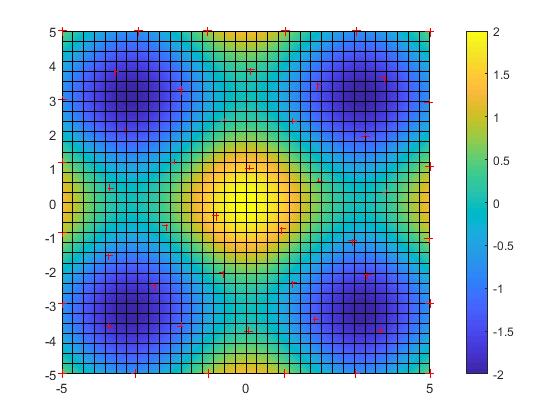}}
\hfill
\subfigure[The solution points along with the associated circles plotted on the surface contour.]{\label{fig:sub6}\includegraphics[width=0.48\linewidth]{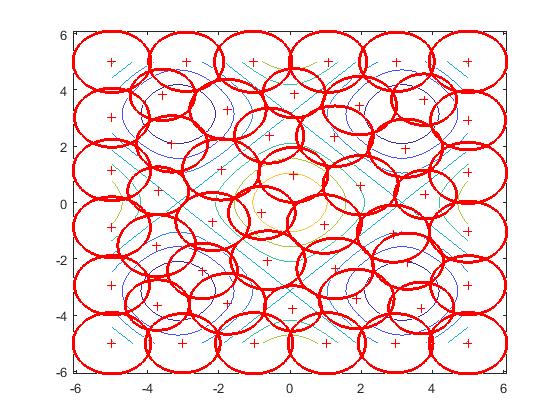}}
\caption{\textit{The circles and the centres of the best solution for the \textbf{MultiStart} solver are plotted for  $n = 48$.}}
\label{fig:50}
\end{figure}

\begin{figure}[htb!]
\centering
\subfigure[The solution points plotted on the surface.]{\label{fig:sub5}\includegraphics[width=0.48\linewidth]{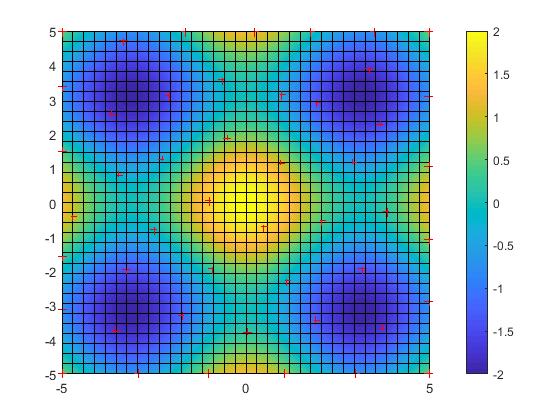}}
\hfill
\subfigure[The solution points along with the associated circles plotted on the surface contour.]{\label{fig:sub6}\includegraphics[width=0.48\linewidth]{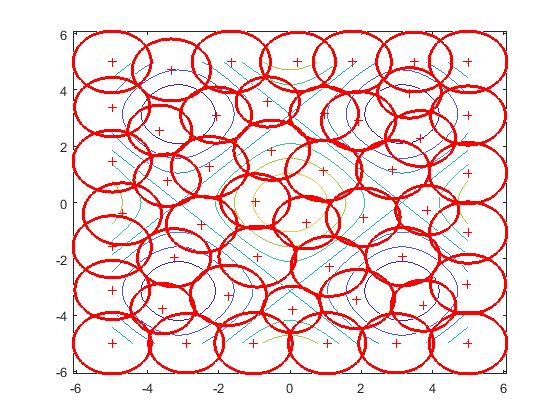}}
\caption{\textit{The circles and the centres of the solution for the \textbf{GlobalSearch} solver are plotted for  $n = 48$.}}
\label{fig:51}
\end{figure}

\section{Comparison of Cost Functions}\label{section:5.2}
Different cost functions were proposed in section \ref{section:2}. Optimization with the cost functions $F_1$, $F_2$, $F_3$, $F_5$ were compared. Along with these, two additional cost functions were compared, for which the coverage penalty was based on actual area of the circles.
\begin{equation}
    G_1(X) = \mathop{\mathlarger{\mathlarger{\sum}}\limits_{i = 1}^{N}  \mathlarger{\mathlarger{\sum}}\limits_{j = 1}^{N}}_{i < j}  \bigg[r(X_i) + r(X_j) - \norm{X_i - X_j}\bigg]_{+} - \mathlarger{\mathlarger{\sum}}\limits_{i = 1}^{N} \pi\left[r(X_i)\right]^2
\end{equation}

\begin{equation}
    G_2(X) = 5\cdot\mathop{\mathlarger{\mathlarger{\sum}}\limits_{i = 1}^{N}  \mathlarger{\mathlarger{\sum}}\limits_{j = 1}^{N}}_{i < j}  \bigg[r(X_i) + r(X_j) - \norm{X_i - X_j}\bigg]_{+} - 0.5\cdot\mathlarger{\mathlarger{\sum}}\limits_{i = 1}^{N} \pi\left[r(X_i)\right]^2
\end{equation}

The table \ref{tab:2} shows the area covered, in \% of the whole area, and the total overlap among the circles, computed for different cost functions. The solutions considered are global and computed by the \textit{GlobalSearch} solver. Here, surface $S$ is given by $z = f(x,y) = cos(x)+cos(y)$; the imaging height is $d = 3$, the useful angular FOV is $\epsilon = 10^{\circ}$ and the surface bounds are given as, $-5 \leq x \leq 5$ and $-5 \leq x \leq 5 $.

\begin{table}[hbt!]
\centering
\begin{tabular}{ccccccc}
\hline
\multicolumn{1}{|c|}{No. of Circles} & \multicolumn{6}{c|}{Area Covered (\%)} \\ \hline
\multicolumn{1}{|c|}{} & \multicolumn{1}{c|}{F1} & \multicolumn{1}{c|}{F2} & \multicolumn{1}{c|}{G1} & \multicolumn{1}{c|}{F3} & \multicolumn{1}{c|}{G2} & \multicolumn{1}{c|}{F5} \\ \hline
\multicolumn{1}{|c|}{20} & \multicolumn{1}{l|}{41.93} & \multicolumn{1}{l|}{56.32} & \multicolumn{1}{l|}{56.32} & \multicolumn{1}{l|}{56.31} & \multicolumn{1}{l|}{56.31} & \multicolumn{1}{l|}{56.50} \\ \hline
\multicolumn{1}{|c|}{30} & \multicolumn{1}{c|}{65.30} & \multicolumn{1}{c|}{75.56} & \multicolumn{1}{c|}{75.28} & \multicolumn{1}{c|}{76.54} & \multicolumn{1}{c|}{75.88} & \multicolumn{1}{c|}{80.78} \\ \hline
\multicolumn{1}{|c|}{40} & \multicolumn{1}{c|}{78.75} & \multicolumn{1}{c|}{84.51} & \multicolumn{1}{c|}{81.82} & \multicolumn{1}{c|}{80.58} & \multicolumn{1}{c|}{83.99} & \multicolumn{1}{c|}{93.40} \\ \hline
\multicolumn{1}{|c|}{50} & \multicolumn{1}{c|}{78.71} & \multicolumn{1}{c|}{93.39} & \multicolumn{1}{c|}{93.53} & \multicolumn{1}{c|}{83.33} & \multicolumn{1}{c|}{81.98} & \multicolumn{1}{c|}{99.52} \\ \hline
\multicolumn{1}{l}{} & \multicolumn{1}{l}{} & \multicolumn{1}{l}{} & \multicolumn{1}{l}{} & \multicolumn{1}{l}{} & \multicolumn{1}{l}{} & \multicolumn{1}{l}{} \\ \hline
\multicolumn{1}{|c|}{No. of Circles} & \multicolumn{6}{c|}{Overlap} \\ \hline
\multicolumn{1}{|l|}{} & \multicolumn{1}{l|}{F1} & \multicolumn{1}{l|}{F2} & \multicolumn{1}{l|}{G1} & \multicolumn{1}{l|}{F3} & \multicolumn{1}{l|}{G2} & \multicolumn{1}{l|}{F5} \\ \hline
\multicolumn{1}{|c|}{20} & \multicolumn{1}{c|}{0} & \multicolumn{1}{c|}{0} & \multicolumn{1}{c|}{0} & \multicolumn{1}{c|}{0} & \multicolumn{1}{c|}{0} & \multicolumn{1}{c|}{0} \\ \hline
\multicolumn{1}{|c|}{30} & \multicolumn{1}{c|}{0} & \multicolumn{1}{c|}{11.77} & \multicolumn{1}{c|}{11.87} & \multicolumn{1}{c|}{0} & \multicolumn{1}{c|}{0} & \multicolumn{1}{c|}{3.12} \\ \hline
\multicolumn{1}{|c|}{40} & \multicolumn{1}{c|}{5.57} & \multicolumn{1}{c|}{37.22} & \multicolumn{1}{c|}{34.02} & \multicolumn{1}{c|}{6.02} & \multicolumn{1}{c|}{5.82} & \multicolumn{1}{c|}{11.91} \\ \hline
\multicolumn{1}{|c|}{50} & \multicolumn{1}{c|}{24.50} & \multicolumn{1}{c|}{58.27} & \multicolumn{1}{c|}{49.63} & \multicolumn{1}{c|}{26.18} & \multicolumn{1}{c|}{24.83} & \multicolumn{1}{c|}{22.12} \\ \hline
\end{tabular}
\caption{Comparison of Area Covered(\%) and Overlap for different cost functions and for different values of $N$.}
\label{tab:2}
\end{table}
It can be observed that cost functions $F_3$ and $G_2$, with penalties based on equal weightage to both coverage and overlap have far less overlap in comparison to cost functions $F_2$ and $G_1$, which have penalties based on more weightage to coverage than overlap. Although, the area covered by $F_2$ and $G_1$ are significantly greater than $F_3$ and $G_2$. Also, it can be noted that using relative radii ($r(X_i)/R$) instead of $r(X_i)$ in cost functions $F_2$ and $F_3$, does not improve the area covered significantly, although the overlaps decrease a little. 
Also, it can be noted that cost function $F_5$ provides the best performance among all the ones considered. While the overlap is relatively small, the area covered is significantly larger than the other ones. For our purpose, area coverage has more priority than overlap, therefore, cost function $F_5$ can be inferred to be the best among the ones proposed. However, the computation time for $F_5$ is much greater than any of the other ones, and hence it should be avoided when faster calculations are prioritized. 

The following figures \ref{fig:52} and \ref{fig:53} give the comparison of the different cost functions for the same surface and parameters as mentioned for table \ref{tab:2}; and the number of circles are taken from $20$ to $50$ at an interval of $n = 2$.

\begin{figure}[!htb]
  \centering
  \includegraphics[width=1\linewidth]{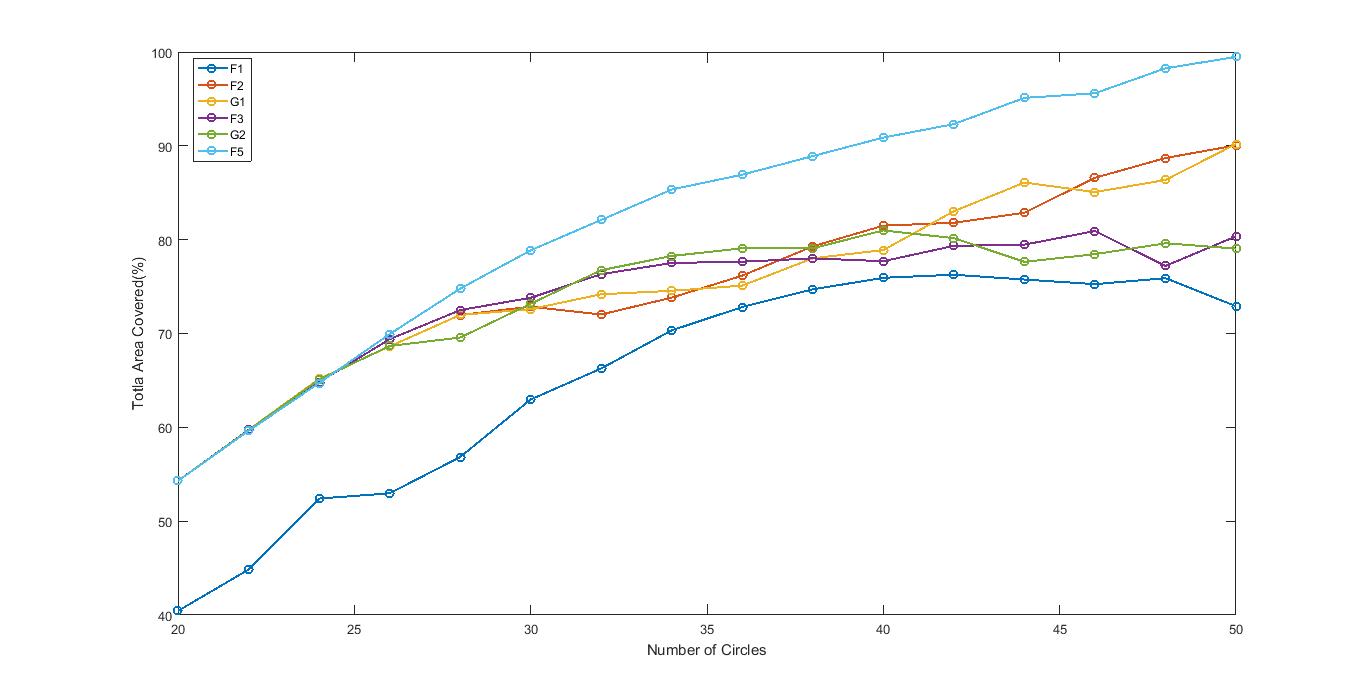}
  \caption{\textit{The area covered(\%) for different cost functions with the variation in number of capture points is computed and plotted in \textsc{Matlab}.}}
  \label{fig:52}
\end{figure}
It can be observed from figure \ref{fig:52} that cost $F_5$ (plotted in light blue) outperforms other costs in terms of area coverage. Also, the figure shows that for higher coverage, costs functions with greater penalty on coverage ($F_2$ and $G_1$) should be preferred. 

It can also be observed that for cost functions with penalties based on equal priority to coverage and overlap or completely based on overlap, the area coverage saturates after a point. It is because higher the number of circles, higher is the chance of overlap, and hence to minimize the overlap, the cost function has to decrease the area(or radius) of the circles. Hence, the coverage per circle decreases and total coverage remains approximately constant.

\begin{figure}[!htb]
  \centering
  \includegraphics[width=1\linewidth]{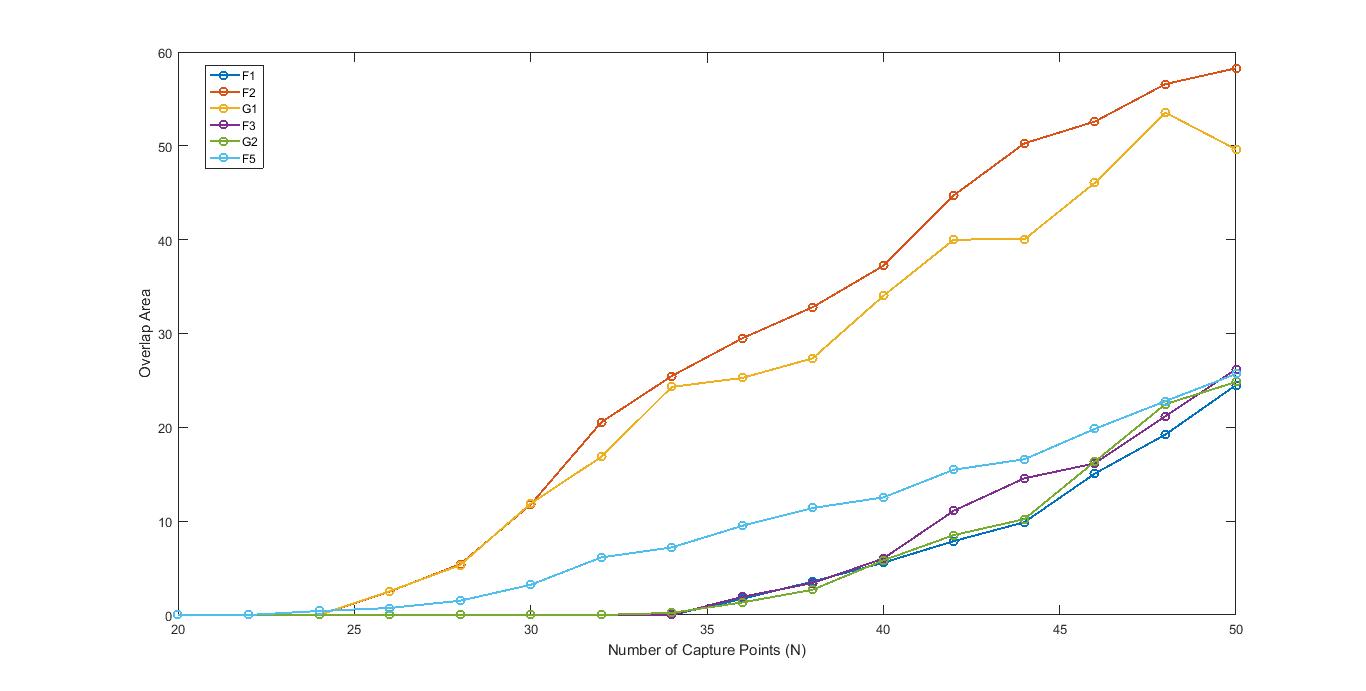}
  \caption{\textit{The overlap area for different cost functions with the variation in number of capture points is computed and plotted in \textsc{Matlab}.}}
  \label{fig:53}
\end{figure}

It can be observed from figure \ref{fig:53} that overlap among the circles are much higher for cost functions $F_2$ and $G_1$ as compared to the rest. These two cost functions give less priority to overlap as stated before. On the other hand, for costs $F_1$, $F_3$ and $G_2$, the overlap is much less, as expected. For function $F_5$, the cost is in between these two categories of costs, but still much less than the first category. 

Hence, considering both the objectives- coverage and overlap, $F_5$ is the best suggestion, if computation time is not an issue. If computation time is to be kept low, then $F_2$ is the better choice, where coverage is priority and $F_3$ is better where reducing overlap is the priority.

\section{Existence of Pareto Fronts}\label{section:5.3}
In section \ref{section:2}, two primary objectives were set for the problem of this thesis - maximizing the coverage of the orthographic regions and minimizing the overlap among them. As per the observations so far, it can be safely inferred that these two objectives result in a trade-off. If a solution, leads to greater coverage, then the overlap is high and vice versa. The same can be observed from figures \ref{fig:52} and \ref{fig:53}. The cost functions $F_2$ and $G_1$ lead to greater area coverage but tend to have the greatest overlap while the opposite happens for costs $F_3$ and $G_2$. Thus it may be of interest to study these two objectives and the relation among them.

By definition, for a nontrivial multi-objective optimization problem, no single solution exists that simultaneously optimizes each objective. In that case, the objective functions are said to be conflicting, and there exists a (possibly infinite) number of Pareto optimal solutions. So, if a multi-objective optimization problem is formulated based on the primary objectives, instead of obtaining a single solution, a set of pareto optimal solutions can be generated. If both the objectives are given equal priority, then all the pareto-optimal solutions can be considered to be 'equally good'. 

Let the multi-objective optimization problem be formulated as,
\begin{equation}
    \min_{X \in \mathcal{X}}\left(f_1(X), f_2(X)\right)
\end{equation}
where, $\mathcal{X}$ is the feasible set of decision vectors. Here the two objective functions can be given as,
\begin{equation*}
\begin{aligned}
    f_1(X) & = \mathop{\mathlarger{\mathlarger{\sum}}\limits_{i = 1}^{N}  \mathlarger{\mathlarger{\sum}}\limits_{j = 1}^{N}}_{i < j}  \bigg[r(X_i) + r(X_j) - \norm{X_i - X_j}\bigg]_{+}, \\
    f_2(X) & = \mathop{\mathlarger{\mathlarger{\sum}}\limits_{i = 1}^{N}  \mathlarger{\mathlarger{\sum}}\limits_{j = 1}^{N}}_{i < j}  \bigg[r(X_i) + r(X_j) - \norm{X_i - X_j}\bigg]_{+} - \mathlarger{\mathlarger{\sum}}\limits_{i = 1}^{N} \pi\left[ \frac{r(X_i)}{R}\right]^2.
\end{aligned}
\end{equation*}
$f_1(X)$ is the objective that minimizes overlap and $f_2(X)$ maximizes coverage, as discussed before. Here a feasible solution lies within the boundary constraints of the surface. For the above multi-objective optimization problem, the set of feasible solutions are generated and plotted in \textsc{Matlab} (fig \ref{fig:54}). The set of solutions form a pareto front as shown in the figure below. The surface taken in this function is $z = f(x,y) = cos(x) + cos(y)$, and the parameters $d = 3$, $N = 20$,and $\epsilon = 10^{\circ}$.

\begin{figure}[!htb]
  \centering
  \includegraphics[width=0.7\linewidth]{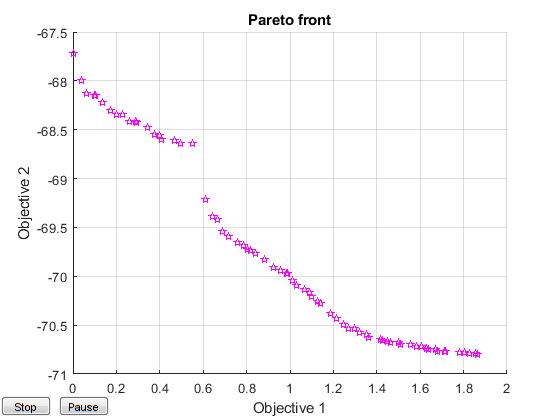}
  \caption{\textit{The pareto optimal solution front or \textbf{pareto front} formed by the solutions shown in pink. Here $Objective 1 = f_1$ and $Objective 2 = f_2$.}}
  \label{fig:54}
\end{figure}

Figure \ref{fig:54} demonstrates the existence of a pareto front of solutions for the present optimization problem, and hence it bolsters the claim that solutions are not unique for the two primary objectives to be optimized. If no priority was given among the objectives, all solutions on the front are equally preferable.

\section{Comparison of Circle Filling Algorithms}\label{section:5.4}
Two circle filling algorithms were stated in section \ref{section:3} of the previous chapter- \textit{\textbf{Batch Filling}} and \textit{\textbf{Sequential Filling}}. The results stated in this chapter so far have been for the \textit{batch filling} algorithm. The \textit{sequential filling} algorithm has been implemented in \textsc{Matlab} and the following figures \ref{fig:55} and \ref{fig:56} are the plot of area covered and overlap with the gradual filling of circles. 

Here the surface $S$ is taken to be $z = f(x,y) = cos(x)+cos(y)$, the imaging height is taken to be $d = 3$, the effective angular FOV is $\epsilon = 10^{\circ}$ and the surface bounds are $-5 \leq x \leq 5$ and $-5 \leq y \leq 5$. Also, at each step the initialization is done randomly within the bounds. Also, two different cost functions have been used for demonstration - $F_2$ (objective based on higher weightage for coverage) and $F_4$ (objective based on varying weightage to overlap and coverage). For $F_4$, the initial priority is given to coverage and as the number of circles increases, the weightage for coverage decreases, ($w_1 = 5$, $w_2 = \frac{0.5}{N}$, ref \ref{section:2}). For the optimization step, the \textit{GlobalSearch} solver was used for each additional increment in $N$. 

\begin{figure}[!htb]
  \centering
  \includegraphics[width=0.6\linewidth]{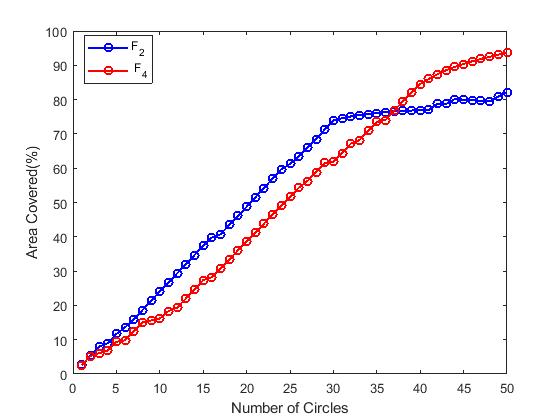}
  \caption{\textit{The area covered(\%) for cost functions $F_2$ and $F_4$ for the sequential filling algorithm for $N = 1$ to $50$ and step size $n=1$.}}
  \label{fig:55}
\end{figure}

\begin{figure}[!htb]
  \centering
  \includegraphics[width=0.6\linewidth]{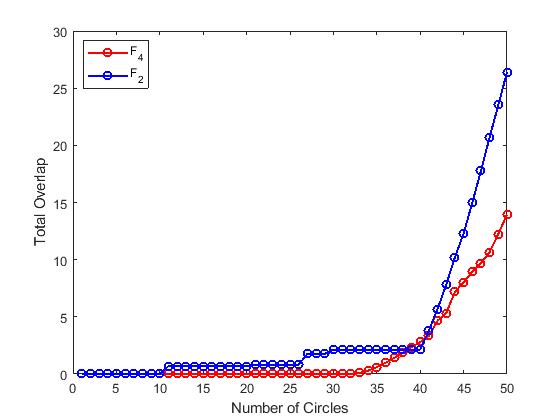}
  \caption{\textit{The total overlap for cost functions $F_2$ and $F_4$ for the sequential filling algorithm for $N = 1$ to $50$ and step size $n=1$.}}
  \label{fig:56}
\end{figure}
\begin{figure}[!htb]
  \centering
  \includegraphics[width=0.6\linewidth]{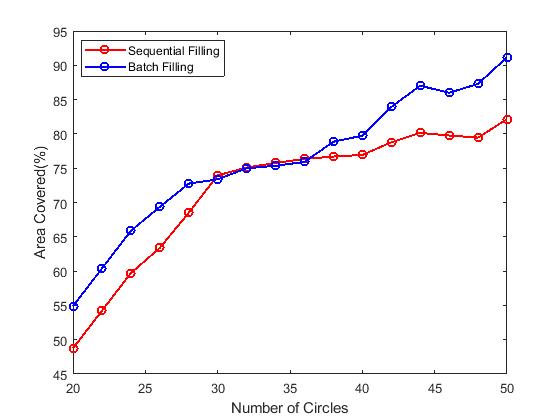}
  \caption{\textit{The total area(\%) for cost functions $F_2$ for the sequential filling and batch filling algorithms for $N = 20$ to $50$ and step size $n=2$.}}
  \label{fig:57}
\end{figure}
\begin{figure}[!htb]
  \centering
  \includegraphics[width=0.6\linewidth]{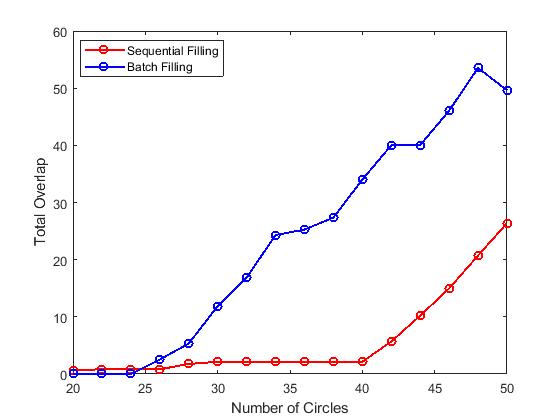}
  \caption{\textit{The total overlap for cost functions $F_2$ for the sequential filling and batch filling algorithms for $N = 20$ to $50$ and step size $n=2$.}}
  \label{fig:58}
\end{figure}

\pagebreak
For comparison, the area covered(\%) and total overlap of the circles for the cost function $F_2$ evaluated by the \textit{Batch Filling} and \textit{Sequential Filling} algorithms are plotted together in figures \ref{fig:57} and \ref{fig:58}.

As can be observed from the figures, \textit{batch filling}(BF) outperforms \textit{sequential filling}(SF) at every step of the evaluation in terms of area coverage. This is due to the fact that in BF, the optimization is done for all the circles at each step, whereas in SF, the optimization is done only for the new added circle(s) while the previously evaluated centers are fixed. It is also interesting to note that the overlap is much higher for BF because of the same reason and because cost function $F_2$ is coverage preferential. However, because in BF the optimization is to be carried out for all the circles at each step, the time consumed increases exponentially and is much greater compared to SF. So, in conclusion sequential filling is a better choice if computation time is a concern, otherwise, batch filling is much preferable for better results.

\section{Surface Division}\label{section:5.5}
In section \ref{section:4} of the previous chapter, a method for dividing the surface into orthogonal regions after the evaluation of the circle locations and the circle boundaries was given. Equation 4.8 and 4.9 formulates the method of deciding the region of a point on the surface located in the overlap region of two or more circles. In other words, the decision boundaries need to be calculated for the regions of overlap. 

The following figure \ref{fig:59} shows the decision boundaries calculated for $N = 50$ circular regions evaluated for $f(x,y) = cos(x) + cos(y)$, for an imaging height $d=3$, for $\epsilon = 10^{\circ}$ and for bounds $(x,y) \in [-5,5]$. The \textit{GlobalSearch} solver was used for evaluation. 

\begin{figure}[!htb]
  \centering
  \includegraphics[width=0.8\linewidth]{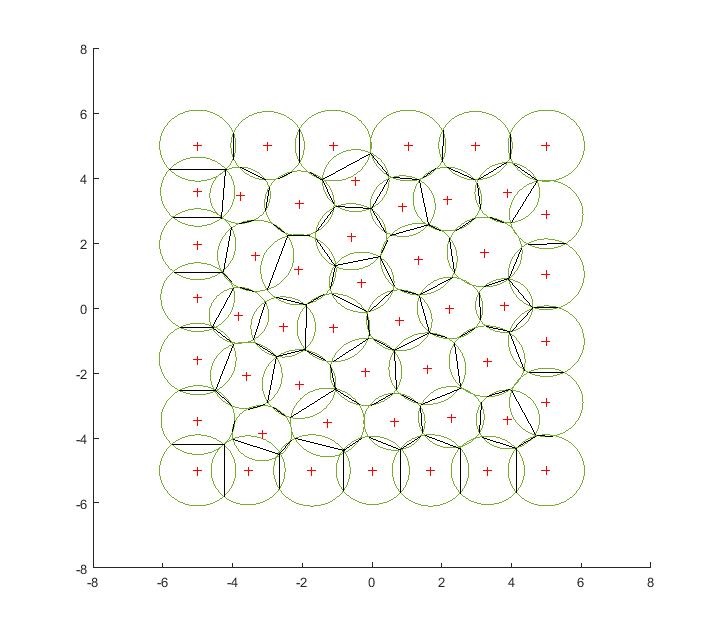}
  \caption{\textit{The evaluated circular boundaries are shown in green, the circle centers are shown in red and the decsion boundaries in the overlapping regions are shown in black.}}
  \label{fig:59}
\end{figure}
\pagebreak

\section{Effect of Imaging Distance}\label{section:5.6}
\begin{figure}[!htb]
\centering
\subfigure[The solution circles for $d = 2$ and $\epsilon = 10^{\circ}$]{\label{fig:sub5}\includegraphics[width=0.48\linewidth]{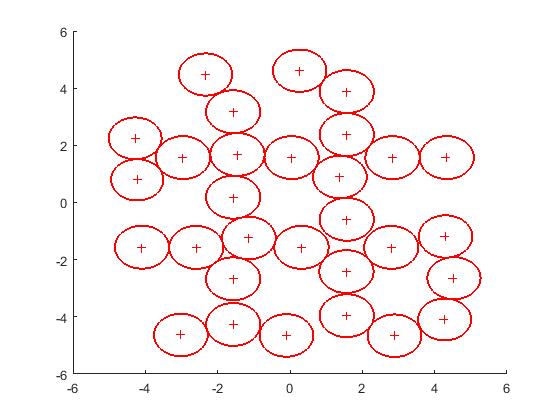}}
\hfill
\subfigure[The solution circles for $d = 3$ and $\epsilon = 10^{\circ}$]{\label{fig:sub6}\includegraphics[width=0.48\linewidth]{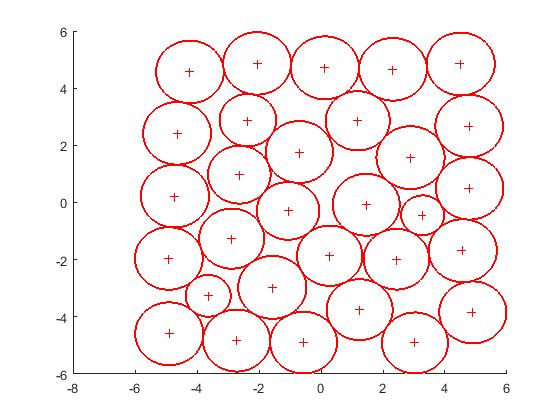}}
\subfigure[The solution circles for $d = 4$ and $\epsilon = 10^{\circ}$]{\label{fig:sub6}\includegraphics[width=0.48\linewidth]{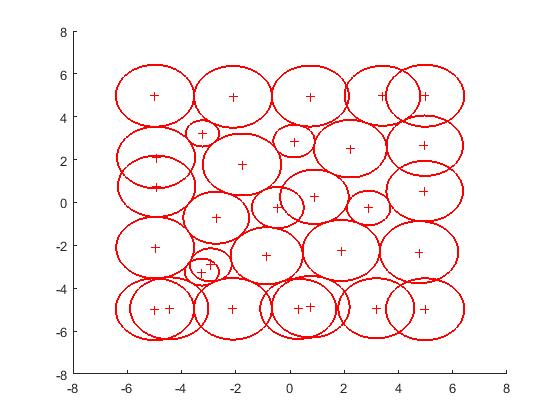}}
\caption{\textit{The solution circles for increasing values of imaging height($d$).}}
\label{fig:61}
\end{figure}
In the figure \ref{fig:61}, the surface $S$ considered for evaluation is $z = f(x,y)$, $(x,y) \in [-5,5]$. The local solutions are obtained using the \textit{fmincon} solver in \textsc{Matlab} for a random initialization and cost function $F_3$ is used. 

As discussed in \ref{section:2}, the maximum boundary circle radius is given by $R = d\cdot tan(\epsilon)$, which increases with the increase in value of imaging height($d$). It can be observed from figure \ref{fig:61} that, as the imaging height($d$) increases, the orthographic regions tend to cover more area, and thus make larger circles. Also, as a consequence, fewer circles are required to cover the entire surface and the net overlap among the regions is higher. But there is an upper limit to $d$ which has been discussed in chapter \ref{Chapter3}.

\section{Effect of $\epsilon$ - Angular FOV}\label{section:5.7}
\begin{figure}[!htb]
\centering
\subfigure[The solution circles for $d = 2$ and $\epsilon = 5^{\circ}$]{\label{fig:sub5}\includegraphics[width=0.48\linewidth]{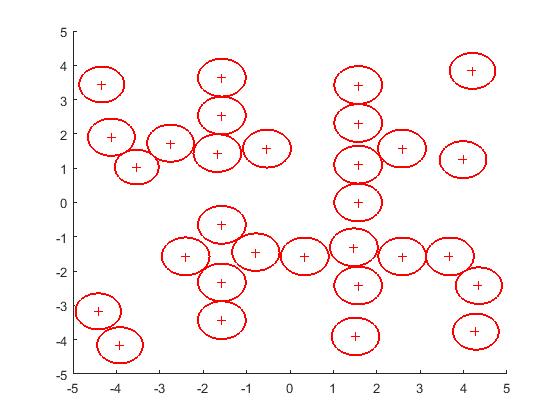}}
\hfill
\subfigure[The solution circles for $d = 3$ and $\epsilon = 10^{\circ}$]{\label{fig:sub6}\includegraphics[width=0.48\linewidth]{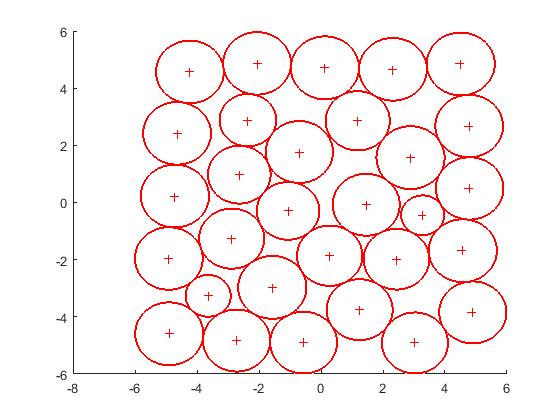}}
\subfigure[The solution circles for $d = 4$ and $\epsilon = 12.5^{\circ}$]{\label{fig:sub6}\includegraphics[width=0.48\linewidth]{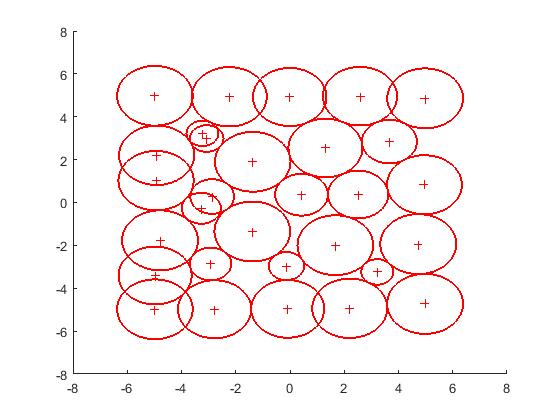}}
\caption{\textit{The solution circles for increasing values of effective angular FOV($\epsilon$).}}
\label{fig:62}
\end{figure}
In the figure \ref{fig:62}, the surface $S$ considered for evaluation is $z = f(x,y)$, $(x,y) \in [-5,5]$. The local solutions are obtained using the \textit{fmincon} solver in \textsc{Matlab} for a random initialization and cost function $F_3$ is used. 

As discussed in \ref{section:2}, the maximum boundary circle radius is given by $R = d\cdot tan(\epsilon)$, which increases with the increase in value of useful angular FOV ($\epsilon$). It can be observed from figure \ref{fig:62} that, as the value of $\epsilon$ increases, the $\epsilon$-orthographic regions tend to cover more area, and thus larger circles. Also, as a consequence, less number of circles are required to cover the entire surface and the net overlap among the regions is higher. But there is an upper limit to $\epsilon$ and it must be kept small, otherwise the view obtained cannot be considered orthographic.

\section{Effect of Curvature}\label{section:5.8}
\begin{figure}[!htb]
\centering
\subfigure[Solution circles for $f(x,y) = \frac{1}{2}\cdot\left(cos(x) + cos(y)\right)$]{\label{fig:sub5}\includegraphics[width=0.48\linewidth]{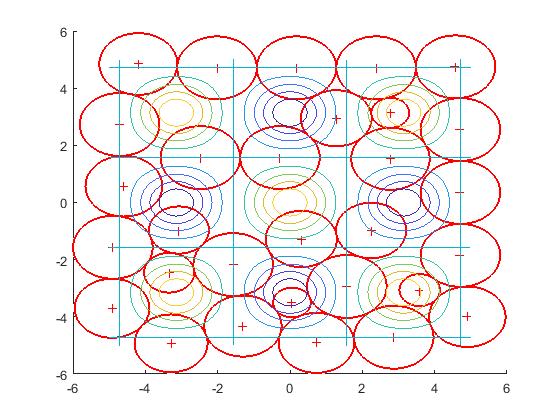}}
\hfill
\subfigure[The solution circles for $f(x,y) = cos(x) + cos(y)$]{\label{fig:sub6}\includegraphics[width=0.48\linewidth]{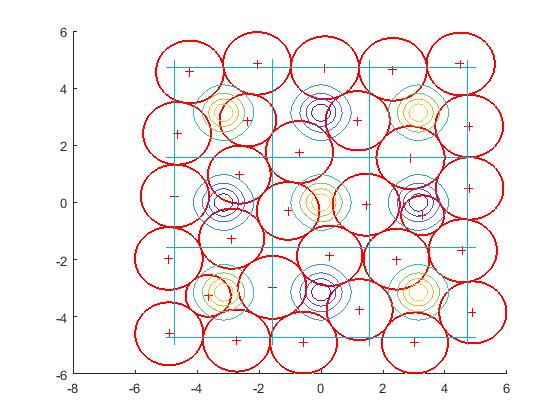}}
\subfigure[The solution circles for $f(x,y) = 2\cdot\left(cos(x) + cos(y)\right)$]{\label{fig:sub6}\includegraphics[width=0.48\linewidth]{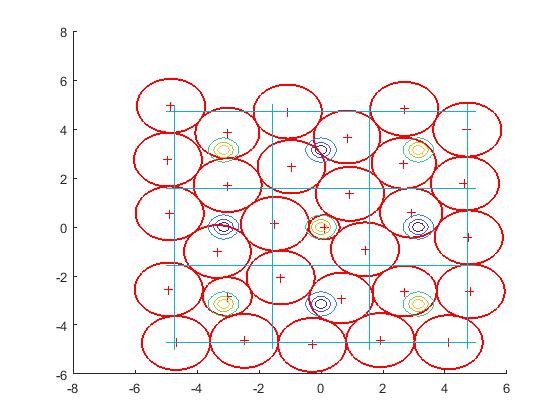}}
\caption{\textit{The solution circles for increasing values of surface curvature plotted along with the gaussian curvature contours.}}
\label{fig:63}
\end{figure}
In the figure \ref{fig:63}, the surfaces $S$ considered for evaluation is $z = f(x,y)$, $(x,y) \in [-5,5]$. The imaging height is set to $d = 3$ and the effective angular FOV is set to $\epsilon = 10^{\circ}$.
The local solutions are obtained using the \textit{fmincon} solver in \textsc{Matlab} for a random initialization and cost function $F_3$ is used. 

As discussed in \ref{section:2}, the approximated boundary circle radius is a function of surface gaussian curvature, or rather the absolute normalized curvature ($\frac{|K(x,y)|}{K_{max}}$). It can be observed from figure \ref{fig:63} that, although the surface curvatures increase by a factor, the absolute normalized gaussian curvatures do not change and hence the circle sizes do not change drastically. However, as the curvatures increase, the peaks become sharper and the flatter regions increase in size. So, more circles can fit in the flat regions of the surface, and hence larger circles are obtained for the function considered. And thus the surface can be filled faster, i.e. with less number of circles. However, if for a function, flat regions are very small, then the circle size decreases with increase in curvature, and more circles will be required to cover the entire surface.  
\chapter{Conclusion and Future work}

\label{Chapter6}

\lhead{Chapter 6. \emph{Conclusion} }

\section{Conclusion}
Orthographic imaging is a crucial tool for terrain survey or terrain mapping. Although technological improvements have been made widely in the devices to capture visual or other sensor data of a surface, a proper and efficient algorithm for reconstructing the surface topography and creating an orthographic projection of the terrain is lacking. This thesis is a compilation of studying and analyzing this problem and proposing novel methods for solving it. A technique for generating topographical surface from elevation maps has been proposed. The effects of imaging height and angular field of view for capturing orthographic views have been formulated and analyzed in detail. A method for calculating orthographic boundaries have been proposed and demonstrated. Different methods of approximating the orthographic boundaries have been proposed and compared. 

Several methods for calculating optimal locations on a surface to capture orthographic views has been formulated and illustrated. Different cost functions for solving the optimization problem has been proposed and they have been compared on the basis of efficiency measures. The different algorithms proposed for computing the optimal points have been analyzed and compared. A method for choosing an algorithm based on priority of objectives, computation time and target application has been proposed. 

Every element of the problem has been mathematically formulated, empirically analyzed and visually demonstrated in \textsc{Matlab}. The obtained results were compared in detail and the choice of algorithms has been suggested on the basis of results.

\section{Future Work}
\begin{itemize}
    \item Better approximations of orthographic boundaries may be explored, and in case of which, how and by how much the results are affected may be studied.
    \item The algorithms can be combined with multiple objectives and how that can facilitate surface reconstruction may be explored.
    \item Methods for computing orthographic views by combining visual data with that from devices not capturing visual data can be explored and incorporated.
    \item If better computational resources are available, the result of the algorithms without boundary approximations may be evaluated and compared with the ones presented in this thesis.
    \item Formulation of faster algorithms for real-time computation and demonstration should be explored.
\end{itemize}

%\input{Chapters/Chapter7} 

%----------------------------------------------------------------------------------------
%	THESIS CONTENT - APPENDICES
%----------------------------------------------------------------------------------------

%\addtocontents{toc}{\vspace{2em}} % Add a gap in the Contents, for aesthetics
%
%\appendix % Cue to tell LaTeX that the following 'chapters' are Appendices
%
% Include the appendices of the thesis as separate files from the Appendices folder
% Uncomment the lines as you write the Appendices

%\input{Appendices/AppendixA}
%\input{Appendices/AppendixB}
%\input{Appendices/AppendixC}

\addtocontents{toc}{\vspace{2em}} % Add a gap in the Contents, for aesthetics

\backmatter

%----------------------------------------------------------------------------------------
%	BIBLIOGRAPHY
%----------------------------------------------------------------------------------------
\nocite{*}
\label{Bibliography}

\lhead{\emph{Bibliography}} % Change the page header to say "Bibliography"

 % Use the "custom" BibTeX style for formatting the Bibliography
% \bibliographystyle{ieeetr} 
\bibliography{bibli} % The references (bibliography) information are stored in the file named "Bibliography.bib"
\bibliographystyle{IEEEtran}

\end{document}